\documentclass[journal]{IEEEtran}

\ifCLASSINFOpdf
\else
\fi
\usepackage{algpseudocode}
\usepackage{algorithm}
\usepackage[dvips]{graphicx}
\usepackage{latexsym}
\usepackage{amssymb}
\usepackage{amsmath}
\usepackage{multirow}
\usepackage{xcolor}
\usepackage{lipsum}
\usepackage{multicol}
\usepackage{enumerate}
\usepackage{algorithm}
\usepackage{algorithmicx}
\usepackage{algpseudocode}
\usepackage{amsmath}
\usepackage{subfig}
\usepackage[utf8]{inputenc}
\usepackage{float}
\usepackage{bm}
\begin{document}
%
\title{Interference Exploitation via Symbol-Level Precoding: Overview, State-of-the-Art and Future Directions}
%
%
%

\author{Ang Li,~\IEEEmembership{Member,~IEEE}, Danilo~Spano,~\IEEEmembership{Member,~IEEE,}
   Jevgenij~Krivochiza,~\IEEEmembership{Student Member,~IEEE,}
    Stavros~Domouchtsidis,~\IEEEmembership{Student Member,~IEEE,}
	Christos~G.~Tsinos,~\IEEEmembership{Member,~IEEE,} Christos Masouros,~\IEEEmembership{Senior Member,~IEEE}, Symeon Chatzinotas,~\IEEEmembership{Senior Member,~IEEE}, Yonghui Li,~\IEEEmembership{Fellow,~IEEE},\\ Branka Vucetic,~\IEEEmembership{Fellow,~IEEE}, and Bj\"{o}rn Ottersten,~\IEEEmembership{Fellow,~IEEE}

\thanks{Manuscript received TIME; revised TIME.}
\thanks{A. Li, Y. Li, and B. Vucetic are with the School of Electrical and Information Engineering, University of Sydney, NSW 2006, Australia. (email: \{ang.li2, yonghui.li, branka.vucetic\}@sydney.edu.au)}
\thanks{D. Spano, J. Krivochiza, S. Domouchtsidis, C. G. Tsinos, S. Chatzinotas and B. Ottersten are with the Interdisciplinary Centre for Security, Reliability, and Trust, University of Luxembourg, 1855 Luxembourg, Luxembourg. (e-mail: \{danilo.spano, jevgenij.krivochiza, stavros.domouchtsidis, christos.tsinos, symeon.chatzinotas, bjorn.ottersten\}@uni.lu)}
\thanks{C. Masouros is with the Department of Electronic and Electrical Engineering, University College London, Torrington Place, London, WC1E 7JE, UK. (email: c.masouros@ucl.ac.uk)}
\thanks{This work was supported in part by the Engineering and Physical Sciences Research Council (EPSRC) Project under Grant project EP/R007934/1, in part by FNR, Luxembourg under the projects INTER CI-PHY and CORE ECLECTIC.}
}



\maketitle

\begin{abstract}
Interference is traditionally viewed as a performance limiting factor in wireless communication systems, which is to be minimized or mitigated. Nevertheless, a recent line of work has shown that by manipulating the interfering signals such that they add up constructively at the receiver side, known interference can be made beneficial and further improve the system performance in a variety of wireless scenarios, achieved by symbol-level precoding (SLP). This paper aims to provide a tutorial on interference exploitation techniques from the perspective of precoding design in a multi-antenna wireless communication system, by beginning with the classification of constructive interference (CI) and destructive interference (DI). The definition for CI is presented and the corresponding mathematical characterization is formulated for popular modulation types, based on which optimization-based precoding techniques are discussed. In addition, the extension of CI precoding to other application scenarios as well as for hardware efficiency is also described. Proof-of-concept testbeds are demonstrated for the potential practical implementation of CI precoding, and finally a list of open problems and practical challenges are presented to inspire and motivate further research directions in this area.
\end{abstract}


%
\IEEEpeerreviewmaketitle

\section{Introduction}

\IEEEPARstart{P}{RECODING} is able to support data transmissions to multiple receivers simultaneously in multi-antenna wireless communication systems, which has attracted significant interest in their development towards 5G \cite{intro-1}. The term `precoding' usually refers to the transmit signal design that directs the desired data symbols to the intended users while limiting the inter-user interference, by exploiting the channel state information (CSI) and potentially the information of the data symbols. In the literature, the dirty paper coding (DPC) technique is known to be capable of achieving the channel capacity theoretically \cite{intro-2}. Despite its optimality, DPC is difficult to implement in practical wireless communication systems, due to (i) the impractical assumption of an infinite source alphabet and (ii) the prohibitive computational complexity incurred by sequential encoding. Therefore, linear precoding methods, where the precoded signals are linear combinations of the information symbols, have become appealing and attracted more research attention because of their low complexity \cite{intro-3}\nocite{intro-4}\nocite{intro-16}\nocite{intro-17}-\cite{intro-5}. In the literature, while the maximum ratio transmission (MRT) precoding offers the lowest computational cost \cite{intro-3}, it does not fully eliminate the multi-user interference, which leads to an error floor at medium-to-high signal-to-noise ratio (SNR) regions. Zero-forcing (ZF) precoding is able to improve the performance of MRT precoding by fully eliminating the multi-user interference via inverting the channel \cite{intro-4}, whose performance can be further improved via the regularized ZF (RZF) precoding by including a regularization factor in the matrix inversion, which alleviates the noise amplification effect that ZF precoding suffers \cite{intro-5}.

In addition to these closed-form precoders, linear precoding methods based on optimization have received increasing research attention recently because of their flexibility to achieve various performance targets, where the most two popular design targets are power minimization (PM) and signal-to-interference-plus-noise ratio (SINR) balancing (SB) \cite{intro-6}\nocite{intro-7}\nocite{intro-8}\nocite{intro-9}\nocite{intro-10}\nocite{intro-11}\nocite{intro-12}\nocite{intro-13}\nocite{intro-14}\nocite{intro-15}-\cite{intro-18}. For unicast applications where the base station (BS) transmits individual information to each receiver, PM aims to minimize the total transmit power at the BS subject to a common minimum SINR target for all the receivers \cite{intro-6} or an individual SINR target for each user \cite{intro-7}, while SB targets at maximizing the minimum SINR for each receiver while satisfying a total transmit power requirement \cite{intro-8} or a per-antenna power constraint \cite{intro-14} at the BS. Given the capability of adaptation to various wireless communication scenarios, optimization-based precoding schemes have been extended to a variety of research areas such as cognitive radio (CR) \cite{cr-1}\nocite{cr-2}\nocite{cr-3}\nocite{cr-4}\nocite{cr-5}\nocite{cr-6}\nocite{cr-7}\nocite{cr-8}\nocite{cr-9}\nocite{cr-10}\nocite{cr-11}\nocite{cr-12}\nocite{cr-13}\nocite{cr-14}\nocite{cr-15}\nocite{cr-16}\nocite{cr-17}\nocite{cr-18}\nocite{cr-19}\nocite{cr-20}\nocite{cr-21}-\cite{cr-22}, simultaneous wireless information and power transfer (SWIPT) \cite{swipt-1}\nocite{swipt-2}\nocite{swipt-3}\nocite{swipt-4}\nocite{swipt-5}\nocite{swipt-6}\nocite{swipt-7}\nocite{swipt-8}\nocite{swipt-9}\nocite{swipt-10}\nocite{swipt-11}\nocite{swipt-12}\nocite{swipt-13}\nocite{swipt-14}\nocite{swipt-15}\nocite{swipt-16}\nocite{swipt-17}\nocite{swipt-18}\nocite{swipt-19}-\cite{swipt-20}, physical-layer (PHY) security \cite{security-1}\nocite{security-2}\nocite{security-3}\nocite{security-4}\nocite{security-5}\nocite{security-6}\nocite{security-7}\nocite{security-8}\nocite{security-9}\nocite{security-10}\nocite{security-11}\nocite{security-12}\nocite{security-13}\nocite{security-14}\nocite{security-15}\nocite{security-16}\nocite{security-17}\nocite{security-18}-\cite{security-19}, full-duplex (FD) communications \cite{full-duplex-0}\nocite{full-duplex-1}\nocite{full-duplex-2}\nocite{full-duplex-3}\nocite{full-duplex-4}\nocite{full-duplex-5}\nocite{full-duplex-6}\nocite{full-duplex-7}\nocite{full-duplex-8}\nocite{full-duplex-9}\nocite{full-duplex-10}\nocite{full-duplex-11}\nocite{full-duplex-12}\nocite{full-duplex-13}\nocite{full-duplex-14}\nocite{full-duplex-15}\nocite{full-duplex-16}\nocite{full-duplex-17}\nocite{full-duplex-18}\nocite{full-duplex-19}\nocite{full-duplex-20}\nocite{full-duplex-21}\nocite{full-duplex-22}-\cite{full-duplex-23}, etc., which will be overviewed in the corresponding chapters in the following.

\begin{figure*}
	\centering
	\includegraphics[scale=0.25]{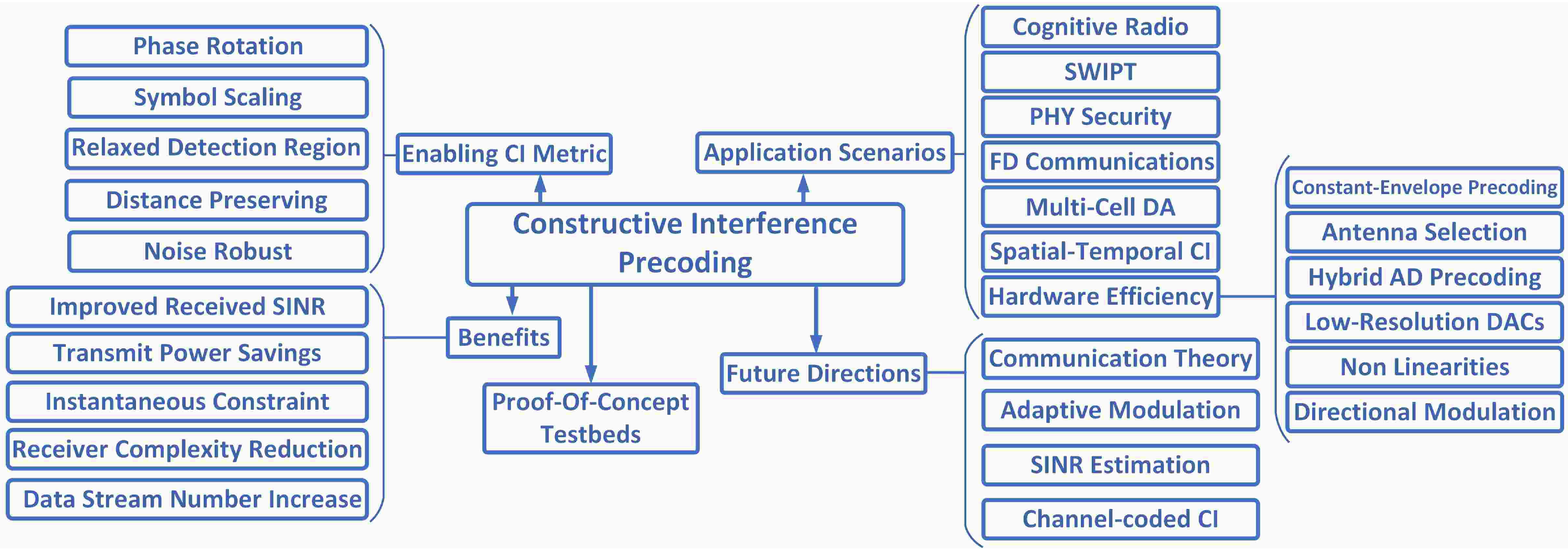}
	\caption{Various aspects of interference exploitation via symbol-level precoding}
\end{figure*}

For both closed-form linear precoding methods \cite{intro-3}-\cite{intro-5} and optimization-based schemes \cite{intro-6}-\cite{intro-14} described above, it is observed that only the information of the channel is exploited for the precoding design, and these precoding methods all treat interference as a detrimental effect. Nevertheless, it has already been observed in non-linear precoding methods such as Tomlinson-Harashima precoding (THP) \cite{intro-thp-1}\nocite{intro-thp-2}-\cite{intro-thp-3} and vector perturbation (VP) precoding \cite{intro-vp-1}\nocite{intro-vp-2}\nocite{intro-vp-3}\nocite{intro-vp-4}\nocite{intro-vp-5}\nocite{intro-vp-6}\nocite{intro-vp-7}-\cite{intro-vp-8} that both the CSI and the data symbols have been included in the symbol-by-symbol precoding design, i.e., the information of the data symbols is also exploited. However, the problem for non-linear precoding schemes is that they are still difficult to be implemented in practical wireless communication systems, due to the complicated encoding and decoding process that leads to unfavorable computational costs. Therefore, it is natural to ask: Is it possible for linear precoding methods to potentially exploit the information of the data symbols as well, or more specifically exploit the interference based on the knowledge of the data symbols to further improve the performance?

To answer the above question, this paper provides a tutorial on a recently proposed concept termed `constructive interference' (CI) and the corresponding CI precoding techniques, as well as their applications to a number of current and future wireless communication scenarios, as illustrated in Fig.~1. Compared with a previous survey paper \cite{ci-24} which focuses on symbol-level precoding (SLP) and its comparison with block-level and multicast precoding, the focus of this tutorial paper is on the definition, characterization and exploitation of the CI effect in a variety of wireless scenarios. We begin with a brief introduction on CI, its potential benefits and current limitations in Section I. Section II then introduces the classification and mathematical characterization of CI for various modulation types, based on which Section III formulates the optimization problems for CI exploitation, whose solution can be obtained via convex optimization tools. Section IV describes the applications of CI exploitation techniques in traditional small-scale multiple-input multiple-output (MIMO) systems, and Section V extends the application to large-scale antenna systems for hardware efficiency. Section VI describe the proof-of-concept testbed for practical implementation of CI exploitation via SLP, developed by University College London and University of Luxembourg, respectively. Section VII discusses some open problems and challenges to be explored, followed by Section VIII that concludes the paper.

For clarity, the following notations are employed in the following chapters of this paper: $a$, $\bf a$, and $\bf A$ denote scalar, column vector and matrix, respectively. ${( \cdot )^*}$, ${( \cdot )^{\text T}}$, ${( \cdot )^{\text H}}$, ${( \cdot )^{-1}}$, and ${( \cdot )^{+}}$ denote conjugate, transposition, conjugate transposition, inverse and pseudo-inverse of a matrix, respectively. ${\text {diag}} \left(  \cdot  \right)$ is the transformation of a column vector into a diagonal matrix, and $\otimes$ is the Kronecker product. $\left|  \cdot  \right|$ denotes the absolute value of a real number or the modulus of a complex number, $\left\|  \cdot  \right\|_2$ denotes the $\ell_2$-norm, and ${\left\| {\cdot} \right\|_\infty }$ denotes the uniform norm. $\Re \left\{ \cdot \right\}$ and $\Im \left\{ \cdot \right\}$ denote the real and imaginary part of a complex scalar, vector or matrix, respectively. ${\bf I}_K$ denotes the $K \times K$ identity matrix, and $j$ denotes the imaginary unit. ${{\cal C}^{n \times n}}$ and ${{\cal R}^{n \times n}}$ represent the sets of $n\times n$ complex- and real-valued matrices, respectively. ${\text {card}}\left\{  \cdot  \right\}$ denotes the cardinality of a set.

\subsection{Interference in Wireless Communications - Is It All Harmful?}
Traditionally, interference is usually viewed as a performance limiting factor in wireless communication systems. In a typical multi-user transmission, the existence of interference is based on the observation that the transmit signals for different users are superimposed in wireless communication channels. Precoding strategies are designed based on the fact that, with CSI known at the BS and potentially with the information of the data symbols as well, multi-user interference is able to be predicted prior to transmission. In fact, information theoretical analysis in \cite{intro-2} shows that when CSI is available at the transmitter, known interference will not affect the capacity of the broadcast channel. More specifically, the DPC method implies that it is optimal to code along interference, instead of attempting to mitigate or cancel interference. Nevertheless, the majority of existing linear precoding schemes still aim to eliminate, avoid or limit the interference \cite{intro-3}-\cite{intro-12}. In these traditional precoding schemes, the precoding matrix is designed based on the CSI only and therefore operate on a block level. In other words, the same precoding matrix is applied across a block of symbols and is updated when the channel changes. This means that only the power of the interference can be controlled, which leads to the statistical view that the effect of interference is similar to noise. On the other hand, if we observe interference from an instantaneous instead of statistical point of view, recent studies have shown that CI precoding via SLP is able to control both the power and the direction of the interfering signals on the received complex plane on a symbol level, such that the interference can act as an additional source of the desired signal power and contribute to the symbol detection, which therefore further improves the system performance \cite{cf-ci-1}\nocite{cf-ci-2}\nocite{cf-ci-3}\nocite{cf-ci-4}\nocite{ci-1}\nocite{ci-2}\nocite{ci-3}\nocite{ci-4}\nocite{ci-5}\nocite{ci-6}\nocite{ci-7}\nocite{ci-8}\nocite{ci-9}\nocite{ci-10}\nocite{ci-11}\nocite{ci-12}\nocite{ci-13}\nocite{ci-14}\nocite{ci-15}\nocite{ci-16}\nocite{ci-17}\nocite{ci-18}\nocite{ci-19}\nocite{ci-20}\nocite{ci-21}\nocite{ci-22}\nocite{ci-23}\nocite{ci-24}\nocite{ci-25}\nocite{ci-26}\nocite{ci-27}\nocite{ci-33}\nocite{ci-28}\nocite{ci-29}\nocite{ci-30}\nocite{ci-31}-\cite{ci-32}. Based on the above description, interference exploitation techniques are foreseen to be most useful in systems where interference can be predicted and manipulated. To motivate the exploitation of interference in precoding designs, we firstly present illustrative examples to demonstrate how instantaneous interference can be divided into CI and destructive interference (DI) below, followed by the systematic CI characterization in Section II.

Let's first consider a simple example where the desired symbol $u$ is from a nominal BPSK constellation, and without loss of generality we assume $u=1$ \cite{ci-book-1}. We express the received signal as 
\begin{equation}
y=u+i+n=r+n,
\label{eq_1}
\end{equation}
where $i$ is the interfering signal, $r$ denotes the received signal excluding noise, and $n$ denotes the additive noise at the receiver side. We consider two distinct cases: (i) $i>0$ and (ii) $i<0$. When $i>0$, the resulting noiseless received signal $r>1$, which means that the interference has pushed $r$ further away from the detection threshold of BPSK, compared to the original data symbol $u$. In this case, the interfering signal $i$ contributes to the useful signal power and is in fact `constructive'. Given a fixed noise power, $y=u+i+n$ is more likely to be correctly detected than the interference-free case $\tilde y=u+n$, and an improved performance can be expected. On the other hand, when $i<0$, the interfering signal causes the received signal $r$ to move closer to the detection threshold, where the interfering signal reduces the useful signal power and is therefore `destructive'. In this case, the noiseless received signal $r=u+i$ is more vulnerable to noise than $\tilde r=u$. 

\begin{figure}[!b]
	\centering
	\includegraphics[scale=0.45]{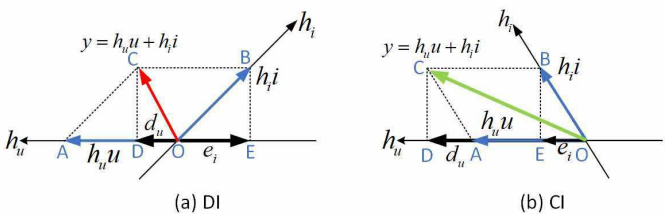}
	\caption{Geometrical representation of CI and DI}
	\label{Geo_CI}
\end{figure}

The above examples have only considered the effect of interfering symbols. To make the concept of interference exploitation more explicit, in the following example we further take the effect of wireless channels into account. In this example, we consider a geometrical representation of an interference scenario with random channels, as shown in Fig.~\ref{Geo_CI} below, where for simplicity we still assume that $u=1$ is the desired data symbol, $i=1$ is the data symbol from the interferer, $h_u$ denotes the wireless channel between the transmitter and the receiver, while $h_i$ is the channel between the interferer and the receiver, respectively. Accordingly, the received signal can be expressed as
\begin{equation}
y=h_u u + h_i i,
\label{eq_2}
\end{equation}
where we have assumed a noiseless case to focus on the effect of interference. In both subfigures of Fig.~\ref{Geo_CI}, $\vec{OA}=h_u u$, $\vec{OB}=h_i i$, the received signal is $\vec{OC}=y$, and its projection on the axis of $h_u$ is denoted by $\vec{OD}=d_u$, whose amplitude directly determines the detection performance. In Fig.~2 (a), we can observe that the amplitude of the projection $d_u$ is smaller than that of the original transmit signal $h_u u$, and consequently the interference is destructive since it reduces the useful signal power. To take a closer look, we can project the interfering signal $\vec{OB}$ onto the axis of $h_u$, and it is then observed that the direction of the effective interfering signal $\vec{OE}=e_i$ is to the opposite side of the desired data symbol, which is similar to the case of (ii) $i<0$ discussed in the previous example and results in DI. In Fig.~2 (b), on the contrary, the interfering signal is constructive for the desired data symbol $u$, since they add up constructively and yield a received symbol whose amplitude is larger than that of the original transmit signal. The above observation implies that for a given data symbol combination, some channel realizations may lead to CI while some other channel realizations may yield DI. 

Based on the above two simple examples, it is important to note that the classification of interference into constructive or destructive depends both on the data symbol combination and the CSI, as will be mathematically shown in Section II.

\subsection{CI Exploitation via Symbol-Level Precoding}
With the two examples illustrated above, we are now able to give the definition of CI: CI is the interference that pushes the received signals away from all of their corresponding detection thresholds of the modulated-symbol constellation, which thus contributes to the useful signal power\footnote{Based on this definition, for multi-level modulations only the outer constellation points can exploit CI, which will be discussed later in Section II-B.}. Moreover, to exploit CI in the precoding design, firstly it should be highlighted that CI-based precoding has to be shifted from block-level operation to symbol-level operation, i.e., SLP is the method to achieve the CI effect. It should be noted that SLP is not limited as a method to exploit CI effects only, but also finds its application in hardware-efficient BS architecture, as will be discussed in Section V.

Early works on CI precoding techniques have focused on the adaptation of simple linear precoding methods such as ZF and RZF for CI exploitation \cite{cf-ci-2}, \cite{cf-ci-3}. In \cite{cf-ci-2}, for the first time the instantaneous interference in a MIMO system is characterized and classified into CI and DI, and a selective precoding is proposed where the CI is retained while the DI is cancelled via ZF. A more advanced approach named correlation-rotation precoding is proposed in \cite{cf-ci-3}, where the DI is manipulated and further rotated to be aligned with the desired data symbols such that DI becomes CI. Compared to the selective precoding that exploits interference only when it is constructive, the correlation-rotation precoding directly controls interference such that all the interference for each user becomes constructive in the system.

\begin{table*}
\label{tab:multicol}
\caption{A summary of major CI precoding outputs in the literature}
\begin{center}
\setlength{\arrayrulewidth}{0.2mm}
\renewcommand{\arraystretch}{1.4}
\scalebox{1.05}
{
\begin{tabular}{| c | c | c | c | }

    \hline
    {\bf Reference} & {\bf Considered Problem} & {\bf CI Metric} & {\bf Considered Modulation} \\
    \hline
    \cite{cf-ci-2} & Selective CI & Strict phase rotation & PSK \\
    \hline
    \cite{cf-ci-3} & Correlation-rotation CI & Strict phase rotation & PSK \\
    \hline
    \cite{ci-3} & CI-VP & Symbol scaling & PSK \\
    \hline
    \cite{ci-4} & CI-based SB & Non-strict phase rotation & PSK + QAM \\
    \hline
    \cite{ci-5}, \cite{ci-6} & CI-MRT + PM + weighted SB & Strict phase rotation & PSK \\
    \hline
    \cite{ci-7}, \cite{ci-8} & CI-based PM + SB & Non-strict phase rotation & PSK \\
    \hline
    \cite{ci-9}, \cite{ci-10} & CI-based PM + weighted SB & Relaxed detection region & PSK \\
    \hline
    \cite{ci-11}, \cite{ci-12} & CI-based PM & Symbol scaling & QAM \\
    \hline
    \cite{ci-13} & CI-based Per-antenna PM & Strict phase rotation & PSK \\
    \hline
    \cite{ci-14} & CI-MMSE precoding & Non-strict phase rotation & PSK \\
    \hline
    \cite{ci-15}, \cite{ci-20} & CI-based PM + SB & Noise-robust CI & PSK \\
    \hline
    \cite{ci-16} & CI-based weighted per-antenna PM & Strict phase rotation & APSK \\
    \hline
    \cite{ci-18}, \cite{ci-19}, \cite{ci-22}, \cite{ci-23} & CI-based PM + SB & Distance preserving CI region & Any constellation \\
    \hline   
    \cite{ci-25}, \cite{ci-26}, \cite{ci-27} & Closed-form CI solutions & Non-strict phase rotation + Symbol scaling & PSK + QAM \\
    \hline
\end{tabular}
}
\end{center}
\end{table*}

The concept of CI has subsequently been applied to the non-linear THP method in \cite{intro-thp-2}, \cite{intro-thp-3} and VP precoding in \cite{ci-3}. The interference-optimized THP (IO-THP) proposed in \cite{intro-thp-2} introduces a complex scaling to the first user such that the interfering signals are better aligned to the symbols of interest, and by optimizing the complex scaling factor to minimize the power of the modified transmit signals, IO-THP reduces the power loss of the conventional THP schemes. As a step further, the power-efficient THP (PE-THP) method proposed in \cite{intro-thp-3} allows complex scaling for a number of users, instead of for the first user only as in \cite{intro-thp-2}. Compared to IO-THP in \cite{intro-thp-2}, the performance improvements come from the fact that PE-THP allows a larger number of variables to be optimized jointly within the constructive area and the signal-to-noise ratio (SNR) threshold, which generally leads to a better and more power-efficient THP solution. \cite{ci-3} proposes CI techniques in the context of VP precoding by substituting the search for the perturbation vectors with a linear scaling precoder, which is the first optimization-based CI technique that involves a linear symbol-scaling operation based on quadratic programming (QP).

More recently, CI-based precoding techniques have been widely combined with optimization to achieve further performance improvements \cite{ci-5}-\cite{ci-16}. In \cite{ci-5} and \cite{ci-6}, the authors firstly propose a CI-MRT precoding method that improves the performance of correlation-rotation precoding by avoiding the ZF operation. In addition, PM optimization and weighted SB optimization based on CI are further discussed in \cite{ci-6}. It is worth mentioning that for CI precoding designs in \cite{ci-5} and \cite{ci-6}, the received signals are forced to be strictly aligned to the desired data symbols with an increase in the amplitude for achieving CI, which follows the CI metric in \cite{cf-ci-3} and is later shown to be sub-optimal and termed `strict phase rotation' in \cite{ci-25} (Fig. 4a). A more advanced CI metric is introduced in \cite{ci-7} and \cite{ci-8}, where the concept of `constructive region' is characterized for PSK constellations, within which all the interference is shown to be constructive. This relaxed CI metric reveals that it is no longer necessary for the interfering signals to be strictly aligned to the symbols of interest, which leads to further performance gains compared to the `strict phase-rotation' CI metric in \cite{cf-ci-3}, \cite{ci-5} and \cite{ci-6}. This advanced CI metric is later termed `non-strict phase rotation' in \cite{ci-25} (Fig. 4b), and is widely adopted in the subsequent precoding designs for CI exploitation \cite{ci-15}, \cite{ci-20}, \cite{ci-25}-\cite{ci-30} and its applications. Meanwhile, a similar and sub-optimal relaxed CI metric is also presented in \cite{ci-9}, \cite{ci-10}, where the `relaxed detection region' metric that is determined by a phase margin related to the SNR target is introduced. 

It should be noted that the above works \cite{intro-thp-2}, \cite{intro-thp-3}, \cite{cf-ci-1}-\cite{cf-ci-3}, and \cite{ci-5}-\cite{ci-10} have all focused on PSK constellations for CI precoding, while the extension to multi-level modulations such as QAM has recently been discussed in \cite{ci-11}, \cite{ci-12}, \cite{ci-26}, \cite{ci-27}, \cite{ci-29} and \cite{ci-30}, where the data symbols for multi-level modulations are divided into outer symbols that can exploit CI and inner symbols that cannot exploit CI. Interestingly, in contrast to claims that CI precoding may not be promising for higher-order QAM modulations since only the outer constellation points benefit from CI, \cite{ci-27} shows that substantial gains can still be observed even for a 64QAM constellation, which will also be numerically shown in Section III. This is because CI exploitation precoding not only allows the outer constellation points to benefit from CI, but more importantly also reduces the noise amplification effect, which is more prominent for a high-order QAM modulation. CI precoding has further been extended to generic two-dimension constellations with any shape and size in \cite{ci-22}, where the CI metric is termed `distance preserving CI region' (DPCIR). 

Meanwhile, a similar concept coined as `directional modulation' \cite{dm-1}\nocite{dm-2}-\cite{dm-3}, which was studied in the past in the context of analog RF and antenna components, has also emerged as a promising hardware-efficient approach, where the phase and amplitude of the transmitting signal on each antenna are directly designed such that multiple interference-free or interference-limited symbols can be transmitted to the receiver, which will be discussed in more detail in Section V-F. Additional studies on CI precoding, which include iterative closed-form CI solutions, multi-group multicasting, symbol error rate (SER) minimization, per-antenna power constraint, non-linear channels, etc. can be found in \cite{ci-13}-\cite{ci-30}, and we summarize the major research outputs on CI precoding in Table I.

\subsection{Benefits of CI and Symbol-Level Precoding}
With the ability to transform the power of the interfering signals into useful signal power without the need of investing additional transmit power, CI precoding has become an appealing technique for modern and future wireless system design, where energy efficiency has become increasingly important. Obviously, the most prominent advantage for CI precoding over conventional precoding is the significant performance improvements in terms of error rate performance and transmit power savings. By carefully designing the precoding matrix based on the CSI as well as information of the data symbols, interference that is usually regarded as a detrimental effect and needs to be mitigated now becomes beneficial and further contributes to the increase in the useful signal power \cite{ci-8}, which has also been mathematically studied in \cite{ci-12} using an equivalent PHY multicasting model.

In addition to the above most significant advantage, some additional benefits that CI precoding enjoys should also be highlighted in particular.

\subsubsection{Instantaneous Constraints}
Conventional precoding methods operate on a block level, and the constraints such as the SINR target for the PM problems or the transmit power requirement for the SB optimizations are met averaged over a transmission block. While these constraints are indeed satisfied from a statistical point of view, during actual transmission these constraints may be violated for some data symbol combinations while over-satisfied for other data symbol combinations within each transmission block.

As a comparison, CI precoding does not have this issue and guarantees that the constraint is strictly satisfied for each data symbol combination during transmission, because CI precoding works on a symbol level and accordingly the constraints are enforced on a symbol level. This can be mathematically observed in \cite{ci-8} as well as in Section III for optimization problems formulated based on CI precoding, where the data symbols have been included in the constraints.

\subsubsection{Receiver Complexity}
CI precoding provides additional complexity reduction at the receiver side. When traditional block-level optimization-based PM or SB precoding is adopted, since the optimization only focuses on the power of the useful signals without considering the phase-rotation effect, each receiver therefore needs to perform phase estimation and compensate the phase-rotation effect of the channel, before the symbols can be correctly decoded, which may suffer from channel estimation errors \cite{cr-10}.

In contrast to these traditional precoding strategies, the received symbols for CI precoding are all located in the constructive area, and therefore the phase-compensation process is not required any more, when PSK constellations are employed. This means that the symbol decoding process does not require the estimated CSI at the receivers, as only a simple decision stage is sufficient, which therefore avoids the effect of CSI estimation errors on the decoding process. This is particularly important in the case of downlink transmission where the receivers are typically computationally constrained mobile devices. Accordingly, CI precoding approaches may further lead to savings in the training time and overhead for signaling the composite channels to the receivers, as demonstrated in \cite{cr-10} and \cite{ci-book-1}.

\subsubsection{Data Stream Number Increase}
In addition to the above two benefits, another benefit that is recently revealed for CI-based precoding over traditional precoding is the support for an increased number of data symbols that can be simultaneously transmitted \cite{ci-8}, \cite{ci-26}, \cite{ci-27}. For traditional precoding approaches, the number of data streams that can be supported for simultaneous transmission is limited by the number of transmit antennas at the BS in the downlink. While MRT and RZF precoding can theoretically support a larger number of data streams, the resulting uncoded bit error rate (BER) performance is on the order of $10^{-1}$, which is not practically useful. For CI precoding, on the contrary, it is shown in \cite{ci-26}, \cite{ci-27} that it can support a larger number of data streams with a significantly improved BER performance. Specifically, when there are $N_t=8$ transmit antennas at the BS and a total number of $K=9$ users in the system, CI precoding is able to achieve an uncoded BER that is lower than $10^{-4}$ when QPSK modulation is considered \cite{ci-26}. A similar result can be observed in \cite{ci-27} for QAM constellations, while the supported number of data streams becomes fewer compared to PSK modulations.

In addition to the above general benefits, a number of application-particular benefits have also been reported and will be overviewed in the respective sections in the following.

\begin{figure*}[!t]
	\centering
	\includegraphics[scale=0.4]{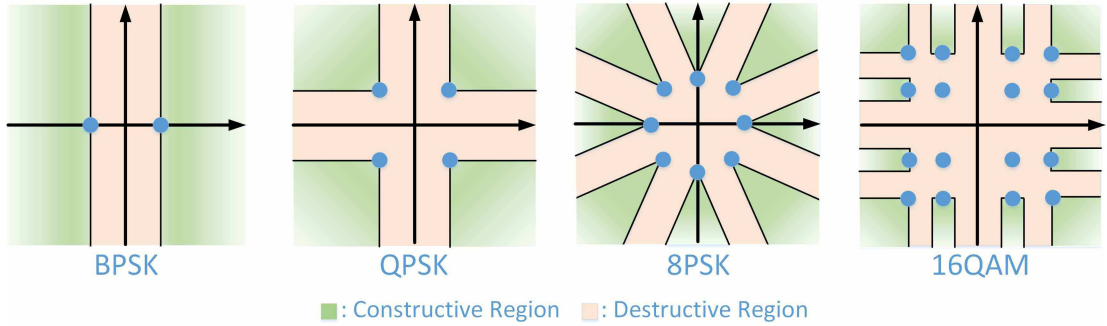}
	\caption{CI region characterization for PSK and QAM constellations}
	\label{CI}
\end{figure*}

\subsection{Current Limitations of CI and Symbol-Level Precoding}
In addition to various benefits CI precoding offers compared to traditional channel-dependent precoders, CI precoding also exhibits some limitations, as discussed below.

\subsubsection{Complexity}
The first and most obvious limitation for CI precoding is its complexity, which also holds for other SLP-based techniques. Compared to traditional block-level precoders where the precoding matrix is updated whenever the channel changes, CI precoding has to update the precoding matrix on a symbol level, which leads to much higher computational costs. Moreover, most CI precoding approaches in the literature are based on optimizations, which means that an optimization problem has to be solved to obtain the desired precoding matrix on a symbol level, which is more demanding than block-level precoders. Both of the above requirements pose significant burden on the hardware components at the BS, especially in practical wireless communication systems that are required to operate on a real-time basis.  

Nevertheless, thanks to recent developments in deriving low-complexity solvers for CI precoding, it has been shown in \cite{ci-25} and \cite{ci-27} that an optimal CI precoding structure exists for the CI-based max-min fair optimization problem, while \cite{ci-23} also presents a closed-form expression for CI-based PM problem by following a similar approach, both of which can greatly alleviate the computational costs for CI precoding. This will be discussed in more detail in Section III-F.

\subsubsection{CI based on Soft Detector}
Another limitation for current CI-based SLP schemes is that they are designed for uncoded communication systems, and its superiority is therefore only guaranteed for uncoded systems. In practical wireless communication systems where channel coding is also adopted, it is shown in \cite{dm-1} that CI precoding with existing channel coding schemes is superior to traditional channel-coded precoding schemes. However, it is still unknown whether this is the optimal performance and whether a joint design of CI precoding and channel coding can lead to further performance improvements. To design the CI-optimal soft detector, the priori probabilities of the input symbols producing a soft output indicating the reliability of the decision should be taken into account, which have not been fully addressed in the literature.

As an initial attempt, \cite{ci-28} conducts performance analysis of CI precoding when PSK signaling is considered, for the first time in the literature, where the achievable rate for each user and the total sum rate upper bound of CI transmission are derived.

\subsubsection{Adaptive Modulation and SINR Estimation}
Current CI precoding approaches are designed for either PSK or QAM constellations, while the extension to a multiple modulation scenario has not been fully discussed. Importantly, since practical wireless communication systems adopt adaptive modulation based on SINR estimation, current CI precoding algorithms designed for single modulation are not directly applicable. Moreover, SINR estimation techniques currently employed may not be suitable for CI-based precoding methods. This will be discussed in Section VII-C.

\section{Constructive Interference Classification and Characterization}
In this section, we characterize the CI condition for several modulations that are typically considered in wireless communication systems, which include PSK, QAM, and APSK constellations. For each modulation type, we first illustrate the CI condition geometrically, based on which we present the mathematical condition for achieving CI. While there also exist several alternative CI metrics in the literature \cite{ci-8}, \cite{ci-10}, \cite{ci-18}, and \cite{ci-22}, in this paper we employ the CI metric that is firstly introduced in \cite{ci-8} and widely adopted in the subsequent studies for CI-based precoding.

For notational convenience, we consider a multi-user multiple-input single-output (MU-MISO) system in the downlink, and we express the received symbol for user $k$ as
\begin{equation}
{y_k} = {\bf{h}}_k^{\text T}{\bf{Ws}} + {n_k},
\label{eq_3}
\end{equation}
where ${y_k}$ is the received signal for user $k$, ${\bf{h}}_k \in {\cal C}^{N_\text{T} \times 1}$ is the channel vector between the BS and user $k$, ${\bf W} = \left[ {{{\bf{w}}_1},{{\bf{w}}_2}, \cdots ,{{\bf{w}}_K}} \right] \in {\cal C}^{N_\text{T} \times K}$ is the precoding matrix, ${\bf s}=\left[ {s_1, s_2, \cdots, s_K} \right]^\text{T} \in {\cal C}^{K \times 1}$ is the transmit data symbol vector from a specific modulation constellation, and $n_k$ is the additive Gaussian noise with zero mean and variance $\sigma^2$. Throughout this section, we consider unit-norm constellations to characterize the CI condition for each modulation type.

Based on the definition of CI and following \cite{ci-8}, \cite{ci-27}, the CI regions for PSK and QAM constellations are shown as the green shaded areas in Fig.~\ref{CI}, where BPSK, QPSK, 8PSK and 16QAM constellations are depicted as representative examples, and their extensions to higher PSK and QAM constellations are trivial. As can be seen, when the received signal is located in the green `constructive region', its distance to all the detection thresholds is increased compared to the nominal constellation point.

\begin{figure*}[!t]
\begin{centering}
\subfloat[QPSK, `strict']
{\begin{centering}
\includegraphics[width=5.2cm]{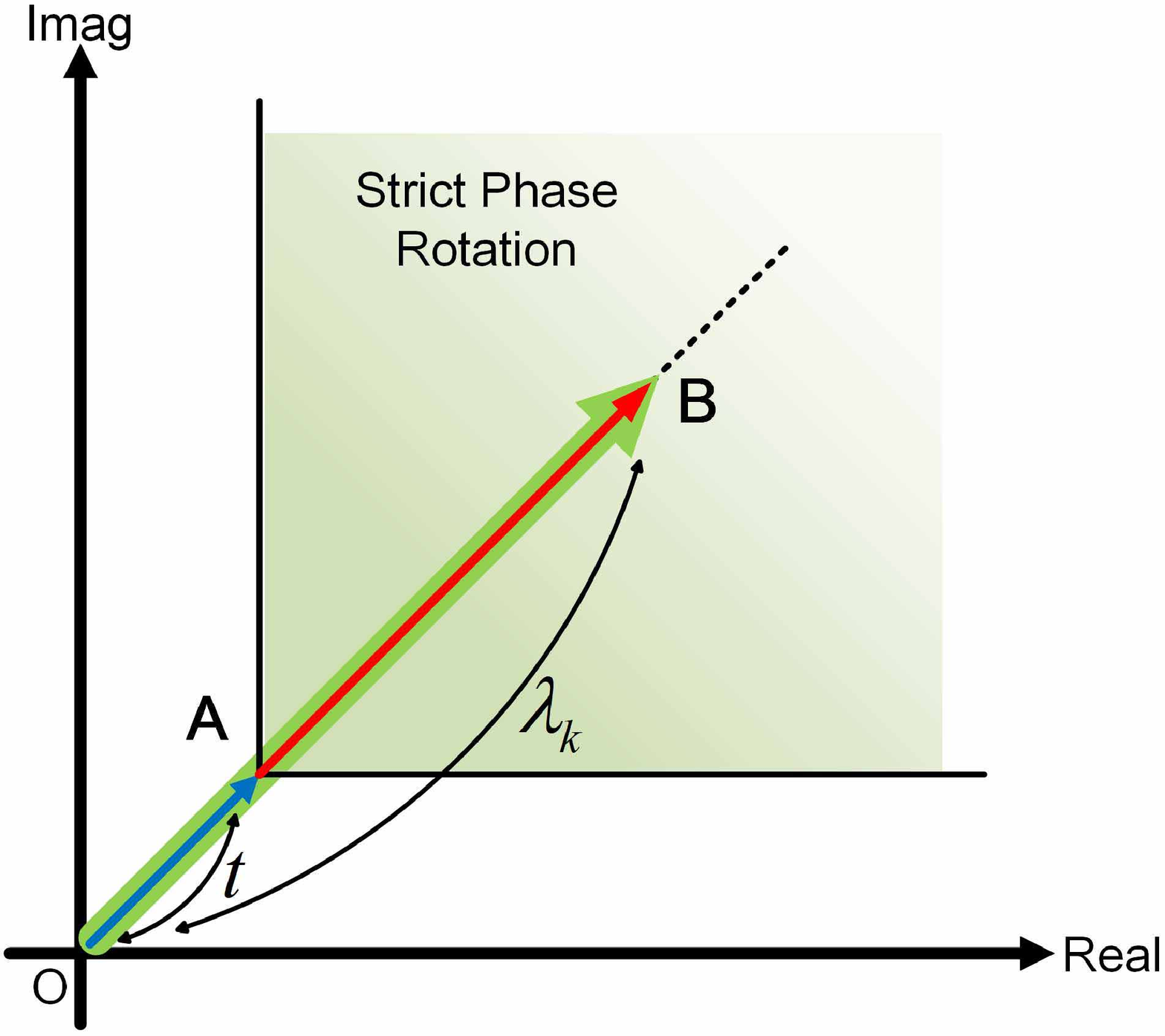}
\label{PSK-strict}
\par
\end{centering}
}
\hspace{0.2cm}
\subfloat[QPSK, `non-strict']
{\begin{centering}
\includegraphics[width=5.2cm]{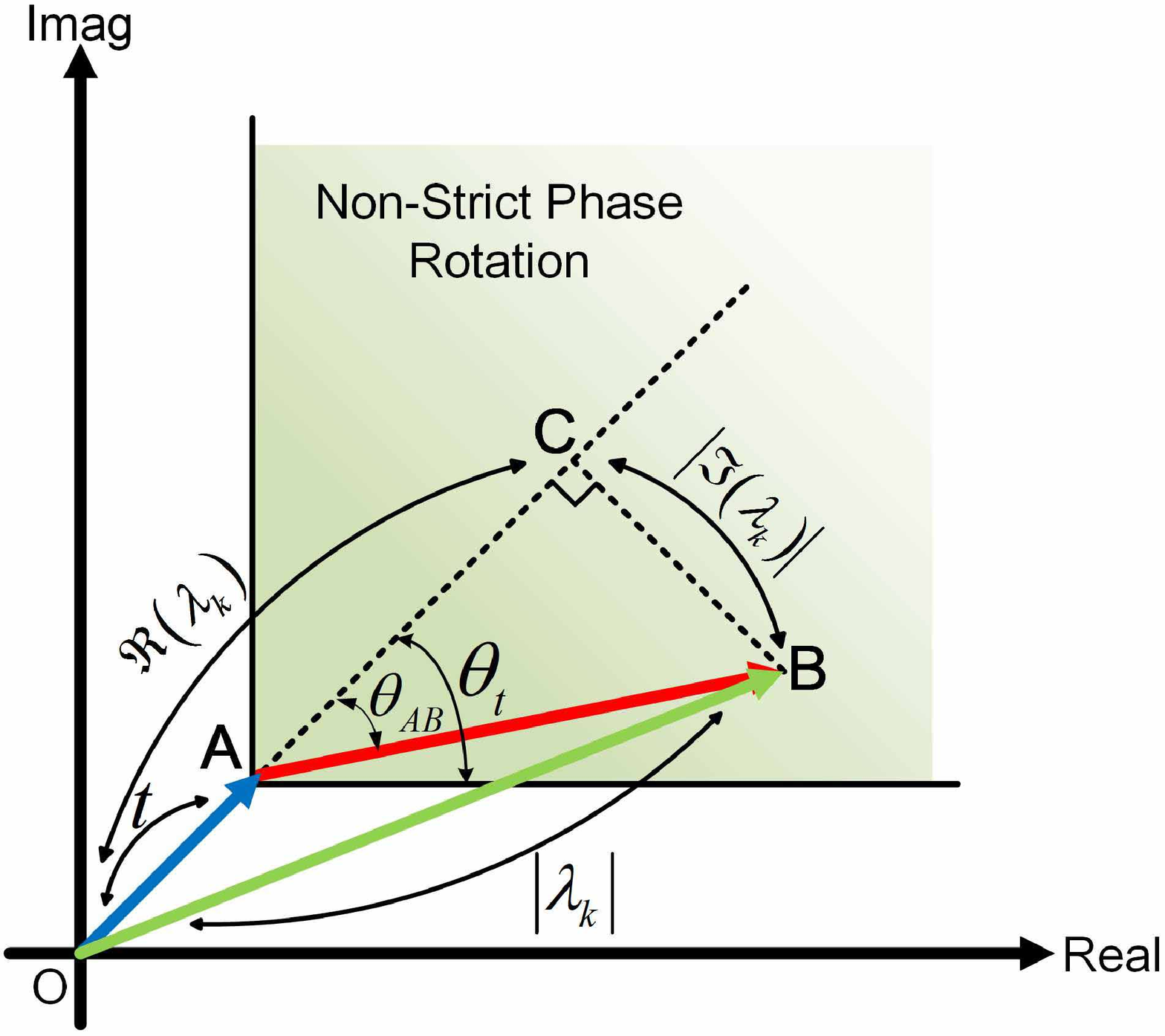}
\label{PSK-non-strict}
\par
\end{centering}
}
\hspace{0.2cm}
\subfloat[QAM, `symbol-scaling']
{\begin{centering}
\includegraphics[width=4.85cm]{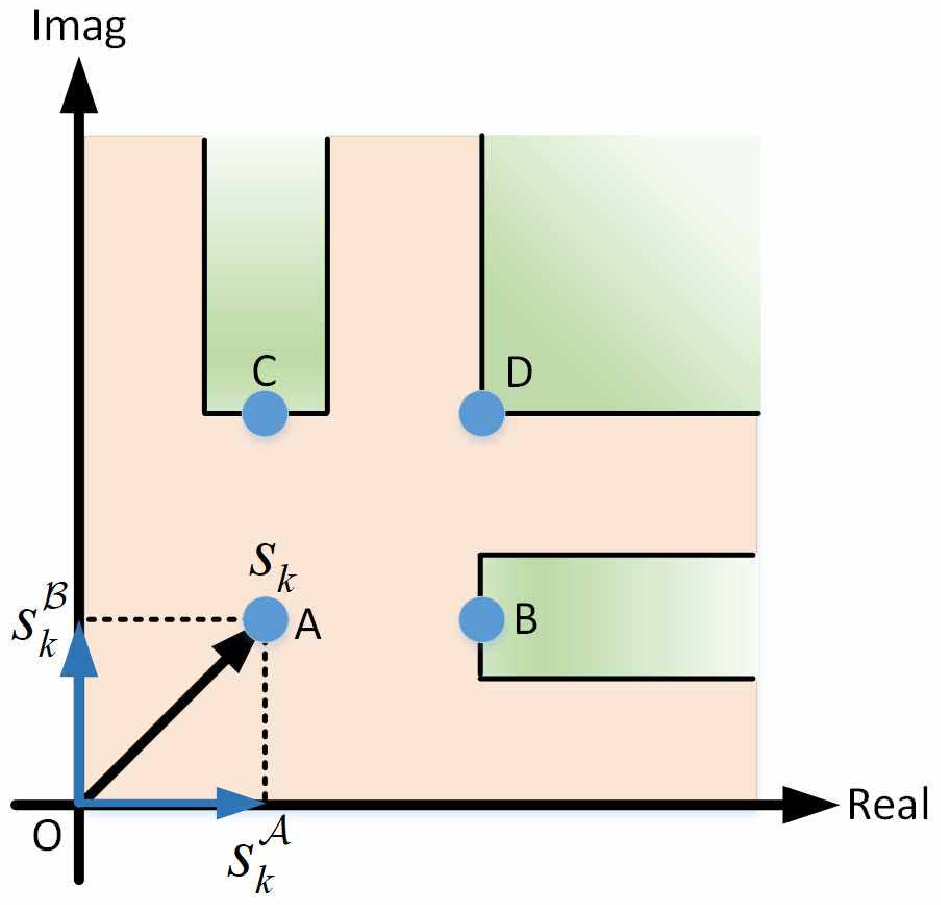}
\label{QAM}
\par
\end{centering}
}
\par
\end{centering}
\caption{\label{fig:PF}CI metric for PSK based on `strict/non-strict phase rotation' and QAM based on `symbol-scaling'}
\end{figure*}

\subsection{Constant-Envelope Constellations (PSK)}
To derive the mathematical formulation of CI conditions for constant-envelope PSK constellations, we consider the `phase-rotation' metric introduced in \cite{ci-8}, and without loss of generality we focus on one quarter of a nominal QPSK constellation, as depicted in Fig.~\ref{PSK-strict} for `strict phase rotation' and Fig.~\ref{PSK-non-strict} for `non-strict phase rotation', where $\vec{OB}$ denotes the noiseless received signal, which leads to $\vec{OB}={\bf{h}}_k^{\text T}{\bf{Ws}}$. $\vec{OA}=t \cdot s_k$ represents the scaled data symbol, and $t= \sqrt{\Gamma_k \sigma^2}$ for PM optimizations, where $\Gamma_k$ is the corresponding SINR target for user $k$ \cite{ci-8}. For SB optimizations, $t$ is the objective function to be maximized \cite{ci-25}. Based on the observation that $\vec{OB}=\vec{OA}+\vec{AB}$, we obtain that $\vec{AB}={\bf{h}}_k^{\text T}{\bf{Ws}}-t \cdot s_k$ is the interfering signals. Following \cite{ci-25}, for notational convenience we introduce a scalar $\lambda_k$ for each data symbol $s_k$, which leads to
\begin{equation}
\vec{OB} = {\bf{h}}_k^{\text T}{\bf{Ws}} = \lambda_k s_k,
\label{eq_4}
\end{equation}
and the value of $\lambda_k$ fully represents the effect of multi-user interference.

For the `strict phase-rotation' CI metric, as considered in \cite{ci-5}, \cite{ci-6} and \cite{ci-25}, it is obvious that each $\lambda_k$ should be purely real to guarantee that the phase of the noiseless received signal for user $k$ is exactly the same as that of $s_k$. Furthermore, the received signals should enhance the useful signal power to achieve CI, and therefore the mathematical condition for the `strict phase-rotation' metric can be expressed as
\begin{equation}
\lambda_k \ge t, {\kern 3pt} \forall k \in {\cal K},
\label{eq_5}
\end{equation}
where ${\cal K} = \left\{ {1,2,\cdots,K} \right\}$. 

For the `non-strict phase-rotation' metric proposed in \cite{ci-8}, $\lambda_k$ can be complex as the phases of the received signals are not necessarily constrained to be strictly aligned to the original symbols of interest. As observed in Fig.~\ref{PSK-non-strict}, as long as the resulting signals are located in the `constructive region', all the interference is constructive and further benefits the detection performance. Based on the geometry, to have the node `B' located in the constructive area is equivalent to
\begin{equation}
\theta_{AB} \le \theta_{\text {th}},
\label{eq_6}
\end{equation}
where $\theta_{\text {th}} = \frac{\pi}{\cal M}$ for a $\cal M$-PSK constellation. Following \cite{ci-8} and \cite{ci-25}, this CI condition in \eqref{eq_6} can be further expressed as a function of the complex scalar $\lambda_k$, given by
\begin{equation}
\left[ {\Re \left( {{\lambda _k}} \right) - t} \right]\tan {\theta _{\text {th}}} \ge \left| {\Im \left( {{\lambda _k}} \right)} \right|, {\kern 3pt} \forall k \in {\cal K},
\label{eq_7}
\end{equation}
which is widely employed in the literature of CI-based precoding \cite{ci-7}, \cite{ci-8}, \cite{ci-15}, \cite{ci-20}, \cite{ci-25}-\cite{ci-30}. Specifically for a BPSK constellation where ${\cal M}=2$, the CI condition in \eqref{eq_7} for the `non-strict phase-rotation' metric is simplified into
\begin{equation}
\Re \left( {{\lambda _k}} \right) - t \ge 0, {\kern 3pt} \forall k \in {\cal K}.
\label{eq_8}
\end{equation}

\subsection{Multi-Level Constellations (QAM, APSK)}
In this section, we describe the CI condition for multi-level modulations, where we consider QAM and APSK as two representative examples. The general observation of CI characterization for multi-level modulations is that CI can be exploited by the outer constellation points, while we consider all the interference for the inner constellation points as destructive. Therefore, the CI condition for multi-level modulations is formulated by decomposing the constellation points into outer constellation points and inner constellation points, as detailed below.

\begin{table*}[t]
\begin{center}
\caption{A summary of mathematical CI conditions for PSK and multi-level modulation}
\setlength{\arrayrulewidth}{0.2mm}
\renewcommand{\arraystretch}{1.4}
\scalebox{1.1}
{
\begin{tabular}{| c | c | c | c | }

    \hline
    {\bf Modulation} & {\bf CI Metric} & {\bf Signal Representation} & {\bf Mathematical CI Condition} \\
    \hline
        PSK & Strict Phase Rotation \cite{ci-6}, \cite{ci-25} & ${\bf{h}}_k^{\text T}{\bf{Ws}} = {\lambda _k}{s_k}$ & $\lambda_k \ge t$ \\
    \hline
        PSK & Non-Strict Phase Rotation \cite{ci-8}, \cite{ci-25} & ${\bf{h}}_k^{\text T}{\bf{Ws}} = {\lambda _k}{s_k}$ & $\left[ {\Re \left( {{\lambda _k}} \right) - t} \right]\tan {\theta _{\text {th}}} \ge \left| {\Im \left( {{\lambda _k}} \right)} \right|$ \\
    \hline
        PSK & Symbol Scaling \cite{ci-3}, \cite{dac-14} & ${\bf{h}}_k^{\text T}{\bf{Ws}} = {\alpha _k^{\cal A}}{s_k^{\cal A}} + {\alpha _k^{\cal B}}{s_k^{\cal B}}$ & $\alpha _k^{\cal A} \ge t$,  $\alpha _k^{\cal B} \ge t$ \\
    \hline
        QAM & Symbol Scaling \cite{ci-27} & ${\bf{h}}_k^{\text T}{\bf{Ws}} = {\alpha _k^{\cal A}}{s_k^{\cal A}} + {\alpha _k^{\cal B}}{s_k^{\cal B}}$ & $\alpha _m^{\cal O} \ge t$,  $\alpha _n^{\cal I} = t$ \\
    \hline
        APSK & Non-Strict Phase Rotation & ${\bf{h}}_k^{\text T}{\bf{Ws}} = {\lambda _k}{s_k}$ & $\left[ {\Re \left( {\lambda _m^{\cal O}} \right) - t} \right]\tan {\theta _{\text {th}}} \ge \left| {\Im \left( {\lambda _m^{\cal O}} \right)} \right|$, $\lambda _n^{\cal I} = t$ \\
    \hline

\end{tabular}
}
\end{center}
\label{tab:multicol}
\end{table*}

\subsubsection{QAM}
QAM modulation is a typical multi-level modulation widely employed in wireless communication systems. CI-based precoding for QAM constellations is firstly considered in \cite{ci-11}, \cite{ci-12}, \cite{ci-29} and \cite{ci-30}, and is also studied in \cite{ci-27} more recently. For QAM modulation, since only the real (or imaginary) part of some outer constellation points can be scaled to exploit CI, as observed in Fig.~\ref{CI} and Fig.~\ref{QAM}, the `phase-rotation' metric in \cite{ci-5}-\cite{ci-8} is no longer applicable to QAM. 

Accordingly, the `symbol-scaling' metric considered in \cite{ci-12}, \cite{ci-27} and \cite{ci-30} is presented here. To be more specific, this metric first performs a signal decomposition of the constellation points as well as the noiseless received signals along the detection thresholds, which mathematically leads to
\begin{equation}
{s_k} = s_k^{\cal A} + s_k^{\cal B}, {\kern 3pt} {\bf{h}}_k^{\text T}{\bf{Ws}} = \alpha _k^{\cal A} s_k^{\cal A} + \alpha _k^{\cal B} s_k^{\cal B} = {\bf \Omega} _k^{\text T} {\bf s}_k, {\kern 3pt} \forall k \in {\cal K}.
\label{eq_9}
\end{equation}
where $s_k^{\cal A}$ and $s_k^{\cal B}$ are parallel to the two detection thresholds for the constellation point $s_k$, ${\bf \Omega} _k = \left[ {\alpha _k^{\cal A}, \alpha _k^{\cal B}} \right]^\text{T}$ and ${\bf s}_k = \left[ {s_k^{\cal A} , s_k^{\cal B}} \right]^\text{T}$. In \eqref{eq_9}, $\alpha _k^{\cal A} \ge 0$ and $\alpha _k^{\cal B} \ge 0$ are two real scalars that jointly represent the effect of interference. Specifically for QAM constellations, we can directly simplify the expression for $s_k^{\cal A}$ and $s_k^{\cal B}$ into
\begin{equation}
s_k^{\cal A}=\Re \left\{ {s_k} \right\}, {\kern 3pt} s_k^{\cal B}=j \cdot \Im \left\{ {s_k} \right\}.
\label{eq_10}
\end{equation}

For notational convenience, we introduce a set $\cal O$ that consists of the real scalars corresponding to the real (or imaginary) part of the outer constellation points that can be scaled to exploit CI, and a set $\cal I$ that consists of the real scalars that correspond to the real (or imaginary) part of the constellation points that cannot be scaled \cite{ci-27}, i.e., in Fig.~\ref{QAM}, the set $\cal O$ includes the scalars for both the real and imaginary part of the constellation point type `D', the real part of `B', and the imaginary part of `C'. Similarly, the set $\cal I$ includes the scalars for both the real and imaginary part of the constellation point type `A', the real part of `C', and the imaginary part of `B'. Accordingly, the set ${\cal O} \cup {\cal I}$ includes all the scaling factors $\alpha_k^{\cal U}$, ${\cal U} \in \left\{ {{\cal A}, {\cal B}} \right\}$, $\forall k \in {\cal K}$, and the CI condition for QAM constellations can be expressed as
\begin{equation}
\alpha _m^{\cal O} \ge t, {\kern 3pt} \alpha _n^{\cal I} = t, {\kern 3pt} \forall \alpha _m^{\cal O} \in {\cal O}, {\kern 3pt} \forall \alpha _n^{\cal I} \in {\cal I}.
\label{eq_11}
\end{equation}
Similar to the case of PSK modulation, $t$ is the objective function for SB optimizations, while $t=\sqrt{\Gamma_k \sigma^2}$ if a PM problem is considered.

It should be noted that while the `symbol-scaling' metric is primarily considered for QAM constellations, it is applicable to PSK modulation as well, as firstly studied in \cite{ci-3} in the context of VP precoding. Nevertheless, for a generic PSK constellation, the expression for $s_k^{\cal A}$ and $s_k^{\cal B}$ is no longer in the form in \eqref{eq_10}, and the corresponding derivations can be found in \cite{dac-14} in detail. 

\subsubsection{APSK}
APSK is another representative example of multi-level modulation, which has been widely considered in satellite communications \cite{APSK-1}, \cite{APSK-2}. Compared to QAM constellations, the advantage for APSK modulation lies in its low peak-to-average power ratio (PAPR) and is therefore more robust against nonlinear channel effects. The constellation of APSK can be viewed as a combination of several PSK constellations with different amplitudes \cite{APSK-1}. 

Similar to the case of QAM, CI can be exploited by the outer constellation points of APSK, while the inner constellation points should not be scaled. Since the outer constellation of APSK can be regarded as a typical PSK constellation, by following \eqref{eq_4}, \eqref{eq_6} and \eqref{eq_7}, the CI condition for APSK can be readily expressed as
\begin{equation}
\left[ {\Re \left( {\lambda _m^{\cal O}} \right) - t} \right]\tan {\theta _{\text {th}}} \ge \left| {\Im \left( {\lambda _m^{\cal O}} \right)} \right|, {\kern 3pt} \lambda _n^{\cal I} = t, 
\label{eq_12}
\end{equation}
where ${\theta _{\text {th}}} = \frac{\pi }{\cal N}$ when the outer ring of the APSK constellation adopts $\cal N$-PSK, $\lambda _m^{\cal O} \in {\cal O}$, and $\lambda _n^{\cal I} \in {\cal I}$. In the case of APSK, the set $\cal O$ includes the complex scalars corresponding to the outer constellation points of APSK, while $\cal I$ includes the complex scalars corresponding to the inner constellation points, respectively, and we have ${\cal O} \cup {\cal I} = \left\{ {\lambda_1, \lambda_2, \cdots, \lambda_K} \right\}$.

We summarize the mathematical CI formulation for the discussed modulation types in Table II.

\section{Symbol-Level Precoding based on Optimization}
In this section, we introduce several optimization-based CI precoding designs in the downlink of a MU-MISO system, which includes CI-based VP precoding, CI-based PM, CI-based SB\footnote{The optimization for the sum-rate maximization is not considered, since the Shannon sum-rate expression is not applicable to CI precoding. This will be discussed in Section VII-A.}, etc., as a foundation for the description of the applications of interference exploitation in Section IV and V. In each subsection, we begin with a brief literature review on these precoding approaches and their corresponding mathematical formulations, followed by their adaptations to CI-based designs.

\subsection{CI-VP based on Symbol Scaling}
Traditional non-linear VP approaches apply a perturbation to the data symbols upon the ZF precoding \cite{intro-vp-1}, such that the resulting transmit signals are better aligned to the eigenvectors of the channel inverse matrix, which leads to a significant decrease in the noise amplification effect. Accordingly, a significantly improved performance is observed for VP precoding over ZF precoding in the high SNR regimes. Meanwhile, since VP approach is highly computationally expensive due to the inclusion of the sphere-search process, a number of studies on VP precoding have focused on the reduction in the computational costs \cite{intro-vp-2}-\cite{intro-vp-4}, \cite{intro-vp-12}\nocite{intro-vp-13}\nocite{intro-vp-14}-\cite{intro-vp-15}. Specifically, this includes the thresholded VP method where the sphere-search process is terminated when a SNR target is met \cite{intro-vp-2}, \cite{intro-vp-3}, and the selective VP approach that only perturbs part of the data symbols for maximizing the energy efficiency \cite{intro-vp-4}. Several studies have also paid their attention to the modulo loss reduction for VP precoding \cite{intro-vp-9}\nocite{intro-vp-10}-\cite{intro-vp-11}. An additional problem for conventional VP precoding in \cite{intro-vp-1} is that it does not apply to multi-modulation scenarios, since the perturbation basis is modulation dependent \cite{intro-vp-1}. To overcome this, several studies have considered multi-modulation VP precoding, which includes the block-diagonalized VP method in \cite{intro-vp-5}, the user-grouping VP method in \cite{intro-vp-6}, and the constellation scaling approach in \cite{intro-vp-7}, \cite{intro-vp-8}. 

The precoded signal vector for a traditional VP precoding can be expressed as
\begin{equation}
{{\bf{x}}_{\text {VP}}} = \frac{\sqrt{P_0}}{\beta_{\text {VP}}}  \cdot {{\bf{H}}^{\text H}}{\left( {{\bf{H}}{{\bf{H}}^{\text H}}} \right)^{ - 1}}\left( {{\bf{s}} + \tau  \cdot {\bf{l}}} \right),
\label{eq_13}
\end{equation}
where $P_0$ is the transmit power, $\tau$ is the modulation-dependent modulo basis, $\bf l$ is the complex-integer perturbation vector, and 
\begin{equation}
\beta_{\text {VP}}=\left\| {{{\bf{H}}^{\text H}}{{\left( {{\bf{H}}{{\bf{H}}^{\text H}}} \right)}^{ - 1}}\left( {{\bf{s}} + \tau  \cdot {\bf{l}}} \right)} \right\|_2
\label{eq_14}
\end{equation} 
is the scaling factor for power normalization, which is also known as the noise amplification factor. For conventional VP precoding, the perturbation vector $\bf l$ is found by the sphere-search algorithm in the whole complex-integer space to minimize the power normalization factor $\beta_{\text {VP}}$, and a modulo operation with basis $\tau$ is also required at the receiver side to remove the perturbation effect.

CI-based VP approach is proposed in the context of complexity reduction for VP precoding \cite{ci-3}. To extend the concept of CI exploitation to VP precoding, it is natural to constrain the search for the optimal perturbation vectors within the constructive area for each constellation point. This then removes the necessity for the modulo operation at the receiver side, since the perturbation vectors are all constructive and contribute to the original data symbols. Moreover, since the perturbation vectors are enforced in the constructive area, their values can be relaxed into complex continuous and are no longer needed to be complex integers. Accordingly, by introducing a diagonal scaling matrix ${\bf \Phi}$ with complex entries, the transmit signal vector for constructive VP (CVP) can be transformed into a linear scaling operation, expressed as
\begin{equation}
{{\bf{x}}_{\text {CVP}}} = \frac{\sqrt{P_0}}{\beta_{\text {CVP}}}  \cdot {{\bf{H}}^{\text H}}{\left( {{\bf{H}}{{\bf{H}}^{\text H}}} \right)^{ - 1}}{\bf \Phi} {\bf{s}},
\label{eq_15}
\end{equation} 
where the power scaling factor $\beta_\text{CVP}$ is obtained as
\begin{equation}
{\beta _{\text {CVP}}} = \left\| {{{\bf{H}}^{\text H}}{{\left( {{\bf{H}}{{\bf{H}}^{\text H}}} \right)}^{ - 1}}{\bf \Phi} {\bf{s}}} \right\|_2.
\label{eq_16}
\end{equation}
By following the decomposition in \eqref{eq_9}, we transform $\bf \Phi s$ into
\begin{equation}
{\bf \Phi s} = {\bf T} {\text {diag}} \left\{{\bf \Phi _E} \right\} {\bf s_E}
\label{eq_17}
\end{equation}
such that the scalars in ${\bf {\Phi _E}}$ are real, where ${\bf T}\in {\cal R}^{K \times 2K}$, ${\bf {\Phi _E}} \in {\cal R}^{2K \times 1}$ and ${\bf s_E} \in {\cal C}^{2K \times 1}$ are given by
\begin{equation}
\begin{aligned}
&{\bf T} = \left[ {\begin{array}{*{20}{c}}
{1}&{1}&0&0& \cdots & \cdots &0&0\\
0&0&{1}&{1}& \ddots & \ddots &0&0\\
 \vdots & \vdots & \ddots & \ddots & \ddots & \ddots & \vdots & \vdots \\
0&0& \cdots & \cdots &0&0&{1}&{1}
\end{array}} \right]={\bf I}_K \otimes \left[{1, 1}\right], \\
&{\bf {\Phi _E}} = {\left[ {\alpha_1^{\cal A},\alpha_1^{\cal B},\alpha_2^{\cal A},\alpha_2^{\cal B}, \cdots ,\alpha_K^{\cal A},\alpha_K^{\cal B}} \right]^{\text T}},\\
&{\bf s_E} = {\left[ {s_1^{\cal A},s_1^{\cal B},s_2^{\cal A},s_2^{\cal B}, \cdots ,s_K^{\cal A},s_K^{\cal B}} \right]^{\text T}}.
\end{aligned}
\label{eq_18}
\end{equation}
Accordingly, the optimization problem that minimizes the power scaling factor $\beta_{\text {CVP}}$ based on CI exploitation is given by \cite{ci-3}
\begin{equation}
\begin{aligned}
&\mathcal{P}_{\text {CVP}}: {\kern 3pt} \mathop {\min }\limits_{\bf \Phi_E}  \left\| {{{\bf{H}}^{\text H}}{{\left( {{\bf{H}}{{\bf{H}}^{\text H}}} \right)}^{ - 1}} {\bf T} {\text {diag}} \left\{{\bf \Phi_E}\right\} {\bf{s_E}}} \right\|_2^2 \\
&{\kern 8pt} {\text {s.t.}} {\kern 14pt} {\rm C1:} {\kern 3pt} \alpha_k^{\cal U} \ge \alpha_0, {\kern 3pt} {\cal U} \in \left\{ {{\cal A}, {\cal B}} \right\}, {\kern 3pt} \forall k \in {\cal K}
\label{eq_19}
\end{aligned}
\end{equation}
which can be further transformed into a QP and solved efficiently. In $\mathcal{P}_{\text {CVP}}$, $\alpha_0>0$ is a lower threshold, whose value will not affect the final solution, and it is convenient to select $\alpha_0=1$. Compared to traditional VP precoding, the CVP method requires less than $10\%$ the complexity when the number of antennas is larger than $K=N_\text{T}=6$, by substituting the sophisticated sphere-search process with a linear scaling operation \cite{ci-3}.

In addition to the formulation in \eqref{eq_15}, an alternative CI-VP approach has recently been presented in \cite{lux-1}, where the data symbol vector $\mathbf{s}$ is replaced with an equivalent symbol vector $\tilde{\mathbf{s}} = \left[{1, 1, \cdots, 1}\right] \in {\cal R}^{K \times 1}$, and the precoded signal vector is given by
\begin{equation}
\mathbf{x}_{\text {CI-VP}} = \frac{\sqrt{P_0}}{\beta_{\text {CI-VP}}} \mathbf{H}^{\text H}  (\mathbf{H} \mathbf{H}^{\text H})^{-1} \mathbf{B}(\tilde{\mathbf{s}} + \tilde{\mathbf{u}}), \label{sec3:slp}
\end{equation}
where $\tilde{\mathbf{u}}$ is the real non-negative perturbation vector and $\mathbf{B}$ is a rotation matrix defined as
\begin{equation}
\label{sec3:Bdef}
\begin{aligned}
\mathbf{B} = 
\begin{bmatrix}
    s_{1}   & 0  & 0 & \cdots & 0 \\
    0 & s_{2} & 0& \cdots & 0 \\
    0 & 0 & s_{3} & \cdots & 0 \\
    \vdots   & \vdots  & \vdots     & \ddots & \vdots \\
    0         & \cdots & 0 & 0 & s_{K}
\end{bmatrix}.
\end{aligned}
\end{equation}
The equivalent symbol vector $\tilde{\mathbf{s}}$ is real-valued rotated versions of the original complex data symbols. The rotation is accounted accordingly in (\ref{sec3:Bdef}) so that users still receive the correct symbols, since ${\bf B}\tilde{\bf s}={\bf s}$. The power scaling factor is calculated as
\begin{equation}
\beta_{\text {CI-VP}} = \left\|{ \mathbf{H}^{\text H} (\mathbf{H} \mathbf{H}^{\text H})^{-1} \mathbf{B}(\tilde{\mathbf{s}} + \tilde{\mathbf{u}}) }\right\|_2 .\label{sec3:bettedef}
\end{equation}
Consequently, the optimization problem can be constructed as
\begin{equation}
\begin{aligned}
\mathcal{P}_{\text {CI-VP}} : \hspace{1mm}  &  \min_{\tilde{\mathbf{u}}} \lVert \mathbf{H}^{\text H}  (\mathbf{H} \mathbf{H}^{\text H})^{-1} \mathbf{B}(\tilde{\mathbf{s}} + \tilde{\mathbf{u}}) \lVert_2\\
{\text {s.t.}} \hspace{5mm}  & \hspace{1mm} \text{C1 : } \tilde{u}_k \geq 0, {\kern 3pt} \forall k \in \mathcal{K}
\label{sec3:slpminproblem}
\end{aligned}
\end{equation}
The above optimization problem \eqref{sec3:slpminproblem} is a non-negative least squares (NNLS) problem and can be solved directly using the Fast NNLS \cite{lux-2} or the closed-form algorithm given by \cite{lux-3}, \cite{lux-4}. The introduction of the equivalent real symbol vector in this CI-VP approach allows to reduce the search space in the optimization problem and decrease its computational complexity by searching only for real-valued and non-negative perturbation vector.

\subsection{Power Minimization}
Traditional optimization-based PM problems aim to minimize the total transmit power subject to a minimum received SINR target at each receiver side \cite{intro-6}-\cite{intro-8}, \cite{intro-10}-\cite{intro-12}. An uplink-downlink duality has been revealed in \cite{intro-7} and \cite{intro-8}, based on which an efficient algorithm is proposed to solve the downlink precoding optimization. Another method to solve PM optimizations is to transform the PM optimization into a semi-definite programming (SDP). Using the semi-definite relaxation (SDR) approach \cite{intro-10}-\cite{intro-12}, the rank-relaxed SDP becomes a convex optimization that can be conveniently solved, and it has been proven that a rank-1 solution always exists when the problem is feasible \cite{intro-10}. A rank-reduction algorithm is further developed in \cite{intro-10} and \cite{intro-11} to effectively reduce the rank of the solution to the relaxed SDP problem, when additional shaping constraints are further included in the PM optimizations. It is further shown in \cite{intro-12} that the exploitation of real-valued orthogonal space time block coding (OSTBC) can effectively increase the degree of freedom in the optimization design, which leads an improved performance.

Mathematically, from a statistical point of view and treating interference as harmful, the SINR expression for user $k$ can be expressed as \cite{intro-6}
\begin{equation}
\gamma_k = \frac{{{{\left| {{{\bf{h}}_k^{\text T}}{{\bf{w}}_k}} \right|}^2}}}{{\sum\limits_{i \ne k} {{{\left| {{{\bf{h}}_k^{\text T}}{{\bf{w}}_i}} \right|}^2}}  + {\sigma ^2}}}, 
\label{eq_20}
\end{equation}
which is obtained based on ${\mathbb E}\left\{ {{\bf{s}}{{\bf{s}}^\text{H}}} \right\} = {\bf{I}}$. Given \eqref{eq_20}, a typical PM problem that targets at minimizing the average transmit power subject to the SINR threshold $\Gamma_k$ for user $k$ can be formulated as
\begin{equation}
\begin{aligned}
&\mathcal{P}_{\text {PM}}: {\kern 3pt} \mathop {\min }\limits_{{{\bf{w}}_k}} \sum\limits_{k = 1}^K {\left\| {{{\bf{w}}_k}} \right\|_2^2}  \\
&{\kern 6pt} {\text {s.t.}} {\kern 10pt} {\rm C1:} {\kern 3pt} \gamma_k \ge \Gamma_k, {\kern 3pt} \forall k \in {\cal K}
\label{eq_21}
\end{aligned}
\end{equation}
The above power minimization problem $\mathcal{P}_{\text {PM}}$ can be efficiently solved either via SDR or via uplink power allocation algorithms by exploiting the uplink-downlink duality \cite{intro-6}.

To apply the concept of CI exploitation to PM problems, firstly it should be noted that the SINR expression in \eqref{eq_20} is no longer valid, since (i) the instantaneous symbol-level interference is related to not only the CSI but also the data symbols, and (ii) the interfering signals become constructive and contribute to the useful signal power. Following the discussion in Section II and recalling \eqref{eq_4} and \eqref{eq_7}, the constructive PM (CPM) optimization for $\cal M$-PSK modulation based on the `non-strict phase-rotation' metric can be expressed as
\begin{equation}
\begin{aligned}
&\mathcal{P}_{\text {CPM}}^{\text {PSK}}: {\kern 3pt} \mathop {\min }\limits_{{{\bf{x}}}} \left\| {{\bf{x}}} \right\|_2^2  \\
&{\kern 8pt} {\text {s.t.}} {\kern 10pt} {\rm C1:} {\kern 3pt} {{\bf{h}}_k^{\text T}}{\bf{x}} = {\lambda _k}{s_k}, {\kern 3pt} \forall k \in {\cal K}\\
&{\kern 30pt} {\rm C2:} {\kern 0pt} \left[ {\Re \left( {{\lambda _k}} \right) - \sqrt {{\Gamma _k}{\sigma ^2}} } \right]\tan \frac{\pi }{\cal M} \ge \left| {\Im \left( {{\lambda _k}} \right)} \right|, {\kern 3pt} \forall k \in {\cal K}
\label{eq_22}
\end{aligned}
\end{equation}
where for convenience we have introduced ${\bf x}={\bf Ws}$ since $\bf Ws$ can be viewed as a single vector for CI-based PM problems, and $\Gamma_k$ is the SINR target for the $k$-th receiver. We further note that since the precoding matrix is symbol-dependent, the objective function also includes the data symbol vector $\bf s$, which guarantees that the transmit power is minimized on the symbol level, as discussed in Section I-C. Compared to traditional PM optimization ${\cal P}_{\text {PM}}$ which is non-convex and has to resort to algorithms such as SDR, we observe that the constraints in ${\cal P}_{\text {CPM}}^{\text {PSK}}$ are linear and the constructed CPM is convex in nature, which can be efficiently solved via off-the-shelf optimization tools \cite{math-1}. By following a similar procedure, the CPM formulation for QAM constellations can be constructed as
\begin{equation}
\begin{aligned}
&\mathcal{P}_{\text {CPM}}^{\text {QAM}}: {\kern 3pt} \mathop {\min }\limits_{{{\bf{x}}}} \left\| {{\bf{x}}} \right\|_2^2  \\
&{\kern 8pt} {\text {s.t.}} {\kern 16pt} {\rm C1:} {\kern 3pt} {\bf{h}}_k^{\text T}{\bf{x}} = {\bf \Omega}_k^{\text T} {\bf s}_k, {\kern 3pt} \forall k \in {\cal K}\\
&{\kern 36pt} {\rm C2:} {\kern 3pt} \alpha _m^{\cal O} \ge \sqrt{\Gamma_k \sigma^2}, {\kern 3pt} \forall \alpha _m^{\cal O} \in {\cal O} \\
&{\kern 36pt} {\rm C3:} {\kern 3pt} \alpha _n^{\cal I} = \sqrt{\Gamma_k \sigma^2}, {\kern 3pt} \forall \alpha _n^{\cal I} \in {\cal I}
\end{aligned}
\end{equation}
Similar to $\mathcal{P}_{\text {CPM}}^{\text {PSK}}$ for PSK constellations, the CPM problem for QAM constellations is also convex and can be readily solved.

To validate the significant transmit power savings that CI precoding exhibits, below in Fig.~\ref{PM} we present a numerical example for CPM optimizations ${\cal P}_{\text {CPM}}$ versus traditional PM formulation ${\cal P}_{\text {PM}}$ that is solved via the SDR approach for several PSK constellations in a $4 \times 4$ MU-MISO system. Significant performance improvements can be observed for  CI-based PM results over the data-independent PM result, with power savings up to 60$\%$ for QPSK, 40$\%$ for 8PSK, and 20$\%$ for 16PSK, which reveals the superiority of interference exploitation in PM problems.

\begin{figure}[!t]
	\centering
	\includegraphics[scale=0.35]{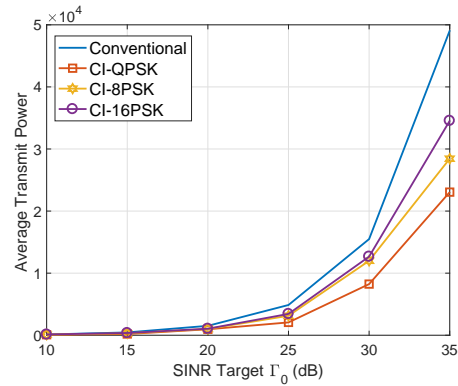}
	\caption{Transmit power v.s. SINR threshold $\Gamma_0$, $\Gamma_k=\Gamma_0$, $\forall k$, $K=N_t=4$, $\sigma^2=1$}
	\label{PM}
\end{figure}

\begin{figure*}[!t]
\begin{centering}
\subfloat[PSK, Uncoded]
{\begin{centering}
\includegraphics[width=4.1cm]{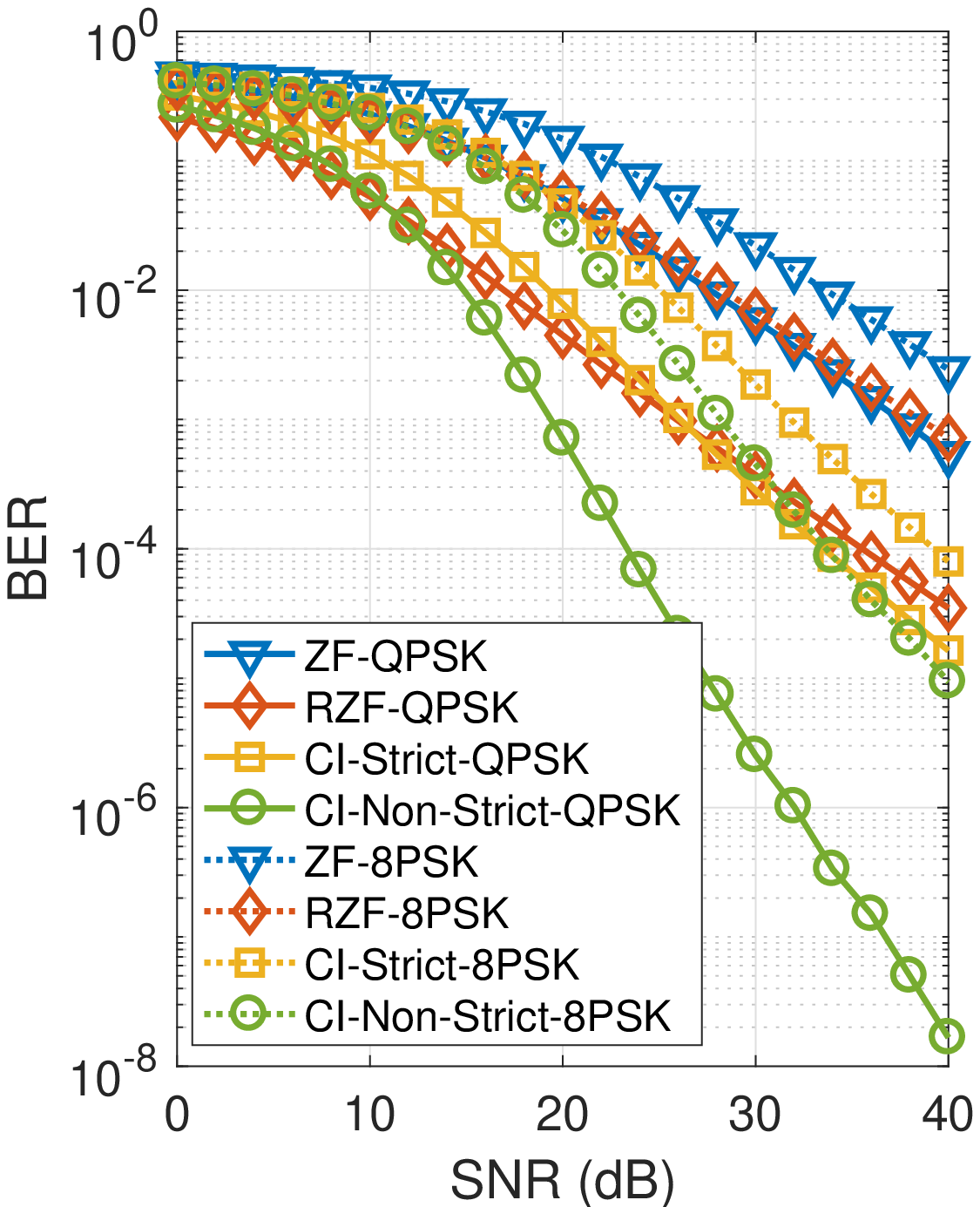}
\label{PSK_uncoded}
\par
\end{centering}
}
\subfloat[QAM, Uncoded]
{\begin{centering}
\includegraphics[width=4.1cm]{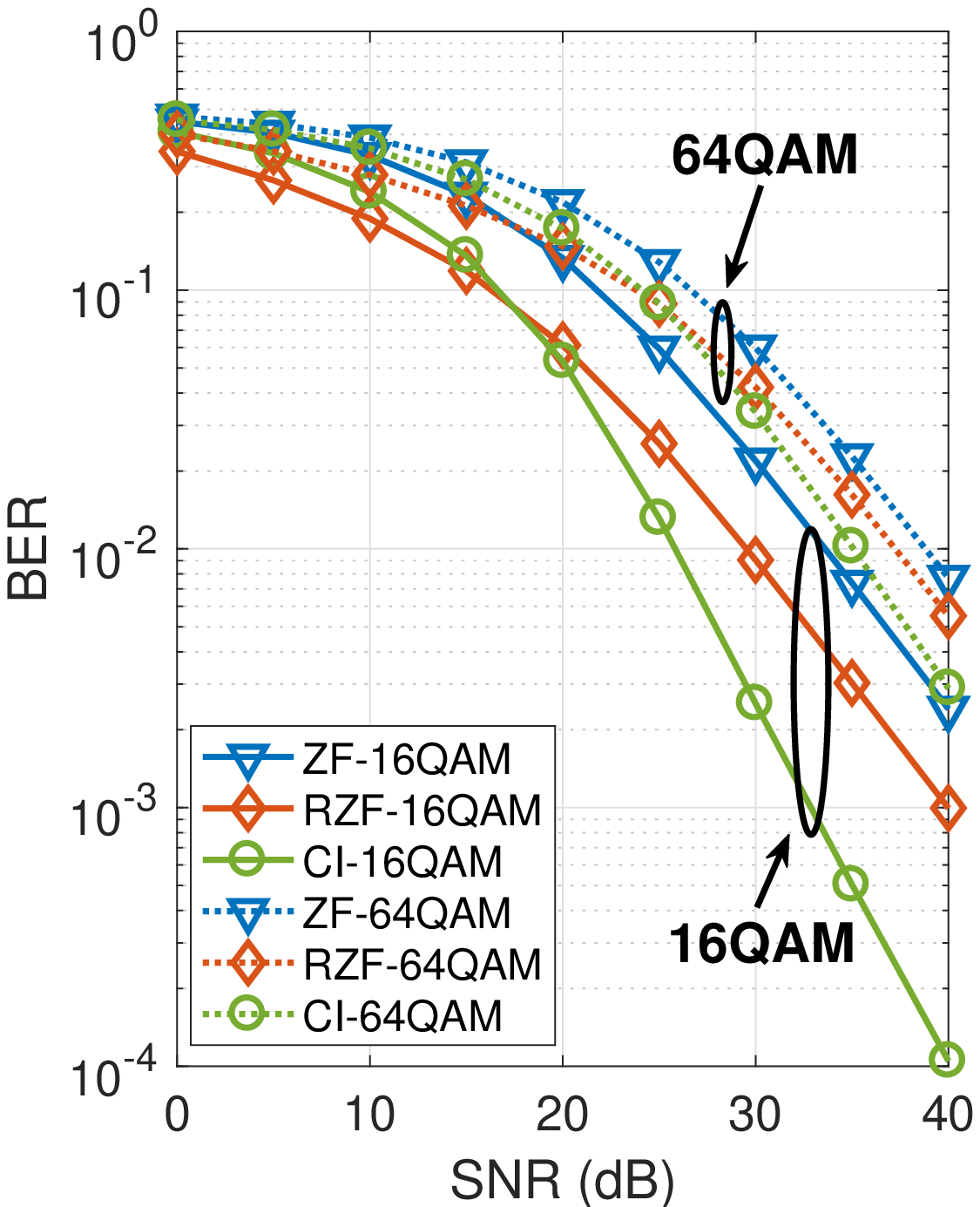}
\label{QAM_uncoded}
\par\end{centering}
}
\subfloat[PSK, Coded]
{\begin{centering}
\includegraphics[width=4.1cm]{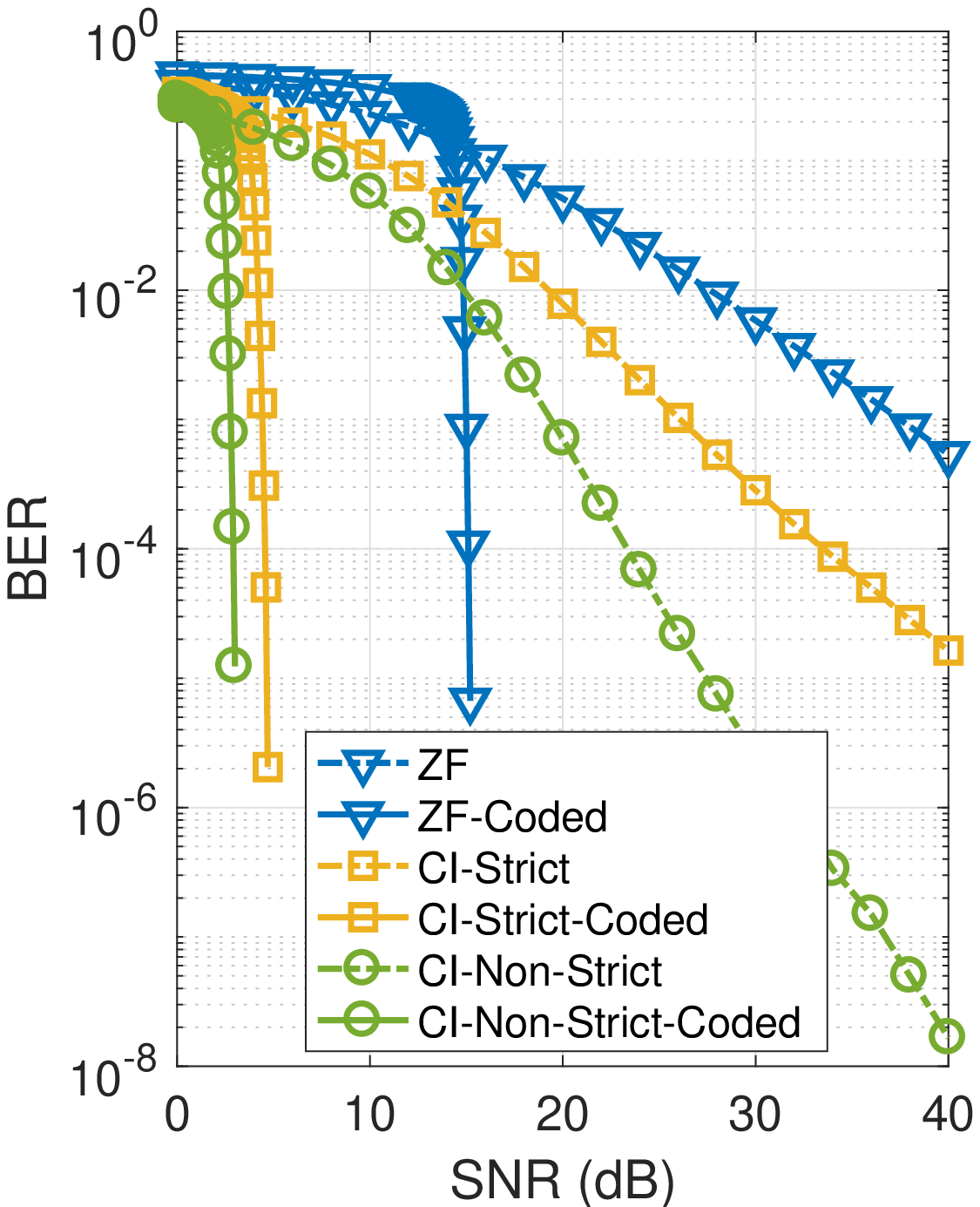}
\label{PSK_coded}
\par
\end{centering}
}
\subfloat[QAM, Coded]
{\begin{centering}
\includegraphics[width=4.1cm]{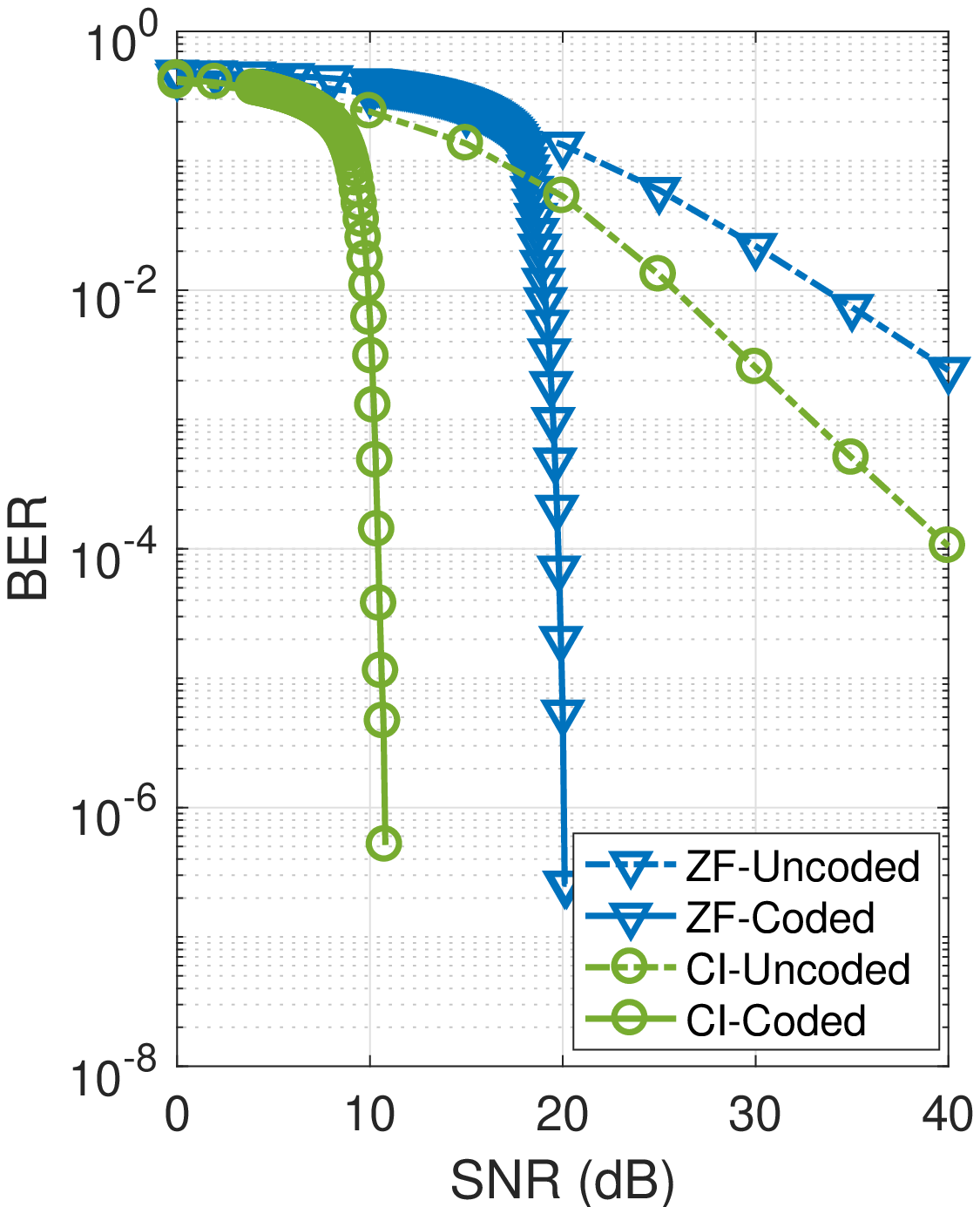}
\label{QAM_coded}
\par
\end{centering}
}
\par\end{centering}
\caption{\label{BER}Uncoded and coded BER v.s. SNR, $K=N_t=12$, $P_0=1$, LDPC, code rate 1/4}
\end{figure*}

\subsection{SINR Balancing}
A typical SB optimization aims to improve the fairness in the wireless communication systems, by maximizing the minimum received SINR among the users subject to a total available transmit power budget at the BS \cite{intro-7}, \cite{intro-8}. A unified analytical framework is developed in \cite{intro-7}, with which the optimal solution to the downlink SB optimization is shown to be equivalent to the optimum of a dual uplink problem. An efficient iterative algorithm that enjoys a fast convergence speed is also developed \cite{intro-7} to obtain this optimal solution. \cite{intro-8} employs the conic optimization approaches to solve the downlink SB optimization, where an inverse problem property is discovered. Based on this property, it is shown in \cite{intro-8} that the SB optimization can be optimally solved by solving a set of PM problems via bisection search. When a per-antenna power constraint instead of the sum-power constraint is considered, it is shown in \cite{intro-14} that the bisection-search framework is still effective to find the globally optimal solution by solving a set of dual per-antenna power balancing problems.

Based on the SINR expression in \eqref{eq_20}, a traditional SB optimization problem subject to a sum-power constraint can be formulated as
\begin{equation}
\begin{aligned}
&\mathcal{P}_{\text {SB}}: {\kern 3pt} \mathop {\max }\limits_{{{\bf{w}}_k}} \mathop {\min }\limits_k {\kern 3pt} {\gamma _k} \\
&{\kern 4pt} {\text {s.t.}} {\kern 13pt} {\rm C1:} {\kern 3pt} \sum\limits_{k = 1}^K {\left\| {{{\bf{w}}_k}} \right\|_2^2}  \le P_0
\label{eq_23}
\end{aligned}
\end{equation}
where $P_0$ is the transmit power allowance. Compared to PM optimizations, SB problems are more difficult to solve. Fortunately, the optimal solution to the SB problems can be obtained either through an iterative algorithm introduced in \cite{intro-7}, or via a bisection-search method proposed in \cite{intro-8} based on the inverse problem property.

To extend CI exploitation to SB optimizations, we follow a similar procedure as done for PM optimizations and formulate a constructive SB (CSB) optimization problem for $\cal M$-PSK constellations, given by
\begin{equation}
\begin{aligned}
&\mathcal{P}_{\text {CSB}}^{\text {PSK}}: {\kern 3pt} \mathop {\max }\limits_{\bf{x}} {\kern 3pt} t  \\
&{\kern 8pt} {\text {s.t.}} {\kern 12pt} {\rm C1:} {\kern 5pt} {{\bf{h}}_k^{\text T}}{\bf{x}} = {\lambda _k}{s_k}, {\kern 3pt} \forall k \in {\cal K}\\
&{\kern 32pt} {\rm C2:} {\kern 3pt} \left[ {\Re \left( {{\lambda _k}} \right) - t } \right]\tan \frac{\pi }{\cal M} \ge \left| {\Im \left( {{\lambda _k}} \right)} \right|, {\kern 3pt} \forall k \in {\cal K} \\
&{\kern 32pt} {\rm C3:} {\kern 3pt} \left\| {{\bf{x}}} \right\|_2^2 \le {P_0}
\label{eq_24}
\end{aligned}
\end{equation}
as demonstrated in \cite{ci-25}, \cite{ci-26}. Recalling Fig.~5, an interesting geometrical observation for ${\cal P}_{\text {CSB}}^{\text {PSK}}$ in \eqref{eq_24} is that the optimization aims to maximize the distance between the `constructive region' and the detection thresholds, such that the received signals that lie in the `constructive region' are pushed as far away from the thresholds as possible. Moreover, an important observation for CPM and CSB optimizations is also revealed in `Corollary 2' of \cite{ci-20}, where it is shown that these two problems are inverse problems as well, which is similar to the case for conventional PM and SB optimizations. Based on this property, efficient algorithms based on the barrier method have been presented in \cite{ci-20}, which are shown to be superior over the gradient descent algorithm that suffers from a low convergence speed. 

The extension of CI-based SB problems to multi-level modulations is trivial, by incorporating the CI conditions for the corresponding multi-level constellations discussed in Section II-B in the formulated optimizations. For example, when QAM constellations are considered, the CSB optimization can be formulated as \cite{ci-27}
\begin{equation}
\begin{aligned}
&\mathcal{P}_{\text {CSB}}^{\text {QAM}}: {\kern 3pt} \mathop {\max }\limits_{\bf{x}} {\kern 3pt} t  \\
&{\kern 9pt} {\text {s.t.}} {\kern 14pt} {\rm C1:} {\kern 3pt} {\bf{h}}_k^{\text T}{\bf{x}} = {\bf \Omega}_k^{\text T} {\bf s}_k, {\kern 3pt} \forall k \in {\cal K}\\
&{\kern 35pt} {\rm C2:} {\kern 3pt} t \le \alpha _m^{\cal O} , {\kern 3pt} \forall \alpha _m^{\cal O} \in {\cal O} \\
&{\kern 35pt} {\rm C3:} {\kern 3pt} t = \alpha _n^{\cal I} , {\kern 3pt} \forall \alpha _n^{\cal I} \in {\cal I}\\
&{\kern 35pt} {\rm C4:} {\kern 1pt} \left\| {{\bf{x}}} \right\|_2^2 \le {P_0}
\label{eq_25}
\end{aligned}
\end{equation}
The optimization problem formulation for APSK can be obtained in a similar way and is omitted for brevity. 

Compared to traditional SB optimizations that are in general difficult to handle, the CI-based SB optimization ${\cal P}_{\text {CSB}}^{\text {PSK}}$ and ${\cal P}_{\text {CSB}}^{\text {QAM}}$ are both convex and can be readily solved via convex optimization tools to obtain their optimal solutions. Moreover, compared to traditional SB optimizations where an average transmit power is maintained, the symbol-level CSB optimization guarantees that the transmit power constraint is strictly met on a symbol-by-symbol basis, as observed in the power constraint which is on a symbol level. In addition, a closer look shows that ${\cal P}_{\text {CVP}}$ that follows the symbol-scaling CI metric in Section III-A and ${\cal P}_{\text {CSB}}^{\text {PSK/QAM}}$ in Section III-C are indeed equivalent problems and return the same precoded signals, based on the observation that the objective function $t$ in the CSB optimization is equal to the inverse of the power scaling factor $\beta_{\text {CVP}}$, and therefore maximizing $t$ is equivalent to minimizing the power scaling factor.

To validate the superior performance of interference exploitation, we present a representative numerical BER result of CI precoding in Fig.~\ref{BER} for a $12 \times 12$ MU-MISO system, where symbol-level ZF precoding and RZF precoding are employed as benchmarks for fairness of comparison. For PSK modulations, we depict the results for both the `strict phase rotation' and `non-strict phase rotation' CI metric, while we employ the `symbol scaling' metric for QAM constellations. In Fig.~\ref{PSK_uncoded}, we observe that the SNR gain for CI precoding based on `non-strict phase rotation' is more than 10dB for both QPSK and 8PSK, when compared to the RZF precoding. When the QAM constellation is employed instead, as shown in Fig.~\ref{QAM_uncoded}, the SNR gain for CI precoding can still be up to 4.5dB for 16QAM and 2.5dB for 64QAM compared to RZF precoding. Additional numerical results presented in \cite{ci-27} show that the SNR gains can become more prominent when the system scales up. It is also observed that the performance gains extend to the case of coded BER results.

\subsection{Physical-Layer (PHY) Multicast Reformulation}
CI precoding has an interesting interpretation from the perspective of PHY multicasting \cite{ci-6}, \cite{ci-8}, \cite{ci-12} and \cite{ci-16}. PHY multicasting refers to the transmission where a common message is transmitted to all the receivers in the network, where no multi-user interference is observed since a single data stream is sent to all receivers \cite{multicast-1}\nocite{multicast-2}\nocite{multicast-3}\nocite{multicast-4}\nocite{multicast-5}\nocite{multicast-6}-\cite{multicast-7}. The typical PM and SB problem for PHY multicasting is studied in \cite{multicast-1}, where both problems can be transformed into SDP forms and solved via SDR. To circumvent the drawback that SDR-based PHY multicasting is only effective when there are not too many users, as demonstrated in \cite{multicast-2}, a stochastic precoding is proposed in \cite{multicast-2}, which is irrespective of the number of users in the system, and the case with finite-alphabet inputs is further studied in \cite{multicast-3}. Precoding for multiple co-channel multicast groups has been considered in \cite{multicast-4}-\cite{multicast-6}, where a total transmit power constraint and a per-antenna power constraint is considered, respectively. The extension of PHY multicasting to a multi-cell network is investigated in \cite{multicast-7}.

From the interference point of view, as firstly revealed in \cite{ci-5}, CI precoding considered in this paper resembles PHY multicasting precoding in a way that there exists no multi-user interference, where we note that CI precoding requires symbol-level operation, whereas multicasting does not. For PHY multicasting this is inherent because of single-stream transmission, while for CI precoding this is achieved by manipulating the interfering signals such that the resulting received signals lie in the constructive area and all the interference contributes to the useful signal power.

Mathematically, it has been shown in \cite{ci-8} and \cite{ci-16} that CPM optimization can be transformed and reformulated into a virtual PHY multicasting optimization. To be more specific, for the ${\cal P}_{\text {CPM}}^{\text {PSK}}$ problem shown in \eqref{eq_22}, by introducing a modified channel vector $\tilde{\bf h}_k=\frac{{{{\bf{h}}_k}}}{{{s_k}}}$ and with the introduced ${\bf x}={\bf Ws}$, ${\cal P}_{\text {CPM}}^{\text {PSK}}$ is equivalent to the following multicast problem:
\begin{equation}
\begin{aligned}
&\mathcal{P}_{\text {CPM}}^{\text {PHY-M}}: {\kern 3pt} \mathop {\min }\limits_{{{\bf{x}}}} \left\| {{\bf{x}}} \right\|_2^2  \\
&{\kern 0pt} {\text {s.t.}} {\kern 7pt} {\rm C1:} {\kern 0pt} \left[ {\Re \left( {{{{\tilde{\bf h}}}_k^{\text T}}{\bf{x}}} \right) - \sqrt {{\Gamma _k}{\sigma ^2}} } \right]\tan \theta_{\text {th}} \ge \left| {\Im \left( {{{{\tilde{\bf h}}}_k^{\text T}}{\bf{x}}} \right)} \right|, {\kern 0pt} \forall k \in {\cal K}
\label{eq_26}
\end{aligned}
\end{equation}
and it has been revealed in \cite{ci-8} that the optimal solution to ${\cal P}_{\text {CPM}}^{\text {PSK}}$ in \eqref{eq_22} and $\mathcal{P}_{\text {CPM}}^{\text {PHY-M}}$ in \eqref{eq_26} has the following connection:
\begin{equation}
{\bf{w}}_k^*{s_k} = \frac{{{{\bf{x}}^*}}}{K} {\kern 3pt} \Rightarrow {\kern 3pt} {\bf{w}}_k^*=\frac{{{{\bf{x}}^*}}}{K \cdot {s_k}}.
\label{eq_27}
\end{equation}
Again, compared to traditional PHY multicasting \cite{intro-9} that is non-convex and needs to be solved via SDR methods, the virtual PHY multicasting formulation for CI precoding is convex and can be readily solved. In addition to the above multicasting reformulation for the CI metric employed in \cite{ci-8}, some similar multicast reformulations can also be found in \cite{ci-6} and \cite{ci-12} for the CI metric considered in the corresponding works.

\begin{figure}[!t]
	\centering
	\includegraphics[scale=0.35]{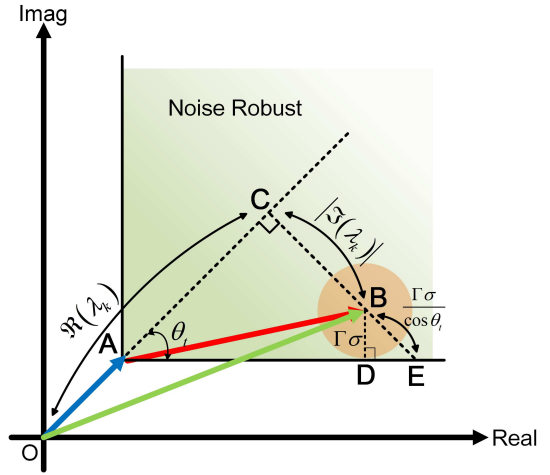}
	\caption{Characterization for noise-robust CI condition}
	\label{noise_robust}
\end{figure}

\subsection{Symbol Error Probability (SEP) Optimization}
It should be noted that many CI-based precoding works \cite{ci-3}-\cite{ci-12} have constructed their CI formulations based on the noiseless received signal ${\bf h}_k^{\text T}{\bf Ws}$, as observed in \eqref{eq_4} where the effect of noise is not taken into account, which may lead to sub-optimal solutions. To fill in this gap, \cite{ci-15}, \cite{ci-20}, \cite{dm-1} and \cite{dm-3} consider the noise-robust CI condition, where given a pre-defined noise variance, the received symbol still falls into the correct detection region of the constellation. As shown in Fig.~\ref{noise_robust}, for a given noiseless received signal $\vec{OB}$, the noise uncertainty region (denoted by the brown zone) needs to be incorporated into the precoding design, such that the received signal including noise still satisfies the SNR target $\Gamma_k$. Specifically, \cite{ci-20} aims to design the precoding matrix such that when the noise is located in the noise uncertainty region, the received signal including noise still falls into the constructive region of the constellations, which can be correctly decoded. Based on the geometry in Fig.~\ref{noise_robust}, the updated noise-robust CI condition for a $\cal M$-PSK constellation can be mathematically expressed as \cite{ci-20}
\begin{equation}
\Re \left( {\lambda_k} \right)\tan \frac{\pi }{\cal M} - \frac{{{\Gamma _k}\sigma }}{{\cos \frac{\pi }{\cal M}}} \ge \left| {\Im \left( {\lambda_k} \right)} \right|, {\kern 3pt} \forall k \in {\cal K}.
\label{eq_28}
\end{equation}
By incorporating the noise-robust CI condition \eqref{eq_28} into the precoding design, the corresponding CPM and CSB can be formulated as
\begin{equation}
\begin{aligned}
&\mathcal{P}_{\text {NR-CPM}}^{\text {PSK}}: {\kern 3pt} \mathop {\min }\limits_{{{\bf{x}}}} \left\| {{\bf{x}}} \right\|_2^2  \\
&{\kern 12pt} {\text {s.t.}} {\kern 18pt} {\rm C1:} {\kern 3pt} {{\bf{h}}_k^{\text T}}{\bf{x}} = {\lambda _k}{s_k}, {\kern 3pt} \forall k \in {\cal K}\\
&{\kern 42pt} {\rm C2:} {\kern 3pt} \Re \left( {\lambda_k} \right)\tan \frac{\pi }{\cal M} - \frac{{{\Gamma _k}\sigma }}{{\cos \frac{\pi }{\cal M}}} \ge \left| {\Im \left( {\lambda_k} \right)} \right|, {\kern 3pt} \forall k \in {\cal K}
\label{eq_NR_1}
\end{aligned}
\end{equation}
and
\begin{equation}
\begin{aligned}
&\mathcal{P}_{\text {NR-CSB}}^{\text {PSK}}: {\kern 3pt} \mathop {\max }\limits_{\bf{x}} {\kern 3pt} t  \\
&{\kern 12pt} {\text {s.t.}} {\kern 18pt} {\rm C1:} {\kern 3pt} {{\bf{h}}_k^{\text T}}{\bf{x}} = {\lambda _k}{s_k}, {\kern 3pt} \forall k \in {\cal K}\\
&{\kern 42pt} {\rm C2:} {\kern 3pt} \Re \left( {\lambda_k} \right)\tan \frac{\pi }{\cal M} - \frac{{{t}\sigma }}{{\cos \frac{\pi }{\cal M}}} \ge \left| {\Im \left( {\lambda_k} \right)} \right|, {\kern 3pt} \forall k \in {\cal K} \\
&{\kern 42pt} {\rm C3:} {\kern 2pt} \left\| {{\bf{x}}} \right\|_2^2 \le {P_0}
\label{eq_NR_2}
\end{aligned}
\end{equation}
respectively. Moreover, a closer observation shows that ${\cal P}_{\text {CSB}}^{\text {PSK}}$ in \eqref{eq_24} and ${\cal P}_{\text {NR-CSB}}^{\text {PSK}}$ are indeed equivalent problems and return the same solution, based on the fact that 
\begin{equation}
\begin{aligned}
&{\kern 3pt} \Re \left( {\lambda_k} \right)\tan \frac{\pi }{\cal M} - \frac{{{t}\sigma }}{{\cos \frac{\pi }{\cal M}}} \ge \left| {\Im \left( {\lambda_k} \right)} \right| \\
\Rightarrow & \left[ {\Re \left( {{\lambda _k}} \right) - \frac{{t\sigma }}{{\sin \frac{\pi }{\cal M}}}} \right]\tan \frac{\pi }{\cal M} \ge \left| {\Im \left( {{\lambda _k}} \right)} \right|
\end{aligned}
\label{eq_NR_3}
\end{equation}
and maximizing $\frac{{t\sigma }}{{\sin \frac{\pi }{\cal M}}}$ is equivalent to maximizing $t$ itself.

\cite{ci-20} also presents an alternative SEP-based CI precoding, where the detection-region based CI constraint in \eqref{eq_28} is replaced by a probabilistic constraint. By defining the SEP as the probability when the received signals including noise falls out of the decision region of a specific constellation point, where incorrect detection occurs, the SEP-based CI condition for a $\cal M$-PSK constellation can be derived as \cite{ci-20}
\begin{equation}
\Re \left( {{\lambda _k}} \right)\tan \frac{\pi }{\cal M} - \frac{{{\rm{er}}{{\rm{f}}^{ - 1}}\left( {1 - 2p} \right)\sigma }}{{\cos \frac{\pi }{\cal M}}} \ge \left| {\Im \left( {{\lambda _k}} \right)} \right|, {\kern 3pt} \forall k \in {\cal K},
\label{eq_29}
\end{equation}
where ${{\rm{er}}{{\rm{f}}^{ - 1}}\left( {\cdot} \right)}$ is the Gaussian error function. $p$ is the target SEP for a PM optimization, while $p$ becomes the objective function when a SB optimization is considered. By incorporating \eqref{eq_29} into the problem formulation, the corresponding SEP-based CPM and CSB optimizations can be constructed accordingly, which is omitted for brevity. The connection between CI precoding based on the noise-robust design in \eqref{eq_28} and CI precoding based on SEP in \eqref{eq_29} is also revealed in \cite{ci-20}.

\subsection{Closed-Form Iterative Precoding}
Compared to traditional block-level precoding approaches that are optimized when the channel changes, it should be noted that CI precoding has to perform optimizations on a symbol level, and the resulting computational complexity becomes an important issue for all CI-based precoding approaches. Therefore, low-complexity solutions and efficient algorithms have become an indispensable part for the realization of CI precoding. In the literature, several efficient algorithms have already been proposed for CI precoding, as detailed in \cite{ci-6}, \cite{ci-8}, \cite{ci-20}, \cite{ci-22}.

Thanks to the simple structures of the objective function as well as the linear constraints for CI-based PM and SB optimizations in Section III-B and Section III-C, recently it has been shown in \cite{ci-25}-\cite{ci-27} that there exist an optimal precoding structure for CI precoding. Based on the introduction of ${\bf x}={\bf Ws}$ and the observation that how the power is distributed among ${\bf w}_i s_i$ does not affect the solution, it is therefore safe to assume that each ${\bf w}_i s_i$ is identical, and the power constraint for the CSB optimization in \eqref{eq_24} and \eqref{eq_25} can be further transformed into \cite{ci-25}
\begin{equation}
\left\| {{\bf{x}}} \right\|_2^2 \le {P_0} \Rightarrow \sum\limits_{i = 1}^K {s_i^*{\bf{w}}_i^{\text H}{{\bf{w}}_i}{s_i}}  \le \frac{{{P_0}}}{K}.
\label{eq_30}
\end{equation}
Based on this equivalent power constraint, by analyzing the optimization problem ${\cal P}_{\text {CSB}}^{\text {PSK}}$ via Karush-Kuhn-Tucker (KKT) conditions and further formulating the dual problem, it is shown in \cite{ci-25} that the optimal precoded signal $\bf x$ for PSK modulation can be expressed as
\begin{equation}
{\bf{x}} ={{\bf{H}}^{\text H}}{\left( {{\bf{H}}{{\bf{H}}^{\text H}}} \right)^{ - 1}}{\text {diag}}\left( {\sqrt {\frac{{{P_0}}}{{{{\bf{u}}^{\text T}}{\bf{S}}{{\bf{T}}^{ - 1}}{{\bf{S}}^{\text T}}{\bf{u}}}}} {\bf{U}}{{\bf{T}}^{ - 1}}{{\bf{S}}^{\text T}}{\bf{u}}} \right){\bf s}.
\label{eq_31}
\end{equation}
We refer the interested readers to \cite{ci-25} for a detailed derivation and the expressions for $\bf S$, $\bf T$, and $\bf U$. Based on this closed-form expression, it is observed in \cite{ci-25} that CI precoding operates on a symbol level, as both $\bf T$ and $\bf u$ are dependent on the data symbol $\bf s$. Another interesting interpretation based on \eqref{eq_31} is that CI precoding can be viewed as the combination of ZF precoding with a pre-scaling operation that is adaptive to the data symbols. In fact, it has been shown in \cite{ci-25} that CI precoding can be regarded as a generalization of ZF precoding, and ZF precoding is a special case of CI precoding when the scaling factors $\lambda_k$ in \eqref{eq_4} are all identical. ${\bf u} \in {\cal R}^{2K \times 1}$ in \eqref{eq_31} needs to be optimized, which can be obtained via the following optimization:
\begin{equation}
\begin{aligned}
&{\cal P}_{\bf u}^{\text {PSK}}: {\kern 10pt} \mathop {\min }\limits_{\bf{u}} {\kern 3pt} {{\bf{u}}^{\text T}}\left( {{\bf{S}}{{\bf{T}}^{ - 1}}{{\bf{S}}^{\text T}}} \right){\bf{u}}  \\
& {\kern 8pt} {\text {s.t.}} {\kern 18pt} {\rm C1:} {\kern 3pt} {{\bf{1}}^{\text T}}{\bf{u}} = 1\\
&{\kern 38pt} {\rm C2:} {\kern 3pt} u_i \ge 0 , {\kern 3pt} \forall i \in \left\{ {1,2,\cdots,2K} \right\}
\label{eq_32}
\end{aligned}
\end{equation}
which is a standard QP optimization over a simplex. 

When a QAM constellation is considered instead, the optimal precoded signal can be obtained in a similar way, which is given by
\begin{equation}
{\bf{x}} = {{\bf{H}}^{\text H}}{\left( {{\bf{H}}{{\bf{H}}^{\text H}}} \right)^{ - 1}}{\bf{U}}\text{diag}\left( {\sqrt {\frac{{{P_0}}}{{{{\bf{u}}^{\text T}}{{\bf{V}}^{ - 1}}{\bf{u}}}}} {{\bf{F}}^{ - 1}}{{\bf{V}}^{ - 1}}{\bf{u}}} \right){{\bf{s}}_{\bf{E}}}.
\label{eq_33}
\end{equation}
The detailed derivations and notations can be found in \cite{ci-27}. The corresponding dual vector $\bf u$ can also be obtained via a QP optimization, while in the case of QAM modulation the QP is no longer optimized over a simplex, since only $u_m$ that corresponds to $\alpha_m^{\cal O} \in {\cal O}$ needs to be constrained as non-negative. By following the steps as in \cite{ci-25}-\cite{ci-27}, the optimal precoding structure for CPM optimizations and the corresponding QP formulations can be derived in a similar way. Meanwhile in the literature, an exact closed-form but sub-optimal solution has also been derived for CPM problems based on the DPCIR metric in \cite{ci-23}.

We further note the complexity reduction with the derived optimal precoding structure. Compared to the original CPM and CSB optimizations that belong to second-order cone programming (SOCP), the formulated optimizations for $\bf u$ are QP problems, which can be more efficiently solved using the interior-point methods, as widely acknowledged in the literature \cite{math-2}\nocite{math-3}-\cite{math-6}. Moreover, for both PSK and QAM constellations, an efficient iterative algorithm is also developed to solve the corresponding QP optimization, as detailed in \cite{ci-25} and \cite{ci-27}. This proposed algorithm allows a closed-form solution within each of its iteration, and obtains the optimal QP solution within only a few iterations. Another important feature for the iterative closed-form algorithm proposed in \cite{ci-25} and \cite{ci-27} is that it returns a feasible precoding matrix after each iteration, which is a great advantage over other efficient algorithms based on gradient descent method \cite{ci-6}, \cite{ci-8}, \cite{ci-22} or barrier method \cite{ci-20}. To be more specific, the iterative closed-form algorithm starts with ZF precoding, and evolves to the optimal CI precoding with the iteration number increasing, which offers a flexible performance-complexity tradeoff compared to other algorithms and makes it most appealing in practical systems, where performance has to be compromised for complexity reduction.

In addition, it should be highlighted that the above optimal precoding structure results and the corresponding QP formulations are based on the conventional case where $K \le N_t$ and ${\left( {{\bf{H}}{{\bf{H}}^{\text H}}} \right)^{ - 1}}$ is applicable. As already mentioned in Section I-C, CI precoding also supports the case of $K > N_t$, and the corresponding precoding structure for PSK and QAM modulation has been derived in \cite{ci-26} and \cite{ci-27}, respectively. By exploiting the singular value decomposition (SVD), the expression for $\bf W$ in the case of $K > N_t$ immediately follows \eqref{eq_31} and \eqref{eq_33} by substituting ${\left( {{\bf{H}}{{\bf{H}}^{\text H}}} \right)^{ - 1}}$ with the pseudo-inverse ${\left( {{\bf{H}}{{\bf{H}}^{\text H}}} \right)^{ +}}$ \cite{math-7}, while it should be noted that the expression for $\bf T$ in \eqref{eq_31} and $\bf V$ in \eqref{eq_33} is different in the case of $K > N_t$, which then leads to a different QP objective function.

\subsection{Robust CI Precoding}
The above studies and results on CI precoding have been derived based on the assumption that perfect CSI is known at the transmitter side, which is however not achievable since perfect CSI is not available in a practical wireless communication system. Therefore, it is important to consider a more realistic scenario and design robust CI-based precoding, when CSI is not perfect at the transmitter. 

In the case of imperfect CSI, the actual channel is usually modeled as
\begin{equation}
{{\bf{h}}_k} = {{\hat{\bf h}}_k} + {{\bf{e}}_k},
\label{eq_robust_1}
\end{equation}
where ${{\hat{\bf h}}_k}$ denotes the estimated CSI known at the BS, and ${{\bf{e}}_k}$ represents the CSI uncertainty. The additive channel estimation errors can be modeled into two different forms, dependent on the duplex mode of the communication system \cite{csi-model}. In the time-division duplex (TDD) mode, the CSI can be directly estimated at the BS using the uplink-downlink reciprocity and is subject to estimation errors. In this case, the entry of ${{\bf{e}}_k}$ can be modeled as a zero-mean Gaussian random variable, whose variance is inversely proportional to the transmit SNR. With this statistical CSI error model, the robust design is usually designed by optimizing the outage performance \cite{robust-4}, \cite{robust-5}. On the other hand, when the frequency-division duplex (FDD) mode is considered, the CSI error ${{\bf{e}}_k}$ is usually modeled as norm bounded to characterize the quantization errors incurred by limited feedback \cite{csi-model}. In this case, the channel uncertainty can be considered as bounded by a spherical region ${{\cal V}_k} = \left\{ {{{\bf{e}}_k} {\kern 2pt} | {\kern 1pt} \left\| {{{\bf{e}}_k}} \right\|_2^2 \le \delta _k^2} \right\}$, and the robust precoding algorithms are usually designed based on the worst-case received SINR \cite{robust-4}, \cite{robust-1}\nocite{robust-3}-\cite{robust-6}. The robust PM optimization in the case of norm-bounded CSI errors can usually be formulated as
\begin{equation}
\begin{aligned}
&\mathcal{P}_{\text {R-PM}}: {\kern 3pt} \mathop {\min }\limits_{{{\bf{w}}_k}} \sum\limits_{k = 1}^K {\left\| {{{\bf{w}}_k}} \right\|_2^2}  \\
&{\kern 10pt} {\text {s.t.}} {\kern 14pt} {\rm C1:} {\kern 3pt} \gamma_k \ge \Gamma_k, {\kern 3pt} \forall {{\bf{e}}_k} \in {{\cal V}_k}, {\kern 3pt} \forall k \in {\cal K}\\
&{\kern 36pt} {\rm C2:} {\kern 3pt} {{\bf{h}}_k} = {{\hat{\bf h}}_k} + {{\bf{e}}_k}, {\kern 3pt} \forall k \in {\cal K}
\label{eq_robust_2}
\end{aligned}
\end{equation}
This conventional robust PM problem $\mathcal{P}_\text{R-PM}$ can be transformed into a SDP form and solved via the SDR approach \cite{robust-9}.

To date, several works have considered the robust design for CI-based precoding \cite{ci-8}, \cite{robust-7}, \cite{robust-8}. In \cite{ci-8}, the robust version of CI precoding for both PM and SB optimization against norm-bounded CSI errors is studied, where the worst-case robust design for CI precoding is formulated based on the multicast formulation. By expanding the complex representation of the considered problem into its real equivalence, the robust version of the CI condition for a generic PSK modulation is derived. Mathematically, this leads to the following CI-based robust precoding design:
\begin{equation}
\begin{aligned}
&\mathcal{P}_{\text {R-CPM}}^{\text {PSK}}: {\kern 3pt} \mathop {\min }\limits_{{{\bf t}_1}, {{\bf t}_2}} \left\| {{\bf{t}}_1} \right\|_2^2  \\
&{\kern 0pt} {\text {s.t.}} {\kern 6pt} {\rm C1:} {\kern 3pt} {\left( {{\hat{\bf h}}_k^{\text{E}}} \right)^{\text T}}{{\bf{t}}_1} - {\left( {{\hat{\bf h}}_k^{\text{E}}} \right)^{\text T}}{{\bf{t}}_2}\tan {\theta _{\text {th}}} + {\delta _k}\left\| {{{\bf{t}}_1} - {{\bf{t}}_2}\tan {\theta _{\text {th}}}} \right\|_2\\
& {\kern 38pt} + \sqrt {{\Gamma _k}{\sigma ^2}} \tan {\theta _{\text {th}}} \le 0, {\kern 3pt} \forall k \in {\cal K} \\
& {\kern 6pt} {\rm C2:} {\kern 3pt} -{\left( {{\hat{\bf h}}_k^{\text{E}}} \right)^{\text T}}{{\bf{t}}_1} - {\left( {{\hat{\bf h}}_k^{\text{E}}} \right)^{\text T}}{{\bf{t}}_2}\tan {\theta _{\text {th}}} + {\delta _k}\left\| {{{\bf{t}}_1} - {{\bf{t}}_2}\tan {\theta _{\text {th}}}} \right\|_2\\
& {\kern 27pt} + \sqrt {{\Gamma _k}{\sigma ^2}} \tan {\theta _{\text {th}}} \le 0, {\kern 3pt} \forall k \in {\cal K} \\
&{\kern 6pt} {\rm C3:} {\kern 6pt} {{\bf{t}}_1} = {\bf \Pi} {{\bf{t}}_2}
\label{eq_26}
\end{aligned}
\end{equation}
where 
\begin{equation}
{\bf t}={\bf Ws}, {\kern 3pt} {{\bf{t}}_1} = {\left[ {{\bf{t}}_\Im ^{\text T},{\bf{t}}_\Re ^{\text T}} \right]^{\text T}}, {\kern 3pt} {{\bf{t}}_2} = {\left[ {{\bf{t}}_\Re ^{\text T}, - {\bf{t}}_\Im ^{\text T}} \right]^{\text T}}, 
\end{equation}
and
\begin{equation}
{\hat{\bf h}}_k^{\text{E}} = {\left[ {{{\left( {{\hat{\bf h}}_k^\Re } \right)}^{\text T}},{{\left( {{\hat{\bf h}}_k^\Im } \right)}^{\text T}}} \right]^{\text T}}, {\kern 3pt} {\bf \Pi}  = \left[ {\begin{array}{*{20}{c}}{{{\bf{0}}_K}}&{{{\bf{I}}_K}}\\ { - {{\bf{I}}_K}}&{{{\bf{0}}_K}}\end{array}} \right].
\end{equation} 
${\cal P}_{\text {R-CPM}}^{\text {PSK}}$ is a standard SOCP formulation and can be solved efficiently \cite{ci-8}. It is shown that compared to traditional PM and SB optimizations \cite{intro-6} and \cite{intro-7}, the transmit power savings for CI-based robust design can be as large as 3dB when the error bound is $\delta _k^2=0.0005$, which can become more prominent when the value of the error bound increases.

In addition, when only quantized CSI is available at the transmitter, \cite{robust-7} divides the interference into predictable interference that can be manipulated to be constructive, and unpredictable interference caused by the quantization in the CSI. Based on the quantized CSI error model, the proposed method in \cite{robust-7} aims to enhance the useful signal power by controlling the predictable interference, while minimizing the effect of uncertainty from unpredictable interference. An upper bound of the unpredictable interference is firstly derived, based on which the formulated optimization problem is transformed into a convex one and an iterative algorithm based on the gradient descent method is further introduced.

Alternative robust CPM optimizations have recently been proposed in \cite{robust-8}, where both the statistical and norm-bounded CSI error model are considered. When a norm-bounded CSI error model is assumed, the worst-case CI constraint is firstly formulated and a robust CPM optimization is proposed, which is similar to \cite{ci-8}. On the other hand, when the statistical CSI errors are assumed, the robust approach is designed based on the probabilistic CI constraints, which is equivalent to designing the precoding matrix such that the probability of violating the CI constraint is below a pre-defined threshold. By applying a decorrelation transformation and employing a lower bound instead, a linear inequality constraint is derived in \cite{ci-8}, based on which the robust CPM optimization against statistical CSI errors can be formulated as a convex optimization problem and solved efficiently.

\section{Applications of CI-based Precoding}
In the previous section, we have shown the CI formulation of conventional PM and SB optimizations, and numerically demonstrated significant performance improvements for CI precoding in terms of error rate and transmit power savings in Fig.~\ref{PM} and Fig.~\ref{BER}. In this section, we discuss how CI precoding can be adapted to other wireless scenarios and the potential gains of exploiting interference in these scenarios.

\subsection{Cognitive Radio (CR)}
Compared to traditional fixed spectrum allocation strategy, CR that enables dynamic spectrum access is an effective way to increase the radio resource utilization and spectral efficiency, which has been extensively studied \cite{cr-1}-\cite{cr-22}. Dependent on the priority of accessing the spectrum, the users are divided into primary users (PUs) and secondary users (SUs) in underlay CR networks, where PUs have the highest priority for the spectrum resources without being aware of the existence of the SUs in the network, while SUs can only access the network under the premise that their interference to PUs is below a certain threshold \cite{cr-4}. Accordingly, a fundamental challenge for CR networks is to enable the opportunistic spectrum access for SUs while guaranteeing PUs' quality-of-service (QoS) requirements, when the CSI of both PUs and SUs is available at the BS.

The tradeoff between throughput maximization and interference minimization for SUs is studied in \cite{cr-5} from an information-theoretic perspective, where the optimal transmission scheme that achieves the capacity of the secondary transmission as well as some sub-optimal algorithms is presented. In addition, precoding designs for CR networks are studied in \cite{cr-6}-\cite{cr-13} for both perfect CSI and imperfect CSI, where a precoding scheme termed MSLNR, which is a combination of the optimal minimum-mean-squared-error (MMSE) receiver and the signal-to-leakage-plus-noise ratio (SLNR) transmitter, is proposed in \cite{cr-6}. A joint downlink precoding and power control optimization is considered in \cite{cr-7} to maximize the weighted sum rate. The formulated non-convex problem is tackled by exploiting the non-negative matrix theory, where a convex approximation as well as a single-input multiple-output (SIMO)-MISO duality is further established. A robust precoding design for a single-SU CR network is considered in \cite{cr-8} in the presence of imperfect CSI, where the service probability of the SU is maximized through an iterative algorithm. Specifically, a closed-form solution can be obtained in the case of only one PU. A robust linear precoding design is further proposed in \cite{cr-9} for an underlay CR network with multiple PUs, where the formulated optimization is transformed into a convex-concave problem based on the uplink-downlink duality, and an iterative SDP-based algorithm is presented. Typically, a max-min fair problem in the CR Z-channel that aims to maximize the minimum SINR subject to average interference power to the PUs and total transmit power constraint at the secondary BS can be formulated as
\begin{equation}
\begin{aligned}
&\mathcal{P}_{\text {CR-SB}}: {\kern 3pt} \mathop {\max }\limits_{{{\bf{w}}_k}} \mathop {\min }\limits_k {\kern 3pt} {\gamma _k}  \\
&{\kern 10pt} {\text {s.t.}} {\kern 16pt} {\rm C1:} {\kern 3pt} \sum\limits_{i = 1}^K {\left| {{\bf{g}}_l^{\text T}{{\bf{w}}_i}} \right|^2}  \le {\varepsilon _l}, {\kern 3pt} \forall l \in {\cal L} \\
& {\kern 38pt} {\rm C2:} {\kern 3pt} \sum\limits_{k = 1}^K {\left\| {{{\bf{w}}_k}} \right\|_2^2}  \le P_0
\label{eq_CR_1} 
\end{aligned}
\end{equation}
where ${\cal L}=\left\{ {1,2,\cdots,L} \right\}$, ${\bf g}_l$ denotes the channel vector between the secondary BS and the $l$-th PU, and ${\varepsilon _l}$ is the maximum allowed interference power. $\rm C1$ therefore guarantees that the interference for the PUs from the secondary BS is below the required threshold. The formulated problem $\mathcal{P}_{\text {CR-SB}}$ can be solved via the bisection-search method and sequential QP \cite{cr-23}\nocite{cr-24}-\cite{cr-25}. Additional robust precoding designs for CR networks against imperfect CSI can be found in \cite{cr-10}-\cite{cr-13}, and CR has also been combined with relay in \cite{cr-14}-\cite{cr-16}.

CI-based precoding has shown to be effective in CR networks \cite{cr-17}-\cite{cr-22}. Early works include \cite{cr-17} and \cite{cr-18} based on selective CI precoding and correlation-rotation CI precoding, where a parallel transmission system aided with a cognitive relay is considered in \cite{cr-17} while an overlay CR network is investigated in \cite{cr-18}. The corresponding performance advantages of CI precoding over ZF precoding have been shown numerically in terms of SER, for both considered scenarios. More recently, the optimization-based CI precoding method proposed in \cite{ci-8} has been applied to the CR Z-channel in \cite{cr-19}, \cite{cr-20} and \cite{relay-1}. The considered optimization aims to minimize the worst SU's SER subject to the total transmit power and the interference to the PUs constraints, formulated as
\begin{equation}
\begin{aligned}
&\mathcal{P}_{\text {CI-CR}}^{\text {PSK}}: {\kern 3pt} \mathop {\min }\limits_{{\bf{x}}, {\kern 1pt}\rho} {\kern 3pt} \rho   \\
&{\kern 6pt} {\text {s.t.}} {\kern 12pt} {\rm C1:} {\kern 3pt} \Pr \left\{ {{s_k}{\kern 3pt}{\text {is incorrectly decoded}} {\kern 2pt} | {\kern 2pt} n_k} \right\} \le \rho, {\kern 2pt} \forall k \in {\cal K}\\
&{\kern 30pt} {\rm C2:} {\kern 4.5pt} {\left| {{\bf{g}}_l^{\text T}{\bf{x}}} \right|^2}  \le {\varepsilon _l}, {\kern 3pt} \forall k \in {\cal L}\\
&{\kern 30pt} {\rm C3:} {\kern 3pt} \left\| {{\bf{x}}} \right\|_2^2  \le P_0
\label{eq_CR_2} 
\end{aligned}
\end{equation}
Particularly, \cite{cr-20} derives the condition under which the formulated probabilistic precoding design $\mathcal{P}_{\text {CI-CR}}^{\text {PSK}}$ can be transformed into a convex deterministic optimization, based on which a simple approximation method that allows a convenient SOCP formulation is further introduced for additional reduction in the computational costs. Compared to traditional interference-reduction max-min fair optimizations, an SNR gain of 10dB can be observed for the CI-based precoding for the underlay CR Z-channel scenario in terms of SER, where the SBS is equipped with 10 transmit antennas with a total number of 8 SUs and 2 PUs in the network, as illustrated in \cite{cr-20}.

\subsection{Simultaneous Wireless Information and Power Transfer (SWIPT)}
Energy harvesting (EH) and wireless power transfer for wireless communication networks have become a new paradigm that allows user equipments (UEs) to prolong the battery life \cite{swipt-1}, \cite{swipt-21}. In wireless communications, the emitted RF signals carry both the information and energy at the same time, and therefore SWIPT techniques that enable the simultaneous transmission of information symbols and energy to the UEs have become particularly appealing \cite{swipt-2}-\cite{swipt-15}. In \cite{swipt-3}, three types of SWIPT scenarios are introduced: (i) separate UEs where each individual UE is acting as either an EH receiver or an information decoding (ID) receiver; (ii) time-switching UEs where the UEs are acting as EH receivers within some symbol durations while as ID receivers within other symbol durations; and (iii) power-splitting UEs where UEs divide the power of the received signals into two parts, one part for ID and the other part for EH.

Some precoding designs for MIMO SWIPT systems have been studied in \cite{swipt-4}-\cite{swipt-7}. \cite{swipt-4} and \cite{swipt-5} consider the separate-UE SWIPT scenario, where in \cite{swipt-4} the ZF precoding method is employed for MIMO SWIPT systems, and it is shown that the harvested energy obtained by the EH receivers can be increased at the cost of an SINR loss of the ID receivers. \cite{swipt-5} designs the precoding approaches which aim to maximize the harvested energy for EH receivers while guaranteeing the SINR target of the IDs, and the globally optimal solutions are obtained via the SDR method. A power-splitting SWIPT scenario is studied for a MISO multicasting system in \cite{swipt-6} and for a unicast system in \cite{swipt-7}, where the joint optimization on the precoding vector and the receive power splitting ratio is investigated. Based on \cite{swipt-7}, when power-splitting UEs are considered, the typical PM formulation for a unicast SWIPT system can be expressed as
\begin{equation}
\begin{aligned}
&\mathcal{P}_{\text {SWIPT}}: {\kern 3pt} \mathop {\min }\limits_{{{\bf{w}}_k}, {\kern 1pt} \rho_k} \sum\limits_{k = 1}^K {\left\| {{{\bf{w}}_k}} \right\|_2^2}  \\
&{\kern 10pt} {\text {s.t.}} {\kern 12pt} {\rm C1:} {\kern 3pt} \frac{{{{\left| {{\bf{h}}_k^{\text T}{{\bf{w}}_k}} \right|}^2}}}{{\sum\limits_{i \ne k} {{{\left| {{\bf{h}}_k^{\text T}{{\bf{w}}_i}} \right|}^2}}  + {\sigma ^2} + \frac{{\sigma _{\rm{C}}^2}}{{{\rho _k}}}}} \ge \Gamma_k, {\kern 3pt} \forall k \in {\cal K}\\
&{\kern 34pt} {\rm C2:} {\kern 3pt} \left( {1 - {\rho _k}} \right)\left( {\sum\limits_{i = 1}^K {{{\left| {{\bf{h}}_k^{\text T}{{\bf{w}}_i}} \right|}^2}}  + {\sigma ^2}} \right) \ge {E_k}, {\kern 3pt} \forall k \in {\cal K} \\
&{\kern 34pt} {\rm C3:} {\kern 5pt} 0 \le \rho_k \le 1, {\kern 3pt} \forall k \in {\cal K}
\label{eq_swipt_1}
\end{aligned}
\end{equation}
where $\sigma_{\rm C}^2$ is the additional noise power introduced in the RF to baseband conversion, $E_k$ is the harvested energy threshold for UE $k$, $\rho_k$ represents the fraction of power for ID, and $\left({1-\rho_k}\right)$ is the fraction of power for EH. $\rm C1$ and $\rm C2$ are the downlink received SINR and harvested energy requirements, respectively. This formulated PM problem, as well as the PM formulation for multicast case in \cite{swipt-6}, can be efficiently solved via the SDR method, which is shown to be tight in both scenarios. In addition to the above studies, the joint information and energy precoding methods for SWIPT are also investigated in MIMO interference channels in \cite{swipt-8}-\cite{swipt-11}, and SWIPT techniques have also been combined with PHY security in \cite{swipt-12}-\cite{swipt-15} by considering the broadcast nature of wireless communication systems.

The concept of CI exploitation has been applied to power-splitting SWIPT systems in \cite{swipt-16}-\cite{swipt-18}. Compared to traditional SWIPT systems that regard multi-user interference as harmful in the ID process, CI-based SWIPT precoding methods take advantages of the CI, which is inherent in the downlink transmission, as an additional power source for both useful information and wireless energy. A joint downlink precoding and receive power splitting optimization is formulated as a PM problem, subject to the QoS target for the ID receivers and EH thresholds for the EH receivers, given by 
\begin{equation}
\begin{aligned}
&\mathcal{P}_{\text {CI-SWIPT}}^{\text {PSK}}: {\kern 3pt} \mathop {\min }\limits_{{{\bf{x}}}, {\kern 1pt} \rho_k} \left\| {{\bf{x}}} \right\|_2^2  \\
&{\kern 0pt} {\text {s.t.}} {\kern 8pt} {\rm C1:} {\kern 3pt} {{\bf{h}}_k^{\text T}}{\bf{x}} = {\lambda _k}{s_k}, {\kern 3pt} \forall k \in {\cal K}\\
&{\rm C2:} {\kern 0pt} \left[ {\Re \left( {{\lambda _k}} \right) - \sqrt {{\Gamma _k}\left( {{\sigma ^2} + \frac{{\sigma _{\rm{C}}^2}}{{{\rho _k}}}} \right)} } \right]\tan {\theta _{\text {th}}} \ge \left| {\Im \left( {{\lambda _k}} \right)} \right|, {\kern 0pt} \forall k \in {\cal K} \\
&{\rm C3:} {\kern 2pt}\left| {{\bf{h}}_k^{\text T}{\bf{x}}} \right| \ge \sqrt {\frac{{{E_k}}}{{1 - {\rho _k}}}}, {\kern 3pt} \forall k \in {\cal K} \\
&{\rm C4:} {\kern 4pt} 0 \le \rho_k \le 1, {\kern 3pt} \forall k \in {\cal K}
\label{eq_swipt_2}
\end{aligned}
\end{equation}
Compared to PM problems for interference-reduction MIMO SWIPT systems where the PM problems can be transformed and readily solved via SDR approaches, CI-based PM problems for MIMO SWIPT systems are more difficult to handle due to the non-convex EH constraints. In \cite{swipt-16}, this is managed by using difference of convex optimization and the successive linear approximation method, which returns a local optimum solution. In \cite{swipt-17} and \cite{swipt-18}, SOCP-based and SDP-based algorithms with polynomial complexity are further introduced, which provide the upper and lower bounds to the optimal solution. By further conducting the asymptotic analysis, the optimality of these algorithms are established when the modulation order and SINR target go to infinity. By allowing a SLP design for MIMO SWIPT systems, the CI-based approach exhibits a transmit power saving as large as 8.6dBw when $E_k=-30$dBm and 9.2dBw when $E_k=-10$dBm, compared to traditional precoding designs based on SDR. 

\subsection{Physical-Layer (PHY) Security}
Compared with wired communications, a wireless communication system is naturally more susceptible and vulnerable to security threats due to its broadcast nature. Traditionally, the security issues are handled at the network layer by key-based cryptographic techniques. Nevertheless, PHY security approaches, which artificially add structured redundancy in the transmit signals in the physical layer such that the legitimate users can correctly decode the confidential information while the eavesdroppers (Eves) cannot, have drawn increasing research attention in the information-theoretic society in recent years \cite{security-1}-\cite{security-17}. By employing PHY transmission schemes that are specifically designed for security using multiple antennas, PHY security techniques can improve the information security and act as an additional security layer on top of the traditional cryptographic approaches \cite{security-1}-\cite{security-4}.

One possible approach for realizing PHY security is through downlink precoding, which is able to direct the signals carrying confidential information to the legitimate users while minimizing the power leakage to the Eves, as studied in \cite{security-5}-\cite{security-7}, where the secrecy rate maximization is discussed when the BS has either perfect or imperfect CSI of the Eves. Another promising approach for PHY security is to send generated artificial noise (AN) signals to interfere the Eves deliberately, as firstly proposed in \cite{security-8} and further investigated in \cite{security-9}-\cite{security-11}. The design for the AN-aided PHY security techniques is largely dependent on whether the BS has the CSI of the Eves. When no CSI of the Eves is available at the BS, a popular design is to generate an isotropic AN that is uniformly spread in the null space of the legitimate channel \cite{security-8}. By doing so, the communication for the legitimate users is not affected by the AN while the reception of the Eves is degraded. If the knowledge of the Eves' CSI is known perfectly or partially at the BS, this information can further be exploited to generate spatially selective AN that is much more effectively than the isotropic AN, as shown in \cite{security-9}-\cite{security-11}. Specifically, when the BS aims to transmit confidential messages to a legitimate user in the presence of $K$ Eves with full CSI available, the AN-aided secure PM optimization subject to QoS constraints can be formulated as \cite{security-16}
\begin{equation}
\begin{aligned}
&\mathcal{P}_{\text {PHY-S}}: {\kern 3pt} \mathop {\min }\limits_{{{\bf{w}}}, {\kern 1pt} {\bf z}} {\kern 3pt} \left\| {\bf{w}} \right\|_2^2 + \left\| {\bf{z}} \right\|_2^2 \\
&{\kern 10pt} {\text {s.t.}} {\kern 20pt} {\rm C1:} {\kern 3pt} \frac{{{{\left| {{\bf{h}}_{\text{L}}^{\text T}{\bf{w}}} \right|}^2}}}{{{{\left| {{\bf{h}}_{\text{L}}^{\text T}{\bf{z}}} \right|}^2} + {\sigma ^2}}} \ge {\Gamma _{\text{L}}} \\
&{\kern 42pt} {\rm C2:} {\kern 3pt} \frac{{{{\left| {{\bf{h}}_{{\text{E}},k}^{\text T}{\bf{w}}} \right|}^2}}}{{{{\left| {{\bf{h}}_{{\text{E}},k}^{\text T}{\bf{z}}} \right|}^2} + {\sigma ^2}}} \le {\Gamma _{{\text{E}},k}}, {\kern 3pt} \forall k \in {\cal K} \\
\label{eq_secure_1}
\end{aligned}
\end{equation}
where ${\bf h}_{\text L}$ is the channel between the legitimate user and the BS, and ${\bf h}_{{\text E},k}$ is the channel between the $k$-th eavesdropper and the BS. $\rm C1$ guarantees that the received SINR for the legitimate user meets a pre-defined threshold for correct detection, while $\rm C2$ ensures that the Eves cannot correctly decode the confidential information. The above optimization $\mathcal{P}_{\text {PHY-S}}$ can readily be solved by the SDR method as shown in \cite{security-9} and \cite{security-11}. Several other representative works on PHY security can be found in \cite{security-12}-\cite{security-14} for two-way relay networks and \cite{security-15} for transmit antenna selection (AS). 

CI-based precoding can be applied to PHY security techniques for additional performance improvements from the following two different perspectives \cite{security-16}-\cite{security-19}. On one hand, when the CSI of the Eves is not known to the BS, the AN design can be shifted from an isotropic null space based method to a CI-based method for transmit power savings, which is similar to the idea of conventional interference exploitation, as shown in \cite{security-18}. On the other hand, in the case where partial/full CSI is available, it allows a more advanced CI-based approach where the AN signals are designed to be constructive to the legitimate users and destructive to the Eves at the same time, which further reduces the required transmit power at the BS while impeding the signal detection at the Eves. The corresponding optimization problem employing PSK modulations can be formulated as \cite{security-18}
\begin{equation}
\begin{aligned}
&\mathcal{P}_{\text {CI-PHY-S}}^{\text {PSK}}: {\kern 3pt} \mathop {\min }\limits_{{{\bf{x}}_{\text L}}, {\bf z}} \left\| {{\bf x}_{\text L}+{\bf z}} \right\|_2^2  \\
&{\kern 0pt} {\text {s.t.}} {\kern 6pt} {\rm C1:} {\kern 3pt} {\bf{h}}_{\text{L}}^{\text T}\left( {{{\bf{x}}_{\text{L}}} + {\bf{z}}} \right) = {\lambda _{\text{L}}}{s_{\text{L}}}\\
&{\kern 10pt} {\rm C2:} {\kern 3pt} {\bf{h}}_{{\text{E}},k}^{\text T}\left( {{{\bf{x}}_{\text{L}}} + {\bf{z}}} \right) = {\lambda _{{\text{E}},k}}{s_{\text{L}}}, {\kern 3pt} \forall k \in {\cal K}\\
&{\kern 10pt} {\rm C3:} {\kern 0pt} \left[ {\Re \left( {{\lambda _{\text{L}}}} \right) - \sqrt {{\Gamma _{\text{L}}}{\sigma ^2}} } \right]\tan {\theta _{\text {th}}} \ge \left| {\Im \left( {{\lambda _{\text{L}}}} \right)} \right|\\
&{\kern 10pt} {\rm C4:} {\kern 0pt} \left[ {\Re \left( {{\lambda _{{\text{E}},k}}} \right) - \sqrt {{\Gamma _{{\text{E}},k}}{\sigma ^2}} } \right]\tan {\theta _{\text {th}}} \le \Im \left( {{\lambda _{{\text{E}},k}}} \right), {\kern 3pt} \forall k \in {\cal K} \\
&{\kern 10pt} {\rm C5:} {\kern 0pt} \left[ {\Re \left( {{\lambda _{{\text{E}},k}}} \right) - \sqrt {{\Gamma _{{\text{E}},k}}{\sigma ^2}} } \right]\tan {\theta _{\text {th}}} \ge - \Im \left( {{\lambda _{{\text{E}},k}}} \right), {\kern 3pt} \forall k \in {\cal K}
\label{eq_secure_2}
\end{aligned}
\end{equation}
where $\rm C3$ guarantees CI for the legitimate user, while $\rm C4$ and $\rm C5$ ensure that the AN signals are destructive to the Eves. $\mathcal{P}_{\text {CI-PHY-S}}^{\text {PSK}}$ is a standard SOCP problem and can be efficiently solved with interior-point based solvers. Similar benefits for CI-based design over interference-reduction designs are also observed when PHY security is combined with SWIPT, as studied in \cite{security-19}.

\subsection{Full-Duplex (FD) Communications}
Traditional wireless communication systems work either in FDD or TDD mode, with both being the half-duplex (HD) model. To meet the high demand for spectral efficiency and QoS requirement of the future wireless systems, FD communications, which allow simultaneous transmission and reception, have been considered as one of the solutions \cite{full-duplex-0}-\cite{full-duplex-4}. While FD transmission can theoretically double the spectral efficiency of the HD systems, the self interference from the transmit antennas to the receive antennas of the FD transceivers can severely affect the quality of communication in practical scenarios. Thanks to the recent major breakthroughs for practical FD systems which allow a self interference cancellation of up to 100dB with the use of additional hardware, as illustrated in \cite{full-duplex-5}, \cite{full-duplex-6}, FD communications have received increasing research attention in recent years \cite{full-duplex-7}-\cite{full-duplex-13}.

In \cite{full-duplex-7}, a pricing-based precoding method specifically tailored for the suppression of self interference is proposed, which guarantees the linearity of the receiver and achieves a spectral efficiency that is nearly 1.8 times of a HD system. In \cite{full-duplex-8}, precoding approaches for a sum-rate maximization in FD systems are designed based on the sequential convex programming, and a joint consideration of forward precoding and self interference cancellation is further studied in \cite{full-duplex-9}. A downlink precoding and uplink multi-user combining problem for sum-rate maximization in FD systems is investigated in \cite{full-duplex-10}. The formulated non-convex optimization is handled by exploiting the uplink-downlink duality, and a minorization-maximization (MM) algorithm is proposed. The practical deployment of a multi-user MIMO system with a FD BS and HD UEs is considered in \cite{full-duplex-11}, where a joint optimization in precoding designs and uplink/downlink user selection is proposed. A multi-objective optimization problem that aims to jointly minimize the total downlink and uplink transmit power for the considered FD system is studied in \cite{full-duplex-12} and \cite{full-duplex-13}, which can be formulated as
\begin{equation}
\begin{aligned}
&\mathcal{P}_{\text {FD}}: {\kern 3pt} \mathop {\min }\limits_{{{\bf{w}}_k},{P_j}} \mathop {\max }\limits_{a = 1,2} {\kern 3pt} \left\{ {{\eta _a}\left( {Q_a^* - {Q_a}\left( {{{\bf{w}}_k},{P_j}} \right)} \right)} \right\}  \\
&{\kern 0pt} {\text {s.t.}} {\kern 10pt} {\rm C1:} {\kern 3pt} \frac{{{{\left| {{\bf{h}}_k^{\text T}{{\bf{w}}_k}} \right|}^2}}}{{\sum\limits_{i \ne k} {{{\left| {{\bf{h}}_k^{\text T}{{\bf{w}}_i}} \right|}^2}}   + {\sigma ^2}}} \ge \Gamma _k^{{\text{DL}}}, {\kern 3pt} \forall k \in {\cal K} \\
&{\rm C2:} {\kern 3pt} \frac{{{P_j}{{\left| {{\bf{u}}_j^{\text T}{{\bf{g}}_j}} \right|}^2}}}{{\sum\limits_{n \ne j} {{P_n}{{\left| {{\bf{u}}_j^{\text T}{{\bf{g}}_n}} \right|}^2}}  + \sum\limits_{k = 1}^K {{{\left| {{\bf{u}}_j^{\text T}{{\bf{H}}_{{\rm{SI}}}}{{\bf{w}}_k}} \right|}^2}}  + {\sigma ^2}\left\| {{{\bf{u}}_j}} \right\|_2^2}} \ge \Gamma _j^{{\text{UL}}}\\
& {\kern 213pt} \forall j \in {\cal J}\\
&{\rm C3:} {\kern 3pt} P_j \ge 0, {\kern 3pt} \forall j \in {\cal J}
\label{eq_FD_1} 
\end{aligned}
\end{equation}
where $\rm C1$ and $\rm C2$ are to guarantee the downlink and uplink SINR requirement, respectively. We refer the interested readers to \cite{full-duplex-12}, \cite{full-duplex-13}, and \cite{full-duplex-22} for a detailed explanation on the notations in $\mathcal{P}_\text{FD}$, where it is also shown that the formulated precoding design for FD communications can be solved via the SDR approach. Additional studies on FD communications include applications to PHY security in \cite{full-duplex-14}-\cite{full-duplex-16}, and FD relays in \cite{full-duplex-17}-\cite{full-duplex-20}.

CI exploitation can also be applied to FD communications for additional performance improvements, as recently studied in \cite{full-duplex-21}-\cite{full-duplex-23} for both PSK and QAM constellations. The CI-based PM problem in a multi-user FD system is considered in \cite{full-duplex-21} and \cite{full-duplex-22}, where a multi-objective optimization via the weighted Chebycheff method is employed to study the tradeoff between the two desirable design objectives, which are the total downlink transmit power at the BS and the total uplink power from the UEs. When PSK constellations are considered, the corresponding CI-based optimization for FD communications can be formulated as \cite{full-duplex-22}
\begin{equation}
\begin{aligned}
&\mathcal{P}_{\text {CI-FD}}^{\text {PSK}}: {\kern 3pt} \mathop {\min }\limits_{{{\bf{x}}},{P_j}, t} {\kern 3pt} t  \\
&{\kern 0pt} {\text {s.t.}} {\kern 10pt} {\rm C1:} {\kern 3pt} {\bf h}_k^{\text T}{\bf{x}} = {\lambda _k}{s_k}, {\kern 3pt} \forall k \in {\cal K} \\
&{\rm C2:} {\kern 0pt} \left[ {\Re \left( {{\lambda _k}} \right) - \sqrt {\Gamma _k^{{\text{DL}}}{\sigma ^2}} } \right]\tan {\theta _{\text {th}}} \ge \left| {\Im \left( {{\lambda _k}} \right)} \right|, {\kern 3pt} \forall k \in {\cal K}\\
&{\rm C3:} {\kern 2pt} \frac{{{P_j}{{\left| {{\bf{u}}_j^{\text T}{{\bf{g}}_j}} \right|}^2}}}{{\sum\limits_{n \ne j} {{P_n}{{\left| {{\bf{u}}_j^{\text T}{{\bf{g}}_n}} \right|}^2}}  + {{{\left| {{\bf{u}}_j^{\text T}{{\bf{H}}_{{\text{SI}}}}{{\bf{x}}}} \right|}^2}}  + {\sigma ^2}\left\| {{{\bf{u}}_j}} \right\|_2^2}} \ge \Gamma _j^{{\text{UL}}}\\
& {\kern 216pt} \forall j \in {\cal J}\\
&{\rm C4:} {\kern 3pt} {\eta _a}\left( {Q_a^* - {Q_a}\left( {{{\bf{x}}},{P_j},t} \right)} \right) \le t, {\kern 3pt} \forall a \in \left\{{1,2} \right\} \\
&{\rm C5:} {\kern 3pt} P_j \ge 0, {\kern 3pt} \forall j \in {\cal J}
\label{eq_FD_2} 
\end{aligned}
\end{equation}
where $\rm C2$ is the CI-adapted SINR requirement for the downlink transmission. $\mathcal{P}_{\text {CI-FD}}^{\text {PSK}}$ is shown to be a convex problem and can be readily solved. Numerical results have shown a 2dBw downlink transmit power saving and a 7dBw uplink transmit power saving by exploiting the CI in a system where FD BS is equipped with 8 antennas with a total number of 6 downlink users and 3 uplink users, which also leads to the substantial reduction in the self interference. Importantly, the multi-objective framework allows the flexible tradeoff between the uplink power savings and the downlink power savings, which further leads to an improvement in the overall energy efficiency. In addition, CI-based FD communications have also been considered in the presence of imperfect CSI in \cite{full-duplex-23}, where a robust design for the joint minimization of the uplink and downlink transmit power and maximization of the harvested energy is studied subject to channel estimation errors. It is shown by numerical results that the gain in the harvested energy can be as large as 5dBm with an increase in the downlink transmit power.

\subsection{Multi-Cell Distributed Antennas (DAs)}
A DA system is a wireless communication architecture where multiple geographically DAs are connected to a central processing unit through distributed remote radio heads (RRHs) \cite{das-0}\nocite{das-00}\nocite{das-1}\nocite{das-2}\nocite{das-3}\nocite{das-4}\nocite{das-5}-\cite{das-6}, where in the literature it is also referred to as `BS cooperation', `network MIMO' or `multi-cell processing', etc. While consuming an increased backhaul requirement compared to traditional centralized multi-antenna BSs, DA systems intuitively have the advantage of greatly reducing the path loss of the communication links, thanks to the reduced distance between the transmitting antennas and receiving UEs. Therefore, the DA system is a promising architecture to reduce the required transmit power of the BS for a fixed channel quality and leads to a more uniform coverage inside the cell \cite{das-1}, \cite{das-2}. Due to these benefits, DA systems have received considerable research attention, which includes the studies on spectral efficiency, energy efficiency and their tradeoffs in \cite{das-3}-\cite{das-5}, power allocation schemes for energy efficiency maximization in \cite{das-6}\nocite{das-7}-\cite{das-8}, precoding designs in \cite{das-8}\nocite{das-9}\nocite{das-10}-\cite{das-11}, and AS strategies in \cite{das-12}. 

More recently, several DA systems have also been proposed in the massive MIMO regime, for example the `cell-free massive MIMO' in \cite{das-13}\nocite{das-14}\nocite{das-15}\nocite{das-16}-\cite{das-17}. Cell-free massive MIMO systems employ a large number of access points (APs) equipped with single antenna or a few antennas geographically distributed over a wide area, which exhibits significant throughput improvements compared to small-cell deployments \cite{das-13}. In a cell-free massive MIMO system, the concept of `cell' does not exist, since the entire area is covered by distributed APs that cooperate phase-coherently via a backhaul network to the central processing unit, which names this DA system \cite{das-13}. One key feature of cell-free massive MIMO systems is that, only local signal processing at each RRH is sufficient without the need for centralized processing, when the MRT precoding is employed \cite{das-13}-\cite{das-15}. In addition to the simple MRT precoding, \cite{das-16} and \cite{das-17} investigate the cell-free massive MIMO systems when ZF precoding is employed, where power control algorithms are presented for SB optimization in \cite{das-16} and for energy efficiency maximization in \cite{das-17}. In addition to cell-free massive MIMO architectures, there is another promising DA system termed `fog massive MIMO' that has recently been proposed in \cite{das-18} and \cite{das-19}, which exploits a small number of RRHs with each RRH deploying a large-scale antenna array.

The concept of CI precoding has been extended to DA systems in \cite{das-20}, where \cite{das-20} focuses on the PHY security enhanced by CI precoding for a user-centric DA system. A joint CI-based secure precoding and a DA selection method is proposed for transmit PM problem, in which the AN signals are designed to be constructive to the UEs while destructive to the Eves. The formulated problems are discussed for the case when only imperfect CSI is known as well as the case when the CSI of the Eves is completely unknown. In addition, CI precoding has also been extended to a multi-cell scenario in \cite{das-21} based on both deterministic optimization and probabilistic optimization, where several coordination schemes are proposed for downlink transmission. A fully-coordinated strategy is firstly considered, where both the intra-cell interference and the inter-cell interference are exploited for performance improvements based on the CI formulation, by sharing the information of both the channel and data symbols among the BSs. A partially-coordinated scheme is also introduced for coordination overhead reduction, where the intra-cell interference is manipulated to be constructive while the inter-cell interference is suppressed by only sharing the CSI among the BSs. In a numerical example where 3 BSs cooperate with 3 users in each cell, \cite{das-21} shows a 4dBm power saving gain based on the deterministic optimization and a 6dBm power saving gain based on the probabilistic optimization.

\subsection{Spatio-Temporal CI: Faster-than-Nyquist Signaling}
Faster-than-Nyquist (FTN) signaling \cite{lux-5}\nocite{lux-6}\nocite{lux-7}\nocite{lux-8}\nocite{lux-9}-\cite{lux-10} is a signal processing scheme allowing a notable improvement of the spectral efficiency of wireless communication systems. The key idea of FTN signaling is a reduction of the time spacing between two adjacent pulses (the symbol period) below the one satisfying the Nyquist condition. In other words, in FTN signaling the data rate is increased by accelerating the transmitted pulses in the temporal dimension (time packing), thus introducing controlled inter-symbol interference (ISI) which needs to be handled. The main problem of FTN signaling is the need to cope with the introduced ISI, which in turn results in complex receivers relying on trellis decoders as well as ad-hoc equalization schemes, whose computational costs are often prohibitive in practical applications. In \cite{lux-11}\nocite{lux-12}-\cite{lux-13}, a novel transmission technique has been proposed, which merges the aggressive frequency reuse relying on precoding, in particular SLP, with FTN signaling. In a generic MU-MISO system, these works extend the concept of SLP at the transmitter side in order to tackle not only the interference in the spatial dimension (the multi-user interference), but also the interference in the temporal dimension (the ISI intentionally introduced through FTN signaling). Such an extension allows FTN signaling in a MU-MISO framework and, at the same time, solves the problem of complex FTN receivers, as the ISI is completely handled at the transmitter. This transmission technique is referred to as spatio-temporal CI, as it enhances the CI both in the temporal and in the spatial dimensions, thus gleaning benefits from both the domains. 

The application of SLP in the context of FTN signaling relies on a new system model, which takes into account the temporal variation of the transmitted streams at each antenna by modeling the pulse shaping filters. Considering a MU-MISO system with $N_\text{T}$ transmit antennas and $K$ single-antenna user terminals, the main idea is to split each data stream in temporal blocks of $S$ symbols. The data symbols for the different users, for a given block, can be represented in a matrix ${\bf S}=\left[ {{\bf s}_1 \ldots {\bf s}_K} \right]^\text{T} \in {\cal C}^{K \times S}$, while the precoded symbol streams can be aggregated in a matrix ${\bf D} = \left[{{\bf d}_1 \ldots {\bf d}_{N_\text{T}}} \right]^\text{T} \in {\cal C}^{N_\text{T} \times S}$. Denoting $T$ as the symbol period and $\mu$ as the oversampling factor, the pulse-shaped transmitted waveform for the generic $n$-th antenna can be represented through its discrete samples, spaced by $t_s = \frac{T}{\mu}$, given by
\begin{equation}
\label{pulse_shaping}
x_n[m] = \sum_{j = 1}^{S}{d_n[j]\alpha[(m-1)t_s - (j-1)T]}, {\kern 3pt} m = 1, \ldots, \mu S,
\end{equation}
where $\alpha(t)$ represents the considered unit energy pulse and $d_n[j]$ is the $j$-th element of the symbol vector ${\bf d}_n$. The output (oversampled) signals from all the antennas can be aggregated in a matrix ${\bf X} = \left[{ {\bf x}_1 \ldots {\bf x}_{N_\text{T}}}\right]^\text{T} \in {\cal C}^{N_\text{T} \times \mu S}$. With this definition, a compact way to represent the pulse shaping operation is ${\bf X} = {\bf D}{\bf A}_\text{TX}$, with ${\bf A}_\text{TX} \in {\cal R}^{S \times \mu S}$ being a block Toeplitz matrix including the filter taps.

\begin{figure}[!t]
\centering
\includegraphics[width=\columnwidth]{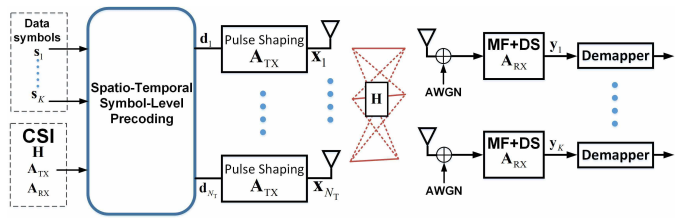}\\
\caption{Block diagram of the SLP approach based on spatio-temporal CI}\label{fig:block_scheme}
\end{figure}

By aggregating the received symbols at the $K$ users in a matrix ${\bf Y} \in {\cal C}^{K \times S}$, the spatio-temporal communication model can be written as \cite{lux-11,lux-13}
\begin{equation}
\label{global_comm_model}
{\bf Y} =  {\bf H}{\bf X}{\bf A}_\text{RX} + \tilde{\bf{Z}}\bf{A}_\text{RX} = \bf{H}\bf{D}\bf{A} + \bf{Z},
\end{equation}
with $\tilde{\bf{Z}}$ being the noise in the oversampled domain, ${\bf A}_\text{RX} \in {\cal R}^{\mu S \times S}$ modeling the matched filtering and downsampling operation performed at each receiver, and ${\bf A} = {\bf A}_\text{TX}{\bf A}_\text{RX} \in {\cal R}^{S \times S}$ representing the combination of the filters at the transmitter and at the receiver. The overall spatio-temporal system model, accounting for the ISI through $\bf{A}$ and the multi-user interference through $\bf{H}$, is represented in the block scheme of Fig. \ref{fig:block_scheme}. By vectorizing the introduced signal matrices over the temporal dimension, the spatio-temporal communication model can be expressed as
\begin{equation}
\label{stacked_global_comm_model}
\bf{y}=\left({{\bf H}\otimes{\bf A}^{\text T}}\right)\bf{d}+\bf{z}=\bf{G}\bf{d}+\bf{z},
\end{equation}
which is formally similar to the spatial model of (3) used in the traditional SLP literature. The matrix ${\bf G} = {\bf H}\otimes {\bf A}^{\text T} \in {\cal C}^{KS \times N_\text{T} S}$ is an equivalent representation of the channel matrix in this spatio-temporal model.

\begin{figure*}
\begin{centering}
\subfloat[Constant-envelope precoding]
{\begin{centering}
\includegraphics[width=3.5cm]{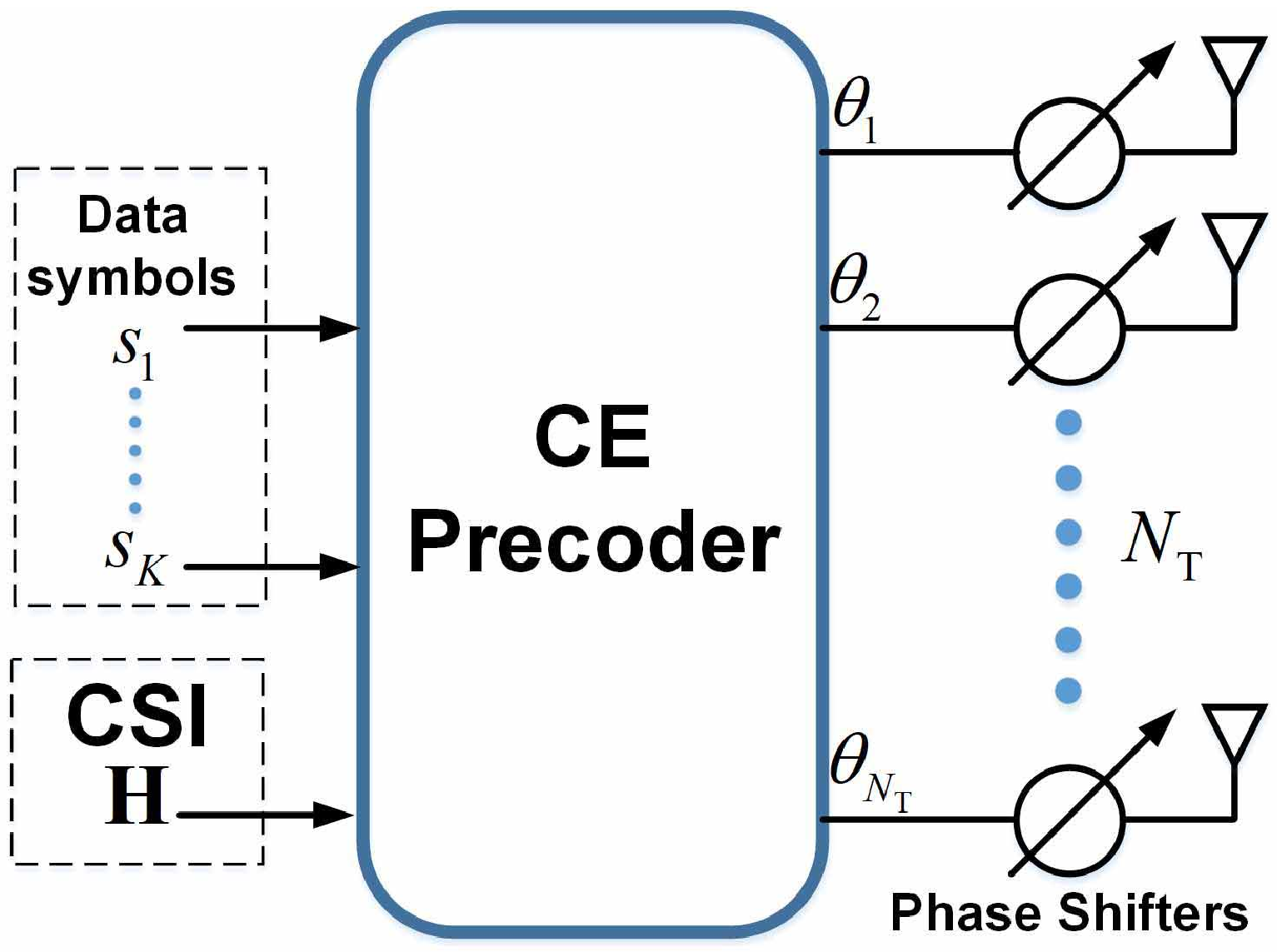}
\label{CEP}
\par
\end{centering}
}
\hspace{0cm}
\subfloat[Antenna selection]
{\begin{centering}
\includegraphics[width=3.9cm]{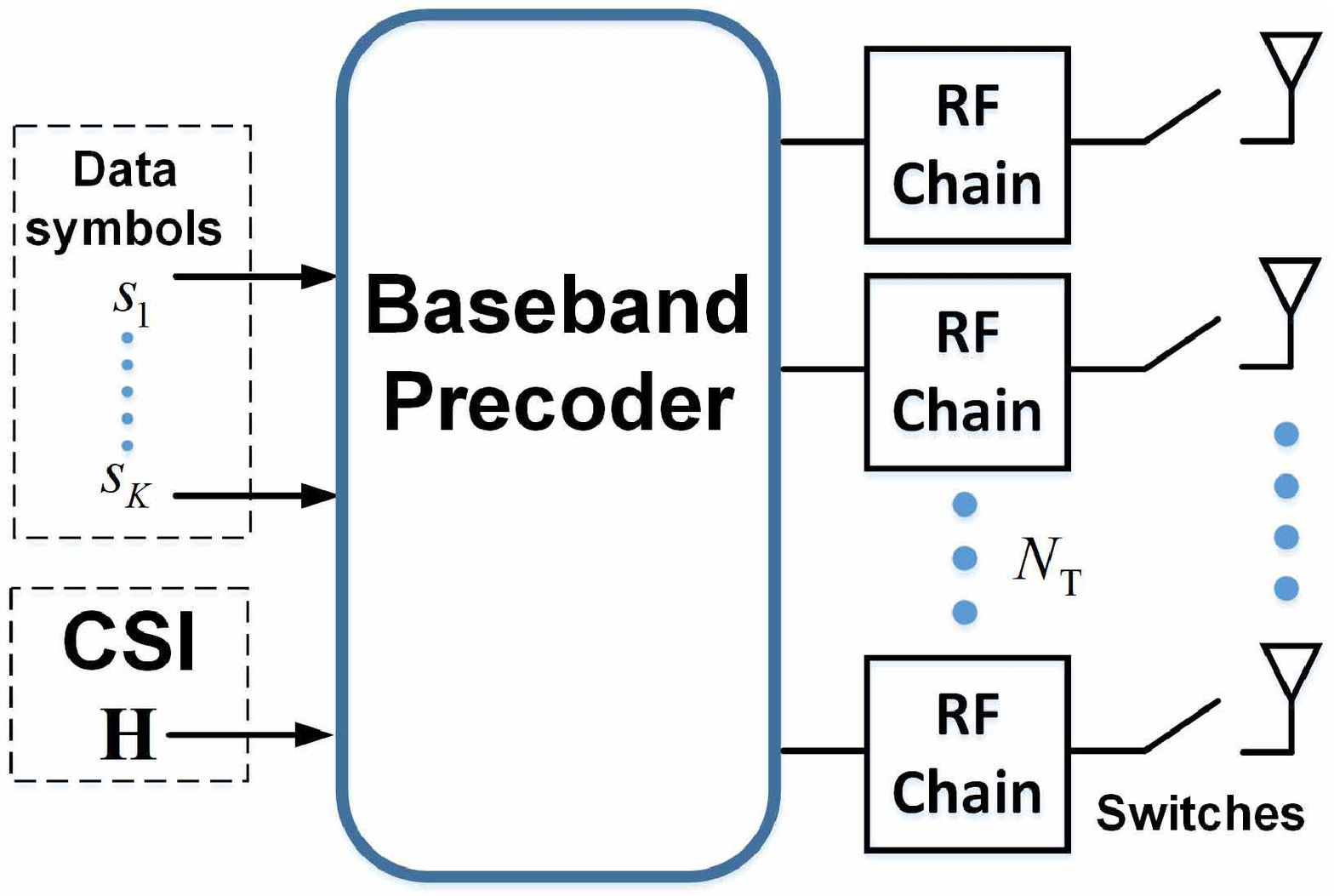}
\label{AS}
\par
\end{centering}
}
\hspace{0cm}
\subfloat[Hybrid precoding]
{\begin{centering}
\includegraphics[width=4.9cm]{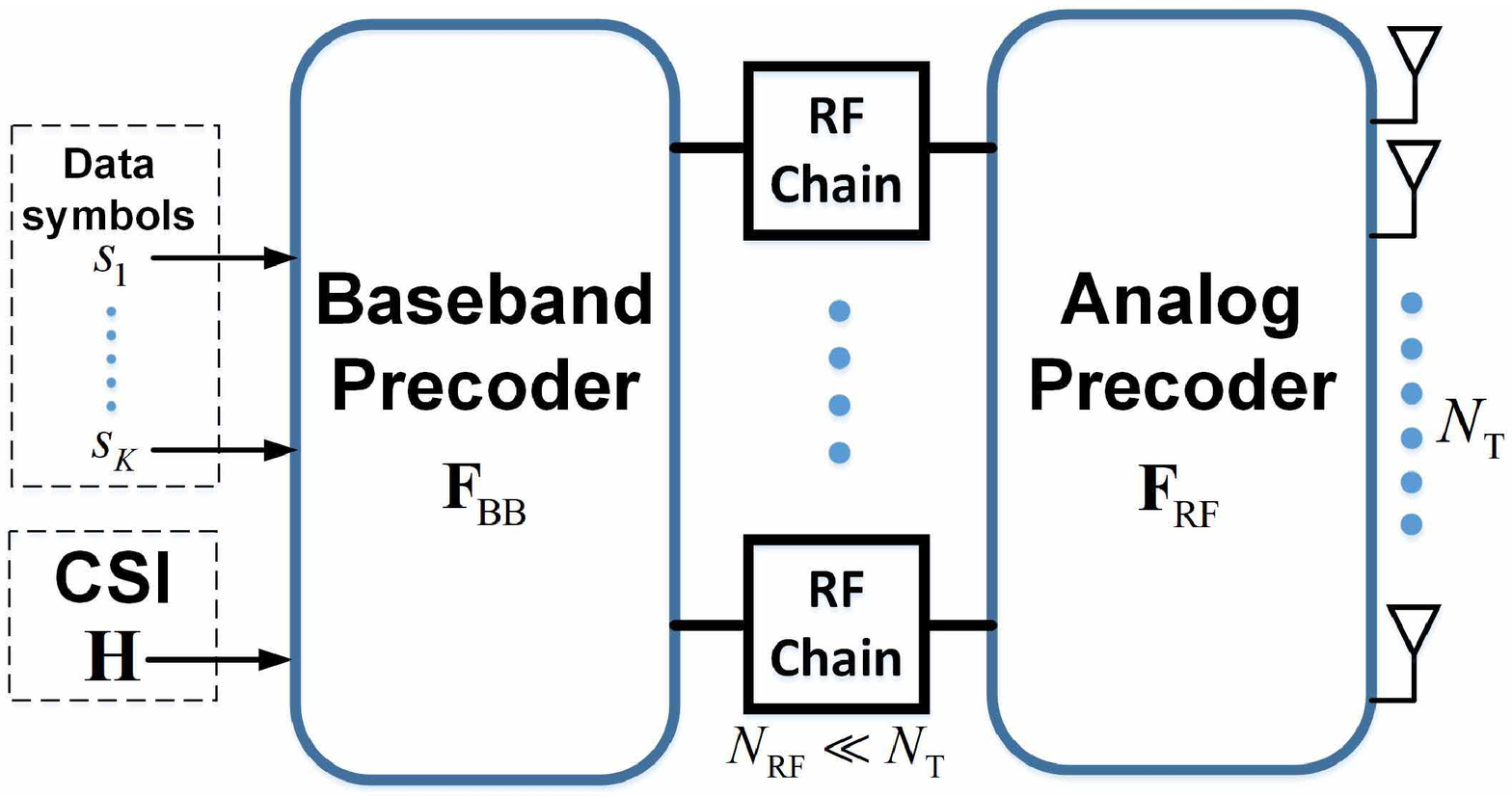}
\label{Hybrid}
\par
\end{centering}
}
\hspace{0cm}
\subfloat[Low-resolution DACs]
{\begin{centering}
\includegraphics[width=3.6cm]{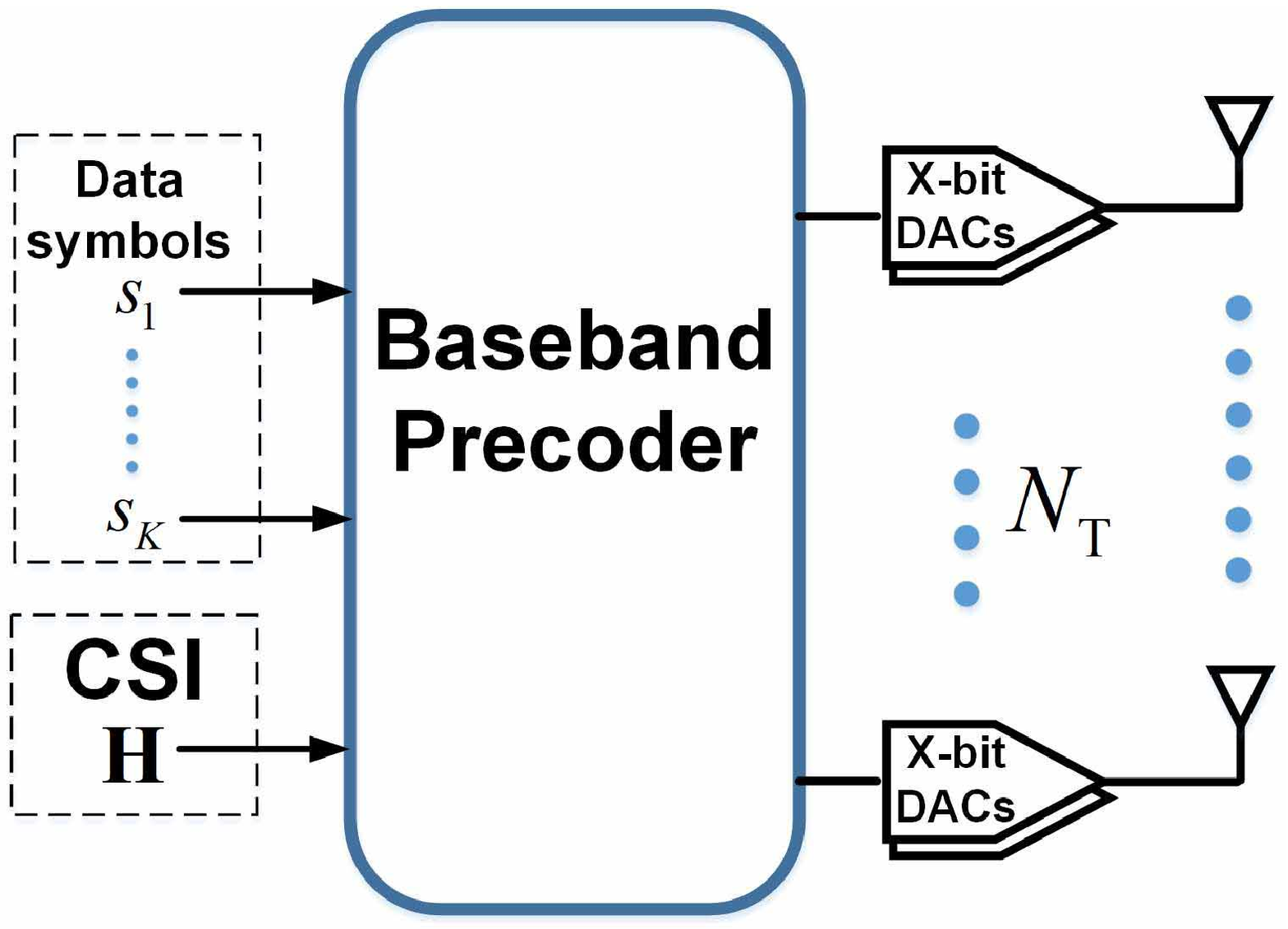}
\label{DAC}
\par
\end{centering}
}
\par
\end{centering}
\caption{\label{he}A variety of hardware-efficient BS architecture}
\end{figure*}

The above spatio-temporal communication model has been used in \cite{lux-11} to formulate a CI-PM problem with QoS constraints, which is mathematically equivalent to the PM problem of \cite{lux-14}. The related problem is convex and can be solved resorting to CVX. Moreover, a sequential approach has been proposed in \cite{lux-11} and \cite{lux-13}, where the residual ISI amongst the subsequent temporal blocks is also tackled. Numerical results have shown that SLP schemes based on FTN can outperform the Nyquist-based counterparts both in terms of energy efficiency and achievable rate, with performance gains up to 25\%.

\section{Symbol-Level Precoding for Hardware Efficiency}
In this section, we further extend the concept of CI exploitation to hardware-constrained large-scale antenna systems for hardware efficiency, which includes constant-envelope precoding (CEP), AS, hybrid analog-digital (AD) precoding, low-resolution digital-to-analog converters (DACs), non linearities, and RF-domain SLP. Aimed at improving both the cost efficiency and energy efficiency of the BS, these schemes attempt to reduce the complexity of the BS architecture and the number of some hardware components, as shown in Fig.~\ref{he} and Fig.~\ref{RF-domain}, respectively.

\subsection{Constant-Envelope Precoding (CEP)}
In recent years, the use of a large-scale antenna array at the transmitter has been shown to offer remarkable benefits compared to a small-scale MIMO system \cite{m-mimo-1}, \cite{m-mimo-2}. Unlike traditional small-scale MIMO systems that employ highly linear and power-inefficient radio frequency (RF) amplifiers, the practical implementation of large-scale MIMO systems requires the RF amplifiers to be power efficient, otherwise the consequent power consumption of the BS would be prohibitively high. Unfortunately, power-efficient RF amplifiers usually experience poor linearity characteristics and therefore further require the input signals to have a low PAPR. Accordingly, CE transmission, which enforces each antenna element to emit CE signals and allows the use of the most power-efficient and cheapest power amplifiers, as shown in Fig.~\ref{CEP}, has become an active research direction \cite{cep-1}\nocite{cep-2}\nocite{cep-3}\nocite{cep-4}\nocite{cep-5}\nocite{cep-6}\nocite{cep-7}\nocite{cep-8}\nocite{cep-9}\nocite{cep-10}\nocite{cep-11}\nocite{cep-12}\nocite{cep-13}-\cite{cep-14}, which holds great potential for the practical implementation of large-scale antenna systems. 

In \cite{cep-1}, a single-user single-stream CE transmission is considered, where it is shown that in this case the noiseless received signal region is doughnut shaped. Based on this observation, a near-optimal capacity-achieving input distribution is derived. It is further revealed in \cite{cep-2} that the inner radius of the doughnut-shaped region has a closed-form expression. A joint optimization of CEP and antenna subset selection is proposed in \cite{cep-3} from a geometric perspective for a single-user MISO system. CEP has been further extended to a multi-user case in \cite{cep-4}, where the CEP problem is formulated as a non-linear least-squares (NLS) optimization to minimize the multi-user interference, given by
\begin{equation}
\begin{aligned}
&\mathcal{P}_{\text {CEP}}: {\kern 3pt} \mathop {\min }\limits_{\bm \theta}  \sum\limits_{k = 1}^K {{{\left| {\sum\limits_{n = 1}^{{N_\text{T}}} {{h_{k,n}}{e^{j{\theta _n}}}}  - \sqrt {{E_k}} {s_k}} \right|}^2}} \\
&{\kern 8pt} {\text {s.t.}} {\kern 10pt} {\rm C1:} {\kern 3pt} \left| {{\theta _n}} \right| \le \pi,
{\kern 3pt} \forall n \in \left\{ {1,2,\cdots,N_\text{T}} \right\}
\label{eq_cep_1}
\end{aligned}
\end{equation}
where $\sqrt{E_k}$ is the magnitude of user $k$'s data symbol $s_k$, $h_{k,n}$ is the $n$-th entry of ${\bf h}_k$, and ${\bm \theta}= \left[ {\theta_1,\theta_2,\cdots,\theta_{N_\text{T}}} \right]^\text{T}$. An iterative algorithm is presented in \cite{cep-4} to efficiently obtain the phases of the CE signals. Building upon this, a cross-entropy optimization (CEO) method is introduced in \cite{cep-5} for $\mathcal{P}_\text{CEP}$, which achieves an improved performance over \cite{cep-4}. Some other works include CEP for frequency-selective channels in \cite{cep-6}, a joint transceiver design for CEP in a point-to-point (P2P) MIMO system in \cite{cep-7}, CEP for MISO multicasting in \cite{cep-8}, \cite{cep-9}, and CEP with quantized phases in \cite{cep-10}.

A closer look at the above studies reveals that CEP has to operate on a symbol level, since the phases of the CE signals are dependent on the information of the data symbols. From this point of view, CI-based precoding that also requires a symbol-level operation can be a perfect match with CEP. The concept of CI exploitation has been applied to CEP firstly in \cite{cep-11} and \cite{cep-12}, where instead of minimizing the multi-user interference, constructive CEP aims to maximize the CI effect subject to transmitting CE signals at each antenna element, which can be formulated as 
\begin{equation}
\begin{aligned}
&\mathcal{P}_\text{C-CEP}^\text{PSK}: {\kern 3pt} \mathop {\max }\limits_{\bf{x}} {\kern 3pt} t \\
&{\kern 10pt} {\text {s.t.}} {\kern 14pt} {\rm C1:} {\kern 3pt} {\bf h}_k^{\text T}{\bf{x}} = {\lambda _k}{s_k}, {\kern 3pt} \forall k \in {\cal K} \\
&{\kern 36pt} {\rm C2:} {\kern 1pt} \left[ {\Re \left( {{\lambda _k}} \right) - t} \right]\tan {\theta _\text{th}} \ge \left| {\Im \left( {{\lambda _k}} \right)} \right|,{\kern 3pt} \forall k \in {\cal K}\\
&{\kern 36pt} {\rm C3:} {\kern 1pt} \left| {{x_n}} \right| =\sqrt {\frac{P_0}{N_\text{T}}}, {\kern 3pt} \forall n \in \left\{ {1,2,\cdots,N_\text{T}} \right\}
\label{eq_cep_2}
\end{aligned}
\end{equation}
where ${\bf x}=\left[ {x_1,x_2,\cdots,x_{N_\text{T}}} \right]^{\text T}$. $\mathcal{P}_\text{C-CEP}^\text{PSK}$ is generally non-convex because of the non-convex constraint $\rm C3$. A CEO-based and a two-step CVX-based approach are further introduced in \cite{cep-12} for $\mathcal{P}_\text{C-CEP}^\text{PSK}$, both of which are shown to achieve significant performance improvements over the traditional CEP methods in \cite{cep-4} and \cite{cep-5} based on interference minimization. While not explicitly mentioned, the CEP approach for MISO multicasting proposed in \cite{cep-8}, which maximizes the minimum scaling effect among the users, is also based on the concept of interference exploitation by employing the symbol-scaling CI metric. In \cite{cep-13}, the CEP problem is studied from a Riemannian manifold perspective. By approximating the real representation of the original problem and mapping it onto a Riemannian manifold, an efficient Riemannian conjugate gradient algorithm is proposed, and additional performance improvements in terms of SER are observed compared to the CEP methods in \cite{cep-12}. The above studies in \cite{cep-11}-\cite{cep-13} show that the concept of interference exploitation can be readily extended to CEP problems and significant performance benefits are observed. When quantized phases are considered, CI-based CEP problems are studied in \cite{cep-14} for both PSK and QAM signaling.

Additionally, the SLP techniques proposed for hardware-efficient architectures in \cite{lux-15} are also candidates for CE transmission. While the proposed architectures include gain control at the transmitter, as contrary to CEP, it is shown that if the gain is kept constant regardless of the channel realization or the transmitted symbols, the derived algorithm that is based on the coordinate descent method is shown to be an efficient solution for CE transmission, which is able to reduce multi-user interference with much fewer transmit antennas than other CEP techniques in the literature. The proposed solution is suitable for PSK as well as multi-level constellations. A more detailed explanation of this work will be presented in Section V-F. 

\subsection{Antenna Selection (AS)}
From the descriptions for CEP schemes, it is observed that CE transmission attempts to reduce the hardware complexity and the consequent power consumption at the BS by employing power-efficient and low-cost RF amplifiers. In addition to the CE transmission, AS is also a low-cost and low-complexity alternative for power-efficient and cost-efficient BSs with large-scale antenna arrays, which has received continuous research attention \cite{as-1}\nocite{as-2}\nocite{as-3}\nocite{as-4}\nocite{as-5}\nocite{as-6}\nocite{as-7}\nocite{as-8}\nocite{as-9}\nocite{as-10}\nocite{as-11}\nocite{as-12}\nocite{as-13}\nocite{as-14}\nocite{as-15}\nocite{as-16}-\cite{as-17}. In AS techniques, only a subset of the entire antenna array is activated for transmission or reception, as shown in Fig.~\ref{AS}, which allows for a reduction in the number of active RF chains and consequently a reduction in the power consumption. In addition, AS also benefits from exploiting the degrees of freedom provided by the excess of antennas, i.e., antenna diversity at the BS \cite{as-1}, \cite{as-2}. 

AS techniques have already been a popular research topic in small-scale MIMO systems \cite{as-3}-\cite{as-9}, which exhibit benefits in terms of power efficiency \cite{as-3}. The initial AS approach is based on the exhaustive search method \cite{as-4}, whose computational cost could become impractically high when the number of antennas scales up. A low-complexity receive AS scheme that maximizes the channel capacity has been introduced in \cite{as-5}, which is aimed at minimizing the performance loss caused by reducing the number of active antennas at the receiver side. Transmit AS methods have been studied in \cite{as-7}-\cite{as-9}, where it is shown in \cite{as-7} that AS can maximize the received SNR when a maximum ratio combining (MRC) detector is considered at the receiver. \cite{as-8} and \cite{as-9} further reveal that the error rate performance can be further improved by transmit AS techniques. More recently, research on AS has been extended to large-scale antenna arrays in \cite{as-10}-\cite{as-12}, where the energy efficiency benefits offered by AS techniques are shown in \cite{as-10} and \cite{as-11}. Energy-efficient AS schemes are studied in \cite{as-12}, where it is shown that a simple random AS scheme can significantly improve the energy efficiency performance of the BS with a large-scale antenna array.

To extend the concept of CI exploitation to AS techniques, a transmit AS scheme has been designed in \cite{as-13} to minimize the error rate, where a partial sub-channel orthogonalization is employed to exploit the constructive part of the existing interference while nullifying the destructive part. A more advanced CI-driven AS technique is introduced in \cite{as-14} and \cite{as-15}, which maximizes the CI effect for the users by selecting the antenna subset that achieves the highest CI effect. With the proposed AS algorithms, it is shown that the combination of CI-based AS with MRT precoding is able to outperform more complicated ZF precoding with traditional computationally expensive AS schemes. More recently, thanks to the developments of optimization-based CI precoding in \cite{ci-7} and \cite{ci-8}, a joint AS and precoding method based on interference exploitation is further proposed in \cite{as-16} and \cite{as-17}. A mix-integer optimization problem is firstly formulated, which maximizes the minimum CI effect among the users by jointly optimizing the transmit AS decision and the precoded signals, given by 
\begin{equation}
\begin{aligned}
&\mathcal{P}_\text{C-AS}^\text{PSK}: {\kern 3pt} \mathop {\max }\limits_{{\bf{x}}, {\bf a}} {\kern 3pt} t \\
&{\kern 8pt} {\text {s.t.}} {\kern 12pt} {\rm C1:} {\kern 3pt} {\bf h}_k^{\text T}{\bf{x}} = {\lambda _k}{s_k}, {\kern 3pt} \forall k \in {\cal K} \\
&{\kern 32pt} {\rm C2:} {\kern 1pt} \left[ {\Re \left( {{\lambda _k}} \right) - t} \right]\tan {\theta _\text{th}} \ge \left| {\Im \left( {{\lambda _k}} \right)} \right|,{\kern 3pt} \forall k \in {\cal K}\\
&{\kern 32pt} {\rm C3:} {\kern 1pt} \left\| {\bf{x}} \right\|_2^2 \le P_0\\
&{\kern 32pt} {\rm C4:} {\kern 1pt} \left| {{\mathop{\text {sgn}}} \left( {\bf{x}} \right)} \right| = {\bf{a}}\\
&{\kern 32pt} {\rm C5:} {\kern 1pt} \sum\limits_{n = 1}^{N_\text{T}} {{a_n}}  = {N_{\rm AC}}, {\kern 3pt} a_n \in \left\{ {0,1} \right\}, \forall n
\label{eq_as_2}
\end{aligned}
\end{equation}
where ${\mathop{\text {sgn}}} \left(  \cdot  \right)$ is the sign function, and ${\bf a}=\left[ {a_1, a_2, \cdots, a_{N_\text{T}}} \right]^{\text T}$. $\rm C4$ and $\rm C5$ jointly guarantee that only ${N_{\rm AC}}$ transmit antennas are active and transmit precoded signals. In \cite{as-17}, three sub-optimal methods are also introduced for the formulated problem $\mathcal{P}_\text{C-AS}^\text{PSK}$ to reduce the computational complexity of the joint approach. Similarly, remarkable performance improvements for this CI-based joint AS and precoding have been observed through extensive numerical results in \cite{as-17}, which demonstrates the superiority of interference exploitation in the area of AS.

Another symbol-level AS scheme that aims to minimize the multi-user interference can be mathematically formulated as the following optimization problem \cite{lux-15}, \cite{lux-16}
\begin{equation}
\begin{aligned}
&\mathcal{P}_{\text {AS}}: {\kern 3pt} \mathop {\min }\limits_{\bf{x}} \left\| {{\bf{Hx}} - \sqrt \gamma   \cdot {\bf{s}}} \right\|_2^2 \\
&{\kern 4pt} {\text {s.t.}} {\kern 14pt} {\rm C1:} {\kern 3pt} \left|\left|\mathbf{x}\right|\right|_0= N_\text{AC}
\label{EQ:P1}
\end{aligned}
\end{equation}
where $\gamma$ is the SNR. $\mathcal{P}_{\text {AS}}$ is a linear least-squares problem with an $\ell_0$-norm constraint, which is generally non-convex and requires an exhaustive search solution. To obtain a feasible solution with a lower computational cost, $\mathcal{P}_{\text {AS}}$ can be reformulated into a regularized least-squares regression problem, as detailed in \cite{lux-15} and \cite{lux-16}.

\subsection{Hybrid Analog-Digital (AD) Precoding}
Compared to AS techniques which reduce the hardware complexity by reducing the number of active antennas and accordingly reducing the number of active RF chains, another potential technique that also employs a reduced number of RF chains is the hybrid AD precoding, which has drawn extensive research attention in the past few years \cite{hybrid-1}\nocite{hybrid-2}\nocite{hybrid-3}\nocite{hybrid-4}\nocite{hybrid-5}\nocite{hybrid-6}\nocite{hybrid-7}\nocite{hybrid-8}\nocite{hybrid-9}\nocite{hybrid-10}\nocite{hybrid-11}\nocite{hybrid-12}\nocite{hybrid-13}\nocite{hybrid-14}\nocite{hybrid-18}\nocite{hybrid-15}\nocite{hybrid-16}\nocite{hybrid-17}-\cite{hybrid-19}. Different from AS techniques, all the antennas are active in hybrid AD structures, and the signal processing is divided into analog part and digital part, as shown in Fig.~\ref{Hybrid}, where the analog part usually consists of low-cost phase shifters \cite{hybrid-3}. Hybrid AD structures have firstly been considered for the future millimeter-wave (mmWave) communications as a promising structure for practical implementation. For mmWave communications, while the small wavelength of mmWave signals allows the use of a large-scale antenna array in a small form factor to combat the severe pathloss \cite{mmwave}, dedicating a single RF chain for each antenna element becomes nearly infeasible for mmWave transceivers, since the hardware components working at mmWave bands are costly and power expensive \cite{hybrid-1}-\cite{hybrid-3}. By reducing the number of RF chains employed at the transceivers, hybrid AD structures are able to greatly reduce the hardware complexity and the corresponding power consumption at the cost of only a slight performance loss, thus achieving an improved balance between performance, complexity and cost.

In \cite{hybrid-4}, a single-user mmWave communication system has been considered,  and the hybrid precoding and combining are joint designed to maximize the spectral efficiency, where an orthogonal matching pursuit (OMP)-based algorithm is proposed. A multi-user transmission has further been considered in \cite{hybrid-6}, where the analog precoder/combiner is designed to maximize the effective channel gain while the digital precoder is designed to mitigate the multi-user interference based on ZF, constituting a two-stage hybrid precoding. The hybrid precoding design for a single-user mmWave transmission in \cite{hybrid-7} considers both the fully-connected and the partially-connected AD structures, where a manifold optimization based algorithm is proposed based on the alternating minimization framework. An important proposition is established in \cite{hybrid-8}, where it is shown that hybrid precoding is able to realize any fully-digital precoding when the number of RF chains is twice the number of data streams. Meanwhile, \cite{hybrid-8} also proposes a near-optimal hybrid precoding design for both single-user and multi-user transmissions, when the number of RF chains becomes fewer. \cite{hybrid-9} has focused on the partially-connected structures in a multi-user scenario and proposed hybrid precoding designs based on the concept of successive interference cancellation (SIC). By assuming the digital precoder to be a diagonal matrix, the total spectral efficiency optimization problem is decomposed into a series of simple sub-rate optimization problems, which can be efficiently solved by the power iteration algorithm. Additional works on hybrid precoding include some low-complexity designs based on MRT in \cite{hybrid-10}, virtual path selection in \cite{hybrid-11}, and SVD in \cite{hybrid-12}. Mathematically, a common multi-user hybrid precoding problem aimed at spectral efficiency maximization can be formulated as
\begin{equation}
\begin{aligned}
&\mathcal{P}_\text{HAD}: \\
&\mathop {\max }\limits_{{{\bf{F}}_{\text {RF}}},{\bf{f}}_k^{\text {BB}},{{\bf{w}}_k}} {\kern 2pt} \sum\limits_{k = 1}^K {{{\log }_2}\left( {1 + \frac{{{{\left| {{\bf{w}}_k^{\text H}{{\bf{H}}_k}{{\bf{F}}_{\text {RF}}}{\bf{f}}_k^{\text {BB}}} \right|}^2}}}{{\sum\limits_{i \ne k} {{{\left| {{\bf{w}}_k^{\text H}{{\bf{H}}_k}{{\bf{F}}_{\text {RF}}}{\bf{f}}_i^{\text {BB}}} \right|}^2}}  + {\sigma ^2}}}} \right)}  \\
&{\kern 16pt} {\text {s.t.}} {\kern 14pt} {\rm C1:} {\kern 3pt} {{\bf{F}}_{\text {RF}}} \in {\cal F}, {\kern 3pt} {{\bf{w}}_k} \in {\cal W}, {\kern 3pt} \forall k \in {\cal K} \\
&{\kern 42pt} {\rm C2:} {\kern 1pt} \left\| {{{\bf{F}}_{{\text{RF}}}}\left[ {{\bf{f}}_1^{{\text{BB}}},{\bf{f}}_2^{{\text{BB}}}, \cdots ,{\bf{f}}_K^{{\text{BB}}}} \right]} \right\|_F^2 = {P_0}
\label{eq_had_1}
\end{aligned}
\end{equation}
which assumes single-stream transmission for each user and analog combining only at the receiver side. The constraint $\rm C1$ is to ensure that the analog precoder ${\bf F}_{\text {RF}}$ and analog combiner ${\bf w}_k$ implemented with phase shifters have constant-envelope entries.

If we assume the analog precoder to be fixed when we design the precoding methods for the digital part, hybrid AD structures are equivalent to a fully-digital MIMO system transmitting through an effective analog channel. From this point of view, interference exploitation techniques can be readily applied to hybrid AD structures for additional performance improvements, as recently shown in \cite{hybrid-13}-\cite{hybrid-18}. In \cite{hybrid-13}, the digital part of hybrid precoding employs CI precoding, and several analog precoding designs particularly tailored for CI-based hybrid precoding are presented and compared. It should be noted that since CI-based precoder is data-dependent, the common spectral efficiency expression is not applicable to CI-based hybrid precoding designs. Accordingly, \cite{hybrid-13} considers the transmit power minimization for CI-based hybrid precoding in a MU-MISO system, formulated as
\begin{equation}
\begin{aligned}
&\mathcal{P}_\text{C-HAD}^\text{PSK}: {\kern 3pt} \mathop {\min }\limits_{{{\bf{F}}_{\text {RF}}},{\bf{x}}} \left\| {{{\bf{F}}_{\text {RF}}}{\bf x}} \right\|_2^2  \\
&{\kern 10pt} {\text {s.t.}} {\kern 10pt} {\rm C1:} {\kern 3pt} {\bf{h}}_k^{\text T}{{\bf{F}}_{{\text{RF}}}}{\bf{x}} = {\lambda _k}{s_k}, {\kern 3pt} \forall k \in {\cal K} \\
&{\kern 32pt} {\rm C2:} {\kern 0pt} \left[ {\Re \left( {{\lambda _k}} \right) - \sqrt {{\Gamma _k}{\sigma ^2}} } \right]\tan {\theta _\text{th}} \ge \left| {\Im \left( {{\lambda _k}} \right)} \right|, {\kern 3pt} \forall k \in {\cal K} \\
&{\kern 32pt} {\rm C3:} {\kern 3pt} {{\bf{F}}_{\text {RF}}} \in {\cal F}
\label{eq_had_2}
\end{aligned}
\end{equation}
where ${\bf F}_\text{RF}$ is updated on a block level and $\bf x$ is updated on a symbol level. This problem is generally non-convex because of the constant-envelope constraint for the entries in ${\bf F}_{\text {RF}}$. Therefore, \cite{hybrid-13} decomposes the joint design into the analog precoding design, followed by the digital precoding design. As a step further, a CI-based hybrid precoding that is specifically designed to be robust against phase errors in the phase shifters is proposed in \cite{hybrid-14}, where the optimal robust digital precoding is obtained based on the cutting plane method and alternating procedure, when the analog part of the hybrid precoder is fixed. Significant performance improvements in terms of SER can be observed for CI-based hybrid designs compared to traditional hybrid methods.

\subsection{Low-Resolution DACs}
In addition to the use of low-cost RF amplifiers as in CEP and the activation of a reduced number of RF chains as in AS and hybrid precoding, another potential technique to reduce the cost and power consumption per RF chain at the BSs is to employ low-resolution DACs \cite{dac-1}\nocite{dac-2}\nocite{dac-3}\nocite{dac-4}\nocite{dac-5}\nocite{dac-6}\nocite{dac-7}\nocite{dac-8}\nocite{dac-9}\nocite{dac-10}\nocite{dac-11}\nocite{dac-12}\nocite{dac-13}\nocite{dac-15}-\cite{dac-16}, as shown in Fig.~\ref{DAC}. It is known that the power consumption of DACs grows linearly with the bandwidth and exponentially with the resolution \cite{dac-1}, and each transmit signal is generated by a pair of DACs connected to the RF chain. Given hundreds of antenna elements at a large-scale antenna array, a large number of DACs are also required, which poses a significant practical challenge if high-resolution DACs are deployed. Therefore, the use of low-resolution DACs, especially the extreme case 1-bit DACs, can greatly simplify the hardware cost and the corresponding power consumption at the BS. In addition, the output signals of 1-bit DACs are CE signals, which allows the use of power-efficient amplifiers to further reduce the hardware complexity.

There have been an increasing number of studies on the downlink transmission design with low-resolution DACs \cite{dac-2}-\cite{dac-13}. Linear precoding methods with few-bit DACs for downlink MIMO systems have firstly been studied in \cite{dac-2}-\cite{dac-4}, where due to the coarse quantization, significant performance degradation is observed compared to the ideally unquantized case, especially when 1-bit DACs are considered. Nonlinear precoding designs, which directly design the precoded signals based on the CSI and the data symbols, have further been studied in \cite{dac-5}-\cite{dac-12}, and the optimization problem can be formulated as
\begin{equation}
\begin{aligned}
&\mathcal{P}_\text{DAC}: {\kern 3pt} \mathop {\min }\limits_{\bf{x}} \left\| {{\bf{s}} - \beta_\text{DAC}  \cdot {\bf{Hx}}} \right\|_2^2 + K{\beta_\text{DAC}^2}{\sigma ^2}  \\
&{\kern 8pt} {\text {s.t.}} {\kern 10pt} {\rm C1:} {\kern 3pt} {\bf{x}} \in {{\cal X}_{{\text{DAC}}}} \\
&{\kern 30pt} {\rm C2:} {\kern 3pt} \beta_\text{DAC}>0
\label{eq_dac_1}
\end{aligned}
\end{equation}
which aims to minimize the MSE between the transmit data symbols and the received symbols. ${\cal X}_{\text {DAC}}$ is the set consisting of the output signals for low-resolution DACs, and specifically in the case of 1-bit DACs, ${\cal X}_{\text {DAC}}=\left\{ { \pm \sqrt {\frac{{{P_0}}}{{2{N_\text{T}}}}}  \pm \sqrt {\frac{{{P_0}}}{{2{N_\text{T}}}}}  \cdot j} \right\}$. In \cite{dac-5}, a non-linear precoding method based on the biconvex relaxation framework is proposed for $\mathcal{P}_\text{DAC}$, which achieves a promising performance with a low computational cost. Its corresponding very large-scale integration (VLSI) design architectures have further been illustrated in \cite{dac-6} to showcase the efficacy. \cite{dac-7} proposes several 1-bit precoding schemes based on SDR, sphere encoding, and squared $\ell_\infty$-norm relaxation. Meanwhile, a 1-bit precoding method is described in \cite{dac-8} based on the branch-and-bound framework, which can theoretically achieve the optimal performance. Some other downlink precoding designs for low-resolution DACs include SER minimization in \cite{dac-9}-\cite{dac-11} and alternating minimization in \cite{dac-12}. A general observation is that non-linear precoding designs can achieve a significantly better performance than the linear methods, when low-resolution DACs are employed at the transmitter.

To achieve a promising error rate performance, the precoding designs for 1-bit DACs have to be non-linear and exploit the information of the data symbols, which creates the opportunity for interference exploitation, as recently studied in \cite{dac-14} and \cite{dac-13}-\cite{dac-16}. \cite{dac-13} considers the transmit signal design for 1-bit massive MIMO system based on CI optimization, where the CI effect is maximized subject to the output constraints of DACs, and the corresponding optimization problem is constructed as
\begin{equation}
\begin{aligned}
&\mathcal{P}_\text{C-DAC}^\text{PSK}: {\kern 3pt} \mathop {\max }\limits_{\bf{x}} {\kern 3pt} t  \\
&{\kern 12pt} {\text {s.t.}} {\kern 12pt} {\rm C1:} {\kern 3pt} {\bf{h}}_k^{\text T}{\bf{x}} = {\lambda _k}{s_k}, {\kern 3pt} \forall k \in {\cal K} \\
&{\kern 36pt} {\rm C2:} {\kern 1pt} \left[ {\Re \left( {{\lambda _k}} \right) - t} \right]\tan {\theta _\text{th}} \ge \left| {\Im \left( {{\lambda _k}} \right)} \right|, {\kern 3pt} \forall k \in {\cal K}\\
&{\kern 36pt} {\rm C3:} {\kern 3pt}{\bf{x}} \in {{\cal X}_{{\text{DAC}}}}
\label{eq_dac_2}
\end{aligned}
\end{equation}
In the case where 1-bit DACs are employed, by expressing $\mathcal{P}_\text{C-DAC}^\text{PSK}$ into a real representation and relaxing the 1-bit constraint, the formulated optimization problem is shown to be a linear programming (LP), which can be efficiently solved, and the final transmit signal is obtained by enforcing an element-wise normalization. \cite{dac-14} and \cite{dac-15} focus on the 1-bit precoding designs for PSK modulations based on the symbol-scaling CI metric, and a refinement process that is applicable upon any 1-bit schemes has been introduced, where additional performance improvements can be observed. The performance improvements are shown to be prominent for low-complexity 1-bit schemes such as 1-bit ZF. A joint consideration of hybrid AD precoding and 1-bit DACs based on interference exploitation has further been studied in \cite{dac-16}, where it is shown that compared to hybrid AD structures with ideal DACs, the number of RF chains in the presence of 1-bit DACs has to be much larger than the number of data streams to achieve a near-optimal performance.

In addition, \cite{lux-17} has discussed the extension from 1-bit DACs to few-bit DACs, where the objective function adopts \eqref{eq_dac_1}. In the case when $B$-bits DACs are adopted at the BS, the set ${\cal X}_\text{DAC}$ can be expressed as
\begin{equation}
{\cal X}_\text{DAC}=\left\{ { \pm \sqrt {\frac{{{P_0}}}{{2N_\text{T}}}} , \cdots , \pm \sqrt {\frac{{{P_0}}}{{{2^{B - 2}}N_\text{T}}}}  \pm \sqrt {\frac{{{P_0}}}{{{2^{B - 1}}N_\text{T}}}} } \right\},
\label{eq_dac_3}
\end{equation}
which is obtained via normalizing the outputs of a uniform quantizer such that the transmission power constraint is always satisfied. Similar to the case of 1-bit DACs, the resulting optimization problem in the presence of $B$-bits DACs is NP-hard, and the optimal solution obtained via an exhaustive search method is prohibitive in terms of computational complexity for large-scale antenna arrays. To obtain a feasible solution, \cite{lux-17} has developed a cyclic coordinate descent (CCD) algorithm \cite{lux-18} with the effort to provide an efficient solution for this problem. It has been shown in \cite{lux-17} that the complexity of the proposed algorithm is only ${\cal O}\left({N_\text{T}^22^B}\right)$ which is much smaller compared to the ${\cal O}\left({N_\text{T}2^{BN_\text{T}}}\right)$ of the exhaustive search approach.

\subsection{Non Linearities}
\begin{figure}[!b]
\centering
\subfloat[AM-AM]{%
  \includegraphics[clip,width=1\columnwidth]{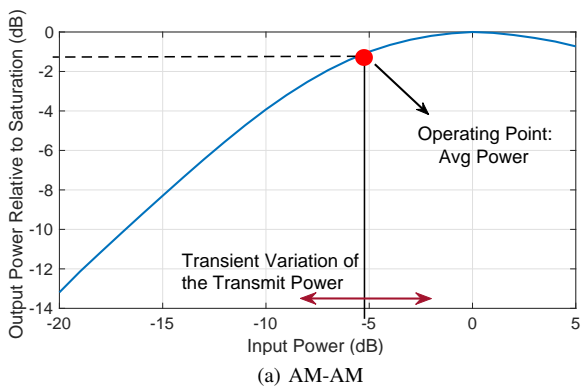}
  \label{fig:AM_AM}
}

\subfloat[AM-PM]{%
  \includegraphics[clip,width=1\columnwidth]{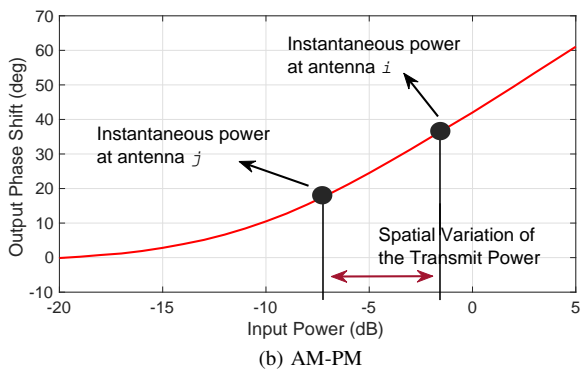}
  \label{fig:AM_PM}
}

\caption{An example of the AM-AM/AM-PM characteristic of a non-linear amplifier}
\end{figure}

As already mentioned in Section V-A, it is common to employ power amplifiers characterized by severe non-linear effects in practical systems relying on large-scale antenna arrays at the transmitter. Therefore, good dynamic properties of the per-antenna transmit power are required in order to limit such effects. In this context, the discussed CEP schemes are an effective strategy, since they achieve the best possible PAPR (unit) at the symbol level. Meanwhile, alternative schemes based on SLP have also been proposed in the literature for non-linear channels in \cite{ci-16}, \cite{lux-19}\nocite{lux-21}\nocite{lux-23}\nocite{lux-30}\nocite{lux-31}\nocite{lux-32}-\cite{lux-33}, where the aim is not to attain a CE transmission, but rather to optimize the power dynamics (such as the dynamic range and PAPR) following an optimization framework in line with \cite{ci-6,lux-14} based on per-user QoS constraints. The main idea of these schemes is to control the instantaneous transmit power and to minimize its peaks, both in the temporal dimension and in the spatial one (i.e. among different antenna elements), so as to limit the performance degradation due to the amplitude-to-amplitude (AM-AM) and the amplitude-to-phase (AM-PM) distortion. An example of AM-AM and AM-PM characteristics of a non-linear amplifier \cite{lux-23} is shown in Figs. \ref{fig:AM_AM} and \ref{fig:AM_PM}, respectively. In particular, Fig. \ref{fig:AM_AM} highlights how the temporal variation of the power around the operating point enhances the distortion. On the other hand, Fig. \ref{fig:AM_PM} shows that the spatial variation of the instantaneous power (across different antennas) also leads to signal deterioration through a differential phase shift. 

A SLP approach for non-linear systems, proposed in \cite{ci-16}, considers a weighted per-antenna PM problem, subject to QoS constraints and a lower bound constraint on the per-antenna transmit power. In this scheme, the imbalances between different RF chains are reduced by constraining the per-antenna transmit power within a specific range. More specifically, the goal is to minimize the maximum power among the different antennas, and meanwhile put a lower bound constraint on such power. The related optimization problem tailored for APSK modulations can be constructed as
\begin{equation}
\begin{aligned}
&\mathcal{P}_\text{NL}^\text{APSK}: {\kern 3pt} \mathop {\min }\limits_{\bf{x}} {\kern 3pt} r  \\
&{\kern 10pt} {\text {s.t.}} {\kern 14pt} {\rm C1:} {\kern 3pt} \alpha {r^2} \le \frac{{{{\left| {{x_i}} \right|}^2}}}{{{p_i}}} \le {r^2}, {\kern 3pt} \forall i \in \left\{ {1,2,\cdots,N_t} \right\} \\
&{\kern 36pt} {\rm C2:} {\kern 2pt} {\left| {{\bf{h}}_k^{\text T}{\bf{x}}} \right|^2} \ge \kappa_k^2\gamma_k\sigma^2, {\kern 3pt} \forall k \in {\cal K}\\
&{\kern 36pt} {\rm C3:} {\kern 3pt} \angle{ \bf h}_k^{\text T} {\bf x} = \angle s_k, {\kern 3pt} \forall k \in {\cal K}
\label{WPPMLB_multi}
\end{aligned}
\end{equation}
where the parameter $\alpha$, chosen such that $0 \leq \alpha \leq 1$, determines the lower bound. The closer $\alpha$ is to 1, the more the power variations will be limited. Nonetheless, the choice of a higher value for $\alpha$ also results in a reduction of the degrees of freedom of the optimization problem. This has been solved through an iterative procedure based on successive convex approximation (SCA). It shall be stressed that this scheme optimizes the power dynamics in the spatial dimension (across different antenna elements) only, while not in the temporal one, as the optimization is performed symbol by symbol. 

A second strategy to improve the spatial dynamics of the transmitted signals is proposed in \cite{ci-16,lux-19}, which performs a minimization on the spatial PAPR, evaluated amongst the transmitting antennas, under QoS constraints. The related optimization problem is formulated as a non-linear fractional programming, given by 
\begin{equation}
\begin{aligned}
&\mathcal{P}_\text{NL}^\text{PAPR}: {\kern 3pt} \mathop {\min }\limits_{\bf{x}} {\kern 3pt} \frac{{\left\| {\bf{x}} \right\|_\infty ^2}}{{\left\| {\bf{x}} \right\|_2^2}} \\
&{\kern 12pt} {\text {s.t.}} {\kern 10pt} {\rm C1:} {\kern 3pt} {\left| {{\bf{h}}_k^{\text T}{\bf{x}}} \right|^2} \ge \kappa_k^2\gamma_k\sigma^2, {\kern 3pt} \forall k \in {\cal K}\\
&{\kern 34pt} {\rm C2:} {\kern 3pt} \angle{\bf h}_k^{\text T}{\bf x} = \angle s_k, {\kern 3pt} \forall k \in {\cal K}
\label{SPAPR_ORIGINAL}
\end{aligned}
\end{equation}
which is tackled by resorting jointly to parametric programming and SCA. 

Both the introduced schemes above have been shown to outperform the state-of-the-art SLP schemes based on QoS constraints optimization, in terms of spatial PAPR, spatial dynamic range, and SER over non-linear channels. While the power minimization scheme with a lower bound constraint has been shown to be more flexible than the spatial PAPR minimization one, the latter is able to achieve a slightly lower SER. 

Finally, a spatio-temporal extension of the SLP PAPR minimization approach has been proposed in \cite{lux-21}, which is based on the system model introduced in Section IV-F for FTN signaling. This allows a minimization of the PAPR at a waveform level, both in spatial domain and in temporal domain. The associated optimization formulation is formally similar to the one in \eqref{SPAPR_ORIGINAL}, which is thus tackled analogously. This spatio-temporal SLP scheme for non-linear channels has been shown to achieve considerable performance gains with respect to the previous ones, in terms of power distribution and SER over non-linear channels.

\subsection{Single-RF MIMO and RF-domain SLP}

\begin{figure}[!t]
\centering
\subfloat[$1$-PS per antenna at the transmitter]{%
  \includegraphics[clip,width=0.8\columnwidth]{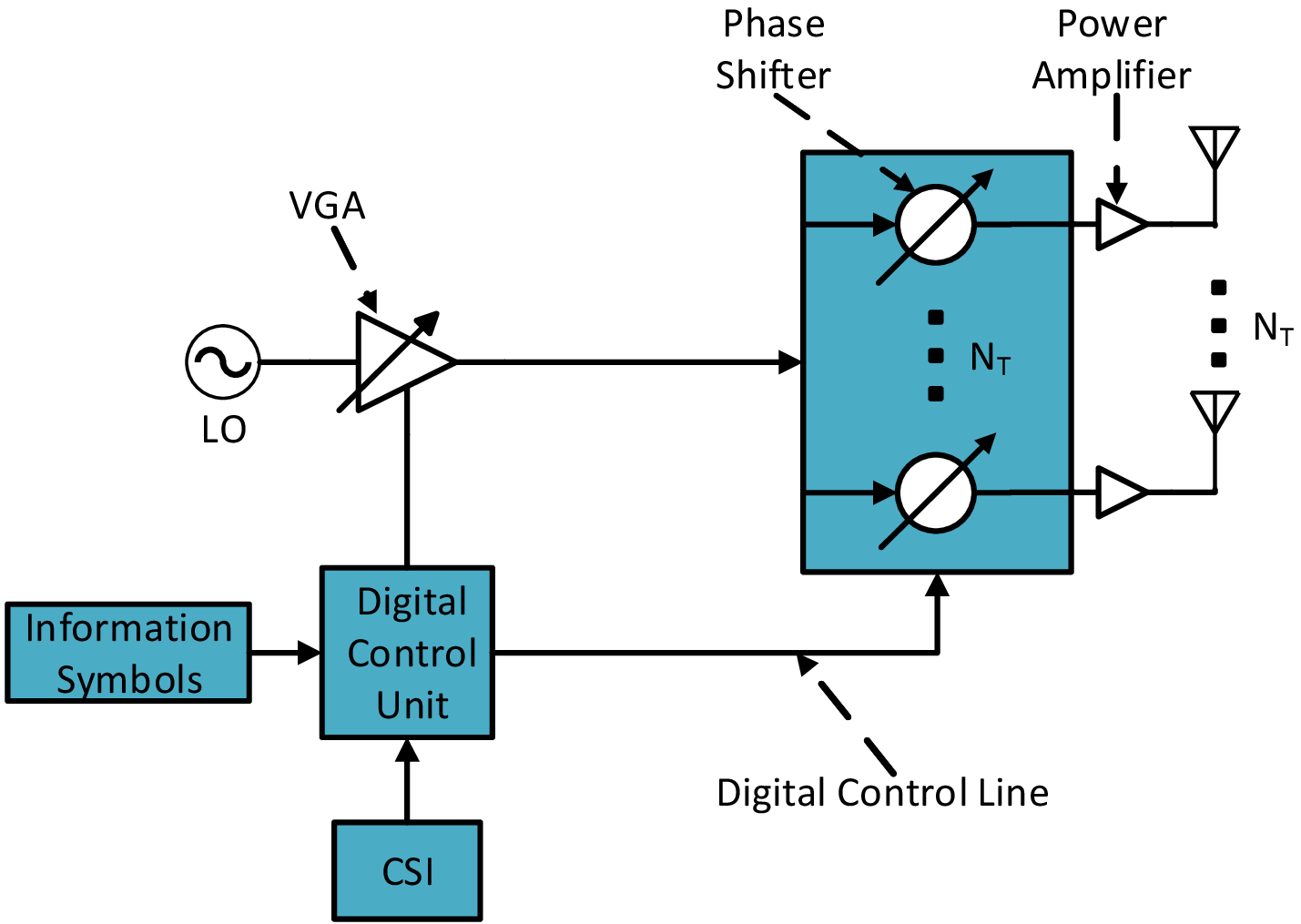}
  \label{Tx1PS}
}

\subfloat[$2$-PS per antenna at the transmitter]{%
  \includegraphics[clip,width=0.8\columnwidth]{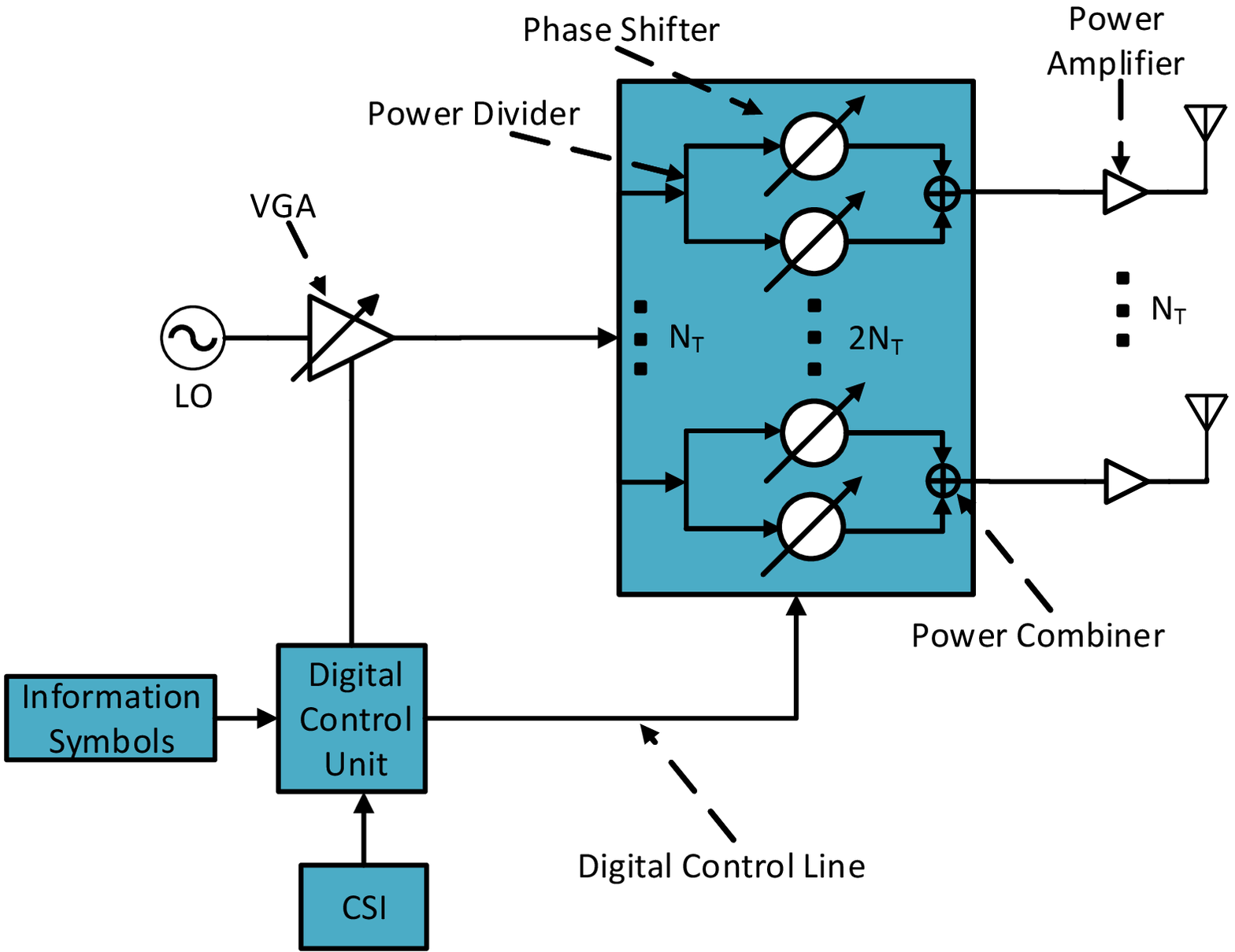}
  \label{Tx2PS}
}

\caption{Block diagram of the RF-domain SLP}
\label{RF-domain}
\end{figure}

In addition to hardware-efficient structures mentioned in the previous sections, more recently \cite{lux-15} has proposed two new transmitter architectures, illustrated in Fig. \ref{Tx1PS} and Fig. \ref{Tx2PS}, respectively. These two hardware structures deal with the increased hardware complexity and power consumption of existing techniques by eliminating the need for DACs and replacing them with analog components. In the proposed transmitter designs, which are referred to as RF-domain SLP, the processing happens only in the RF domain, as the DACs have been eliminated. The transmitted signals are modulated directly on the antennas by analog phase shifters.

The first architecture in Fig. \ref{Tx1PS} includes a single variable gain amplifier (VGA), which controls the amplitude $\alpha$ of the signals that are driven to the transmit antennas, where each antenna is driven by a dedicated phase shifter that changes the phase of the RF signal before transmission. As a result, the transmit signal can be expressed as 
\begin{equation}
x_n=\alpha e^{j \theta_n}, {\kern 3pt} \forall n \in \left\{ {1,2,\cdots,N_\text{T}} \right\}.
\end{equation} 
The purpose of the precoding is to find each phase shift $\theta_n$ and the gain of the VGA $\alpha$ that minimize the Euclidean distance between the received signal at the user side and the information symbol. Accordingly, the problem can be formulated as 
\begin{equation}
\begin{aligned}
&\mathcal{P}_\text{DM}^\text{VGA}: {\kern 3pt} \mathop {\min }\limits_{{\bf{v}}, {\kern 1pt} \alpha } \left\| {\alpha  \cdot {\bf{Hv}} - \sqrt \gamma   \cdot {\bf{s}}} \right\|_2^2 \\
&{\kern 8pt} {\text {s.t.}} {\kern 14pt} {\rm C1:} {\kern 3pt} \left| {{v_n}} \right| = 1, {\kern 3pt} \forall n \in \left\{ {1,2,\cdots,N_\text{T}} \right\}
\end{aligned}
\end{equation}
where $\mathbf{v}$ is an auxiliary variable such that $\mathbf{x}=\alpha \cdot \mathbf{v}$. An iterative solution based on the coordinate descent algorithm has been developed in \cite{lux-15} to solve this non-convex problem, which is shown via numerical results to converge to a local minima. As already mentioned in Section V-A, the derived solution for this architecture is also an efficient solution for CEP, when the VGA gain $\alpha$ is kept constant regardless of the channel realization $\mathbf{H}$ and the information symbols $\mathbf{s}$.

Similar to \cite{doubling-phase-shift}, an alternative transmitter structure has also been proposed in \cite{lux-15}, as shown in Fig. \ref{Tx2PS}, where each antenna element adopts a dual phase shifter structure. With the joint effect of two phase shifters, the amplitude of the transmit signal can also be altered by selecting suitable phases for the two superimposing signals. Therefore, the transmit signal $x_n$ on the $n$-th antenna port can be expressed as
\begin{equation}
x_n=\alpha\left(e^{j\theta_{n,1}}+e^{j\theta_{n,2}}\right),{\kern 3pt} \forall n \in \left\{ {1,2,\cdots,N_\text{T}} \right\}.
\end{equation}
The objective of the precoder remains the same, while the constraint has to adapt to the addition of the second phase shifter. Subsequently, the precoding problem is formulated as 
\begin{equation}
\begin{aligned}
&\mathcal{P}_\text{DM}^\text{DPS}: {\kern 3pt} \mathop {\min }\limits_{{\bf{v}}, {\kern 1pt} \alpha } \left\| {\alpha  \cdot {\bf{Hv}} - \sqrt \gamma   \cdot {\bf{s}}} \right\|_2^2 \\
&{\kern 8pt} {\text {s.t.}} {\kern 12pt} {\rm C1:} {\kern 3pt} {\left\| {\bf{v}} \right\|_\infty } \le 2, {\kern 3pt} \forall n \in \left\{ {1,2,\cdots,N_\text{T}} \right\}
\end{aligned}
\end{equation}
This optimization problem can be solved using an iterative solution based on the coordinate descent method, while with a different update step compared to that of ${\cal P}_\text{DM}^\text{VGA}$, as detailed in \cite{lux-15}. Numerical results show that the addition of a second phase shifter provides an interesting tradeoff, as a small increase in the power consumption is observed due to the increased number of phase shifters, while a decrease in the computational complexity of the solution is also observed, since the set of constraints of the optimization problem now becomes convex.

The proposed RF-domain systems, where DACs are replaced by analog components, are shown to outperform competing fully-digital and hybrid AD structures when the number of transmit antennas is much larger than that of users, thanks to the low power consumption of the analog phase shifters. More specifically, it has been shown in numerical results that for a system with $100$ transmit antennas and $10$ single-antenna users, the RF-domain SLP method with $1$-PS and $2$-PSs can achieve $36\%$ and $11\%$ improvement in terms of power efficiency over the hybrid AD precoding, and $135\%$ and $83\%$ over the fully-digital SLP scheme, respectively.

\begin{figure}[!b]
	\centering
	\includegraphics[scale=0.8]{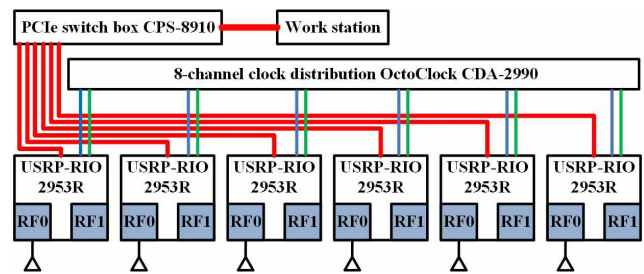}
	\caption{System architecture of the real-time hardware platform}
	\label{fig-ucl-1}
\end{figure}

\begin{figure*}
\setcounter{figure}{13}
	\centering
	\includegraphics[scale=0.15]{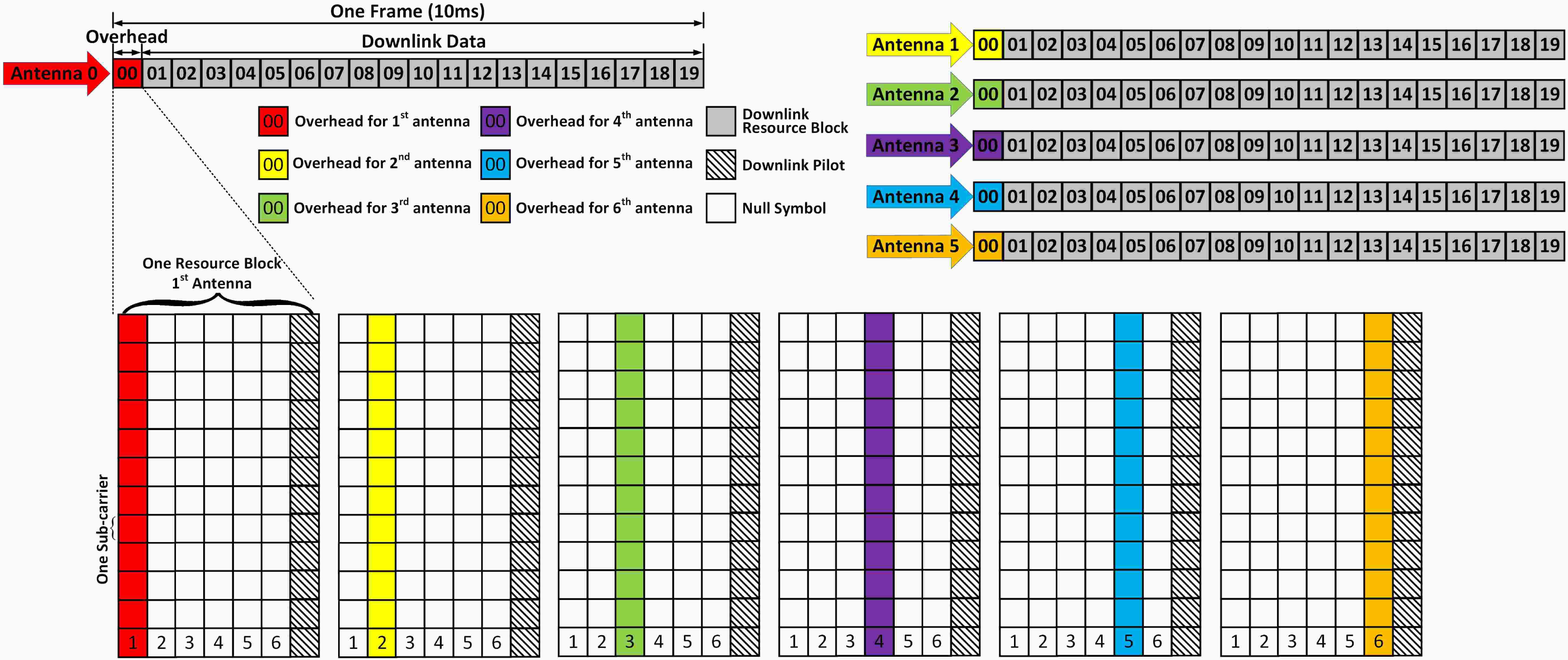}
	\caption{Frame and resource block structure for each antenna element \cite{ucl-2}}
	\label{fig-ucl-2}
\end{figure*}

\section{Proof-of-Concept Testbeds}

\subsection{Multi-Carrier CI Testbed}
A multi-carrier proof-of-concept hardware platform is hosted in University College London (UCL), which employs real-time CI precoding. The platform combines CI precoding with the spectral efficient frequency division multiplexing (SEFDM) \cite{ucl-1}, \cite{ucl-2}. Compared to orthogonal frequency division multiplexing (OFDM), SEFDM breaks the orthogonality by allowing closely-spaced non-orthogonal sub-carriers such that the total occupied bandwidth is reduced, thereby improving the effective spectral efficiency. Meanwhile, this artificial non-orthogonal setting creates inter-carrier interference (ICI), which can be exploited by CI precoding, as studied in \cite{ucl-1}.

\begin{table}[!b]
\caption{Parameters of the Experiment System}
\label{experimentsets} 
\centering
\begin{tabular}{|l|c|}
 \hline
Parameter &  Value \\
\hline 
Number of transmit antennas & 6 \\
Number of users & 2 \\
RF center frequency (GHz) & 2.4 \\
Sampling frequency (MHz) & 1.92 \\
FFT size & 128 \\
Number of guard band sub-carriers & 58 \\
Number of data sub-carriers & 12 \\
Number of cyclic prefix samples & 10 \\
Modulation & QPSK \\
Bandwidth compression ratio ($\upsilon$) \cite{ucl-1} & 0.85 \\
Sub-carrier bandwidth (kHz) & 15 \\
Sub-carrier spacing & 15$\times \upsilon$ \\
Maximum spectral efficiency (bits/s/Hz) & 2/$\upsilon$ \\
 \hline 
\end{tabular}
\label{tab-ucl-1}
\end{table}

\subsubsection{Platform Description}
The system architecture of the real-time hardware platform is presented in Fig. \ref{fig-ucl-1}, where the experiment configurations follow the 3GPP NB-IoT standard \cite{ucl-3}. In this testbed, omni-directional antennas are used for both the transmitter and receiver to validate the feasibility of precoding. A total number of 6 transmit antennas are configured for the purpose of demonstration, and accordingly 6 software defined ratio USRP-RIO 2953R devices are connected as a 6-antenna BS. Each USRP has two separate and independent RF chains, where one RF chain is for signal transmission and the other one is for signal reception, as seen in Fig. \ref{fig-ucl-1}. The raw data stream is generated by the work station, which are then sent to a cabled PCI-Express switch box CPS-8910 via an NI MXI-Express Gen $2\times8$ cable that supports a data rate up to 3.2Gb/s. The switch box separates the raw data stream into 6 data streams, which are delivered to 6 USRPs in parallel also via the NI MXI-Express cable. For the purpose of synchronization, these USRPs are also connected via SMA cables to a CDA-2990 8-channel clock distribution OctoClock module, which can split and amplify a 10MHz reference signal and a pulse-per-second (PPS) signal to support synchronization for a maximum of 8 USRPs.

\begin{figure}[!b]
\setcounter{figure}{12}
	\centering
	\includegraphics[scale=0.2]{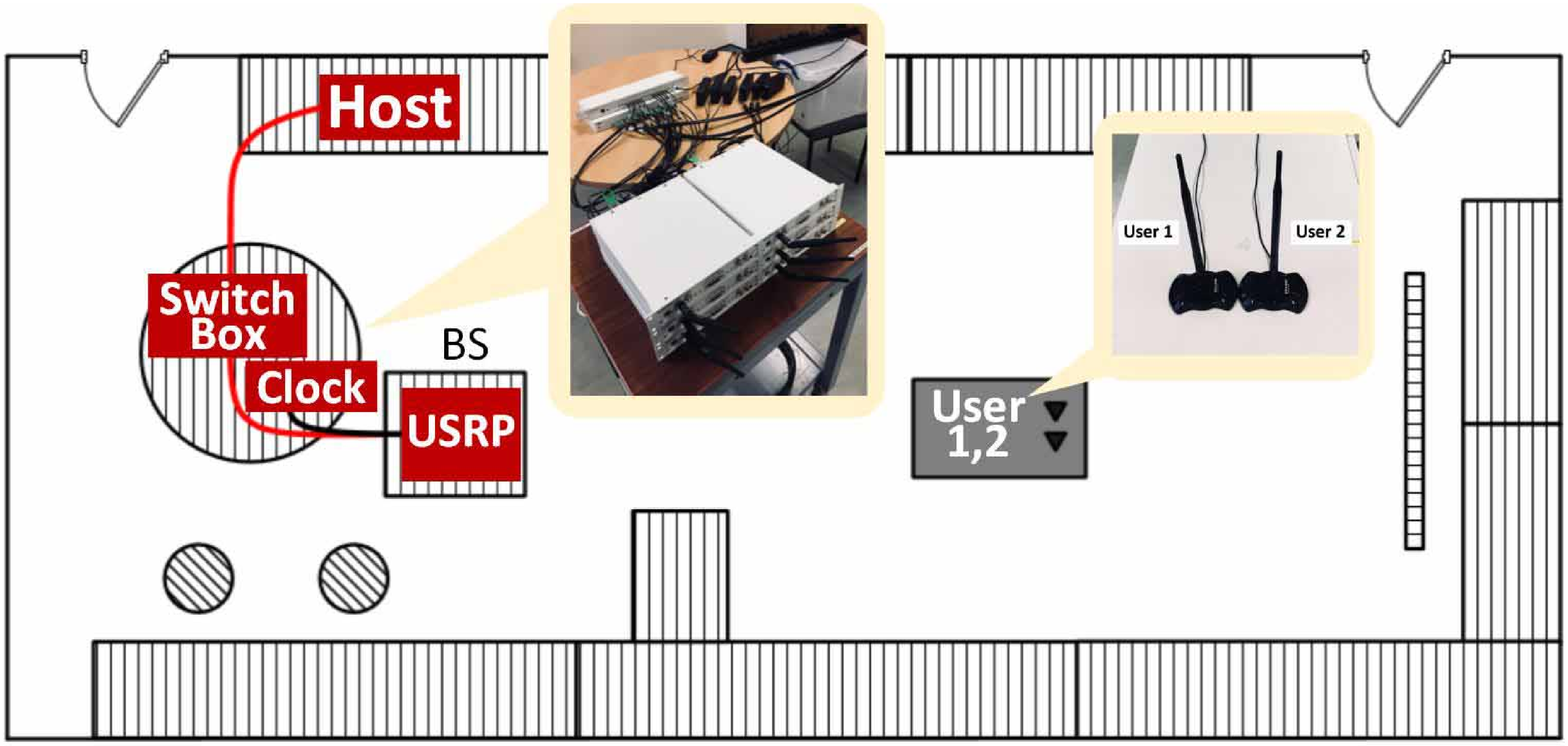}
	\caption{Illustration of the indoor experimental environment}
	\label{fig-ucl-3}
\end{figure}

\begin{figure*}
\begin{centering}
\setcounter{figure}{14}
\subfloat[Unprecoded]
{
\begin{centering}
\includegraphics[width=4.2cm]{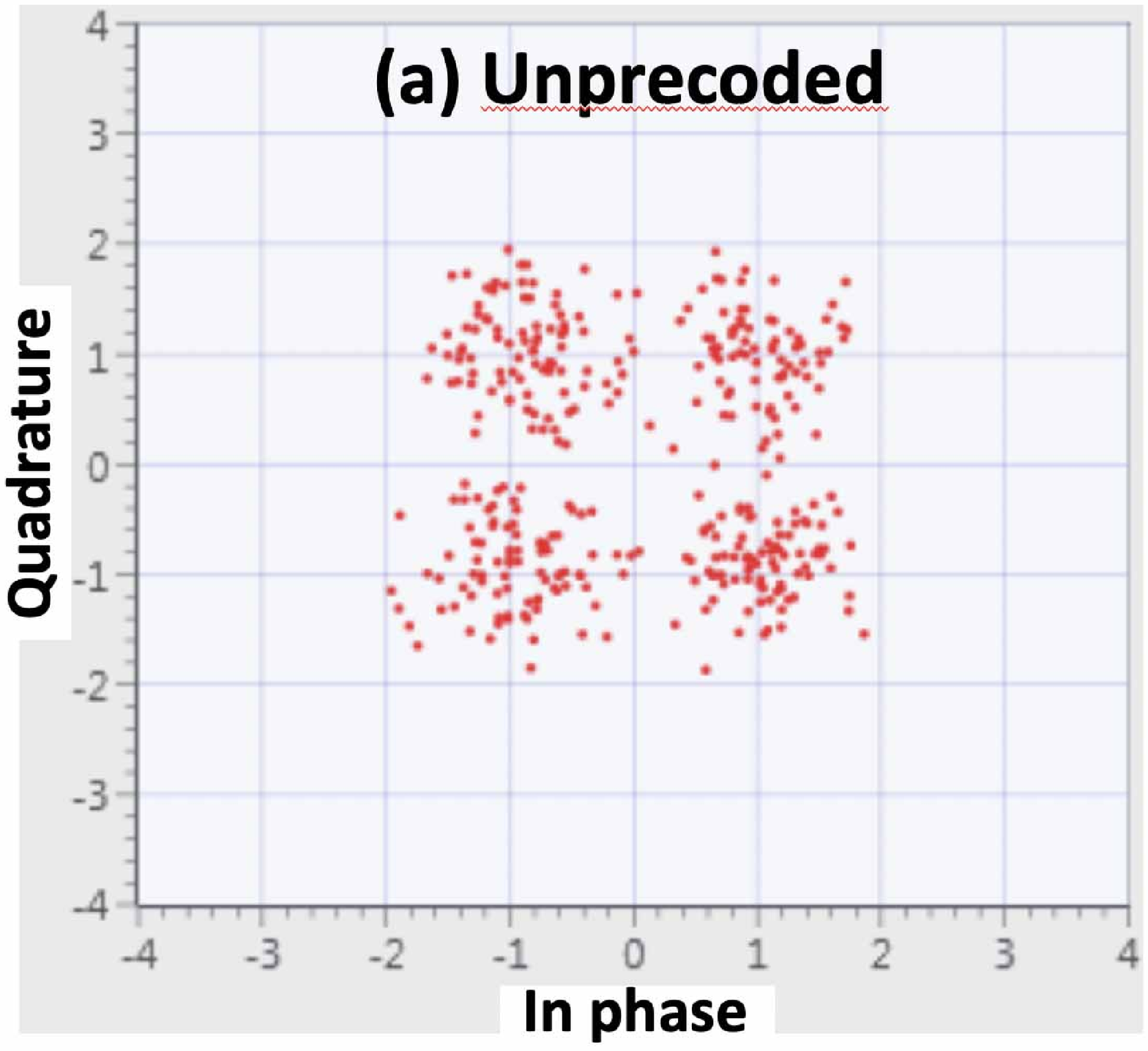}
\label{fig-ucl-3-1}
\par
\end{centering}
}
\subfloat[ZF precoded]
{
\begin{centering}
\includegraphics[width=4.2cm]{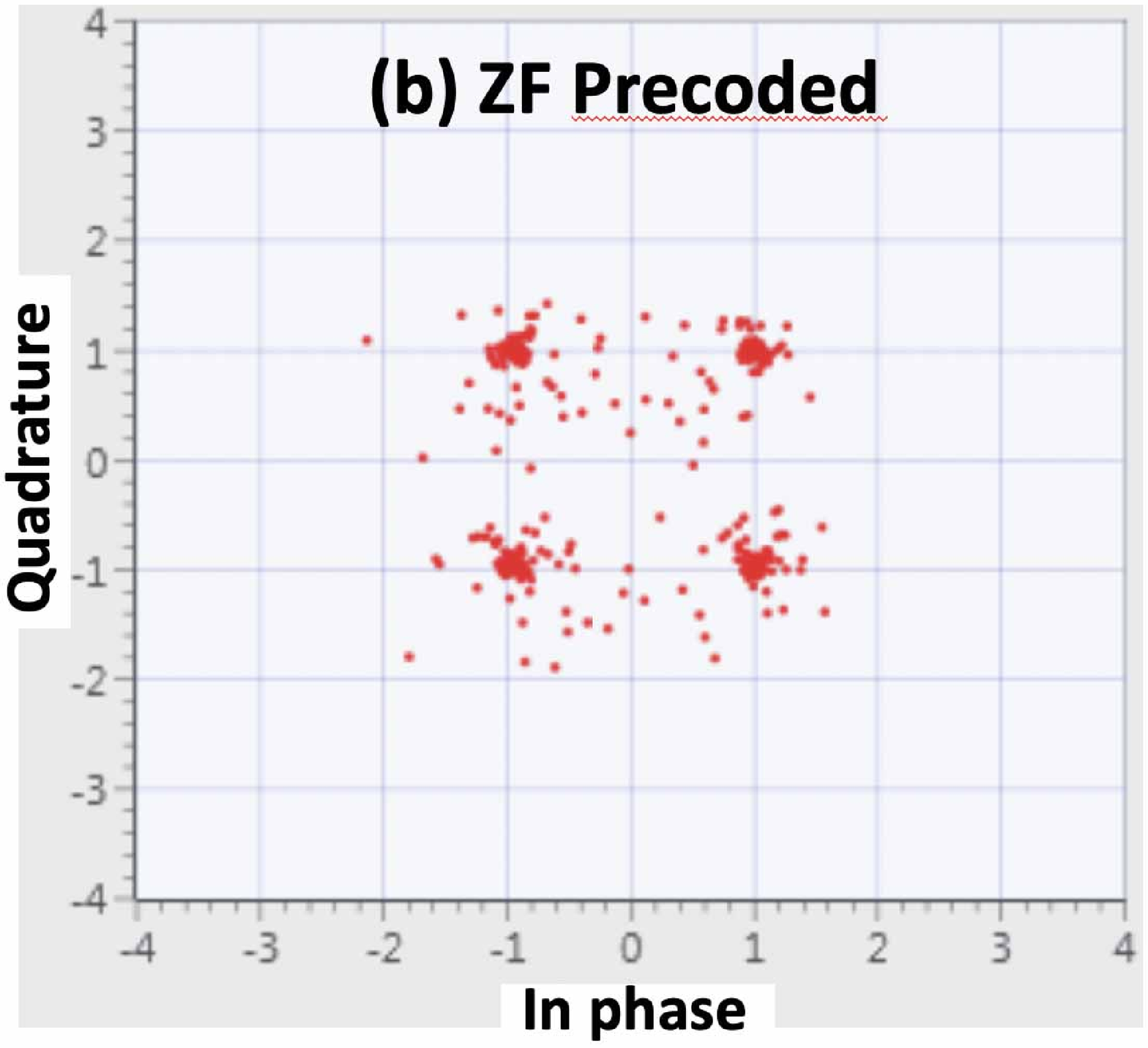}
\label{fig-ucl-3-2}
\par
\end{centering}
}
\subfloat[CI precoded]
{
\begin{centering}
\includegraphics[width=4.2cm]{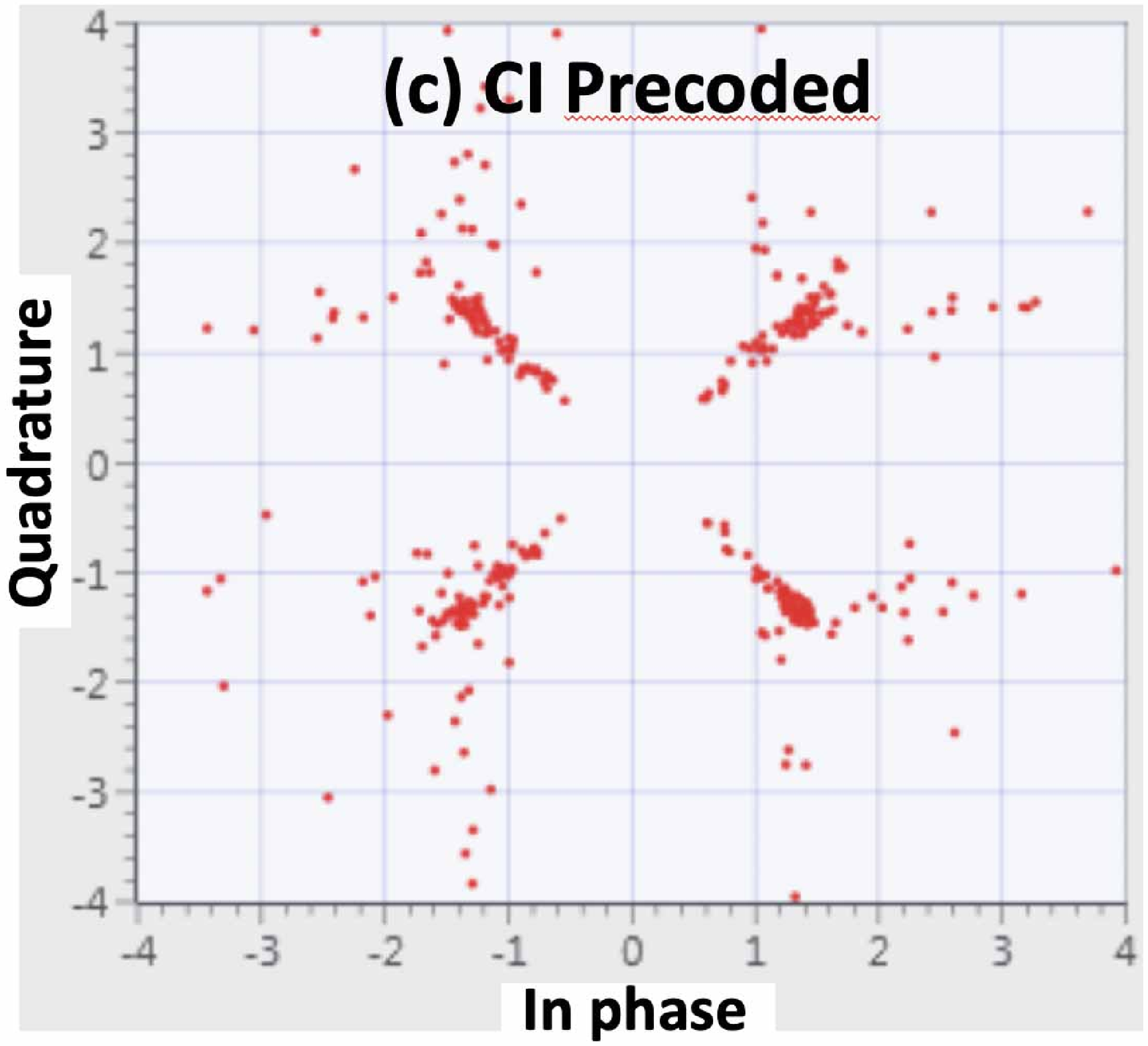}
\label{fig-ucl-3-3}
\par
\end{centering}
}
\par
\end{centering}
\caption{Experimentally obtained received constellation pattern for unprecoded and precoded system}
\end{figure*}

This real-time platform operates at 2.4GHz, and the other key parameters are shown in Table \ref{tab-ucl-1}, where the SEFDM approach with a bandwidth compression ratio $\upsilon=0.85$ is considered \cite{ucl-1}. The experimental environment is an indoor scenario, as shown in Fig.~\ref{fig-ucl-3}, there are a total number of 2 single-antenna receivers that are deployed and placed randomly, where the distance between the BS and the users is flexible and can be up to 9 meters. Due to this limited indoor space, in addition to the over-the-air experiments, a commercialized channel emulator Spirent VR5 \cite{ucl-4} is further adopted in order to compare precoding techniques at different SNR values. Assuming that the noise power mainly comes from the receiver side, the VR5 is connected to the 2 users via two separate cables. By configuring the VR5, the path loss and the power of the additive Gaussian noise can be set for each user. The receiver bandwidth is set to be 180kHz and the bandwidth of the additive noise is set to be 1.5625MHz.

\subsubsection{Frame Structure and CSI Estimation}
The transmit signal is generated at the BS with a designed frame structure and resource block structure \cite{ucl-2}, as shown in Fig. \ref{fig-ucl-2}. The 10ms frame is divided into 20 time slots with each occupying 0.5ms, and each time slot consists of 7 OFDM/SEFDM symbols. The first time slot is reserved for CSI estimation, while all the other time slots are used for data transmission. In this real-time platform, the BS obtains the downlink CSI by feedback from the receivers. In order to obtain an accurate estimate of the downlink spatial CSI for each antenna, time-domain orthogonal sounding reference signals (SRSs) are used in the 1st time slot of each frame. Since the considered system is a $2\times 6$ MU-MISO scenario, each user only needs to feedback 6 CSI coefficients, which is a reasonable overhead length.

To avoid interference between antennas during CSI estimation, the overhead of CSI estimation for each antenna is allocated at different symbol locations in time, as seen in Fig.~\ref{fig-ucl-2}. More specifically, the overhead of CSI estimation for the first antenna occupies the first OFDM/SEFDM symbol in one resource block, while keeping blank for the following 5 OFDM/SEFDM symbols. With other 5 antennas following the same principle, the overlapping interference can be avoided. The last OFDM/SEFDM symbol in this time slot is reserved for the downlink pilots, which are precoded and then sent simultaneously. These pilot signals are used to compensate for imperfect channel issues such as power normalization and imperfect time and phase synchronization \cite{ucl-5}.

\subsubsection{Experimental Validation}
The received constellation symbol results that are measured based on over-the-air transmission using this hardware platform are shown in Fig. 15, where the received symbols for the unprecoded system, ZF precoded system and CI precoded system are presented in Fig. \ref{fig-ucl-3-1}, Fig. \ref{fig-ucl-3-2}, and Fig. \ref{fig-ucl-3-3}, respectively. Without precoding, it is observed that the received symbols are scattered in the constellation due to the existence of self-created ICI by SEFDM signals. When ZF precoding is employed, the distribution of the received symbols becomes more focused at the four nominal QPSK points, as depicted in Fig. \ref{fig-ucl-3-2}, which means that adopting ZF precoding can improve the performance. When CI precoding is employed, as shown in Fig. \ref{fig-ucl-3-3}, a special received symbol pattern is observed, where we observe that the received symbols are pushed away from the detection thresholds, which follows the design principle of CI precoding. A significantly improved error rate performance can therefore be expected for CI precoded scenario over the unprecoded and ZF precoded case.

\begin{figure}[!b]
	\centering
	\includegraphics[scale=0.3]{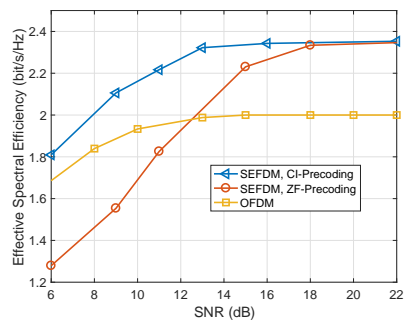}
	\caption{Experimental result for spectral efficiency, bandwidth compression ratio $\upsilon=0.85$}
	\label{fig-ucl-5}
\end{figure}

In Fig. \ref{fig-ucl-5}, the experimental effective spectral efficiency is depicted with an increasing SNR value, where the various SNR values are obtained by tuning the noise power in the VR5 channel emulator. The considered metric, effective spectral efficiency, is calculated as
\begin{equation}
{\text {SE}}_\text{eff}=\frac{1}{\upsilon}\left( {1 - {\text {BER}}} \right){\log _2}{\cal M}
\end{equation}
for $\cal M$-PSK. When QPSK is considered as in the experiment, ${\cal M}=4$ and ${\text {SE}}_\text{eff}$ becomes ${\text {SE}}_\text{eff}=\frac{2}{\upsilon}\left( {1 - {\text {BER}}} \right)$. Based on the result in Fig. \ref{fig-ucl-5}, we observe that in low-to-medium SNR region, the CI precoded SEFDM system achieves a significantly better performance over the ZF precoded and OFDM case. In the high SNR region, the OFDM case reaches its maximum effective spectral efficiency 2bits/s/Hz, because in this case the bandwidth compression ratio is 1. For the precoding-aided SEFDM results, we observe that both ZF precoded and CI precoded SEFDM approach can reach the maximum effective spectral efficiency 2.35bits/s/Hz, both of which outperform the OFDM case.

\subsection{DVB Standard-based CI Testbed}

\subsubsection{Platform Description}
\begin{figure}[!t]
\centering{\includegraphics[trim={0cm 0cm 0cm 0cm},width= 1\columnwidth]{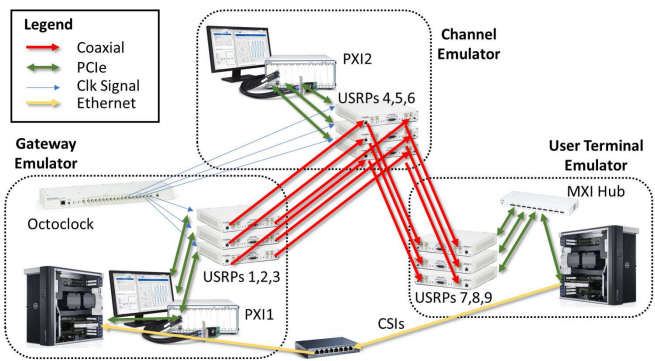}}
\caption{Implementation platform of the hardware demonstrator\label{fig:prinicpal_diagram}}
\end{figure}

A DVB standard fully-compatible hardware platform is hosted in University of Luxembourg (UL), which employs full frequency reuse in wireless communications named SERENADE \cite{lux-24}. The demonstrator uses the DVB-S2X standard \cite{lux-25} for signal transmission and reception, which includes a novel SLP technique in \cite{lux-3} that optimizes the precoding vectors per every modulated symbol vector, and software defined radios are used to build the testbed. This platform enables the design of a scalable architecture of the transmitter, channel emulator and UEs, as shown in Fig. \ref{fig:prinicpal_diagram}. The commercially available software defined radio platform developed by National Instruments (NI) has been employed for this task. The platform consists of two NI PXI (PCI EXtension for Instruments) 1085 chassis, which allow centralized connection of the set of the NI USRP (Universal Software Radio Peripheral) 2954R and FlexRIO (Reconfigurable IO) 7976R. The NI USRP and FlexRIO have integrated FPGA (Field-Programmable Gate Array) module Kintex-7 from Xilinx. 

\begin{table}[!b]
\caption{Parameters of the hardware demonstrator}
\label{experimentsets} 
\centering
\begin{tabular}{|l|c|}
 \hline
Parameter & Value \\
\hline 
Transmitter IQ channels & 6 \\
Sampling frequency & 1 MHz \\ 
Oversampling factor & 4  \\ 
Transmitter TX freq. & 1.21 GHz \\
Channel Emulator RX freq. & 1.21 GHz \\
Channel Emulator TX freq. & 960 MHz \\
User Terminal RX freq. & 960 MHz \\
Filter roll-off factor & 0.2, 0.15, 0.1, 0.05 \\
Forward Error Correction & yes \\
LDPC code rate & 1/2, 2/3, 3/4, 5/6   \\
 \hline 
\end{tabular}
\end{table}

The transmitter simultaneously transmits 6 precoded signals towards 6 user terminals through a $6 \times 6$ multi-beam channel emulator. The channel emulator acquires the transmitted signals, applies the impairments of the communication channel, Gaussian noise, and the multi-beam interference and transmits the signals to the UEs. Each UE estimates ${\mathbf{h}}_k$ based on the orthogonal Walsh-Hadamard pilot sequences and reports the estimated values to the transmitter through a dedicated feedback channel over an Ethernet link. The transmitter uses this CSI and modulated symbols to compute the precoding matrix. In the full frequency reuse scenario, the multi-user interference is mitigated by precoding techniques. Table \ref{experimentsets} summarizes the operational parameters of the demonstrator.

\begin{figure}[!t]
\centering{\includegraphics[trim={0cm 0cm 0cm 0cm},width= 1\columnwidth]{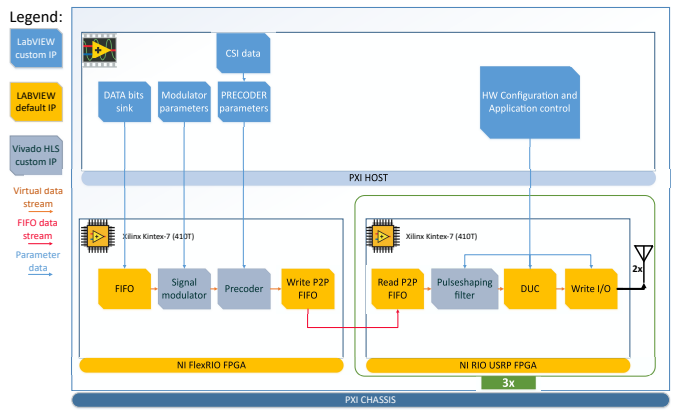}}
\caption{Block diagram of the transmitter
\label{fig:tx_diagram}}
\end{figure}

The transmitter operates with a central NI FlexRIO FPGA and three NI USRP nodes. Fig. \ref{fig:tx_diagram} shows the logical connections between the NI FlexRio, the NI USRP nodes and the controller (NI PXI HOST), where the upper blue section represents the processes implemented in the host computer and the lower yellow section represents the blocks implemented in the FPGA for fast processing. The transmitter transmits symbols modulated according to the DVB-S2X standard \cite{lux-25,lux-28}, and the streams are jointly precoded by the PRECODE FPGA block. The precoder block multiplies 6 symbols from a single time slot with the precoding matrix ${\mathbf W}$ and sends the streams to the NI USRP nodes. Subsequently, each node oversamples the streams and transmits them through digital up converted (DUC) to the RF domain at a desired carrier frequency.

\begin{figure}[!b]
\centering{\includegraphics[trim={0cm 0cm 0cm 0cm},width= 1\columnwidth]{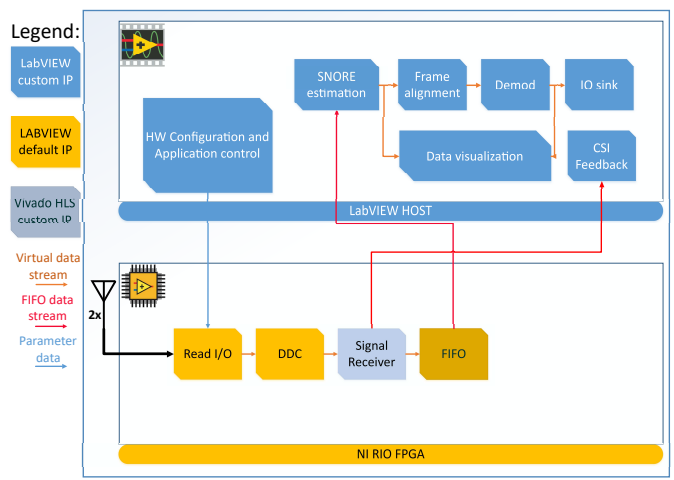}}
\caption{Block diagram of the UE, where two input RF chains are present in a single USRP \label{fig:rx_diagram}}
\end{figure}

Fig. \ref{fig:rx_diagram} shows an architectural block diagram of the UE implementation. A single USRP RIO FPGA unit is capable to simultaneously receive and process two information signals. The processing includes frequency acquisition, matched filtering, time synchronization, frame synchronization, fine phase tracking and CSI estimation. All the frame fields and the CSI information are passed to the host computer for further processing, and the host computer reports the CSI information to the transmitter using a custom feedback channel.  


\begin{table}[!t]
\caption{HDL Core Resource Occupation on Kintex-7 (xc7k410TFFG-2)}
\label{table:fpga_resource}
\centering
\scalebox{0.9}
{
\begin{tabular}{|c||c|c|c|c|}
\hline
$K=N_t$ &  DSP48E & Slices & look-up tables & Effective baud rate\\ 
\hline
2 & 16 & 479 & 216 & 166~MBd \\
6 & 72 & 2019 & 2488 & 498~MBd \\
12 & 288 & 9891 & 9938 & 996~MBd \\
16 & 512 & 11683 & 19010 & 1.33~GBd \\
20 & 800 & 21187 & 27602 & 1.66~GBd \\
\hline
Available & 1540 & 508400 & 254200 & \\
\hline
\end{tabular}
}
\end{table}

\subsubsection{FPGA Accelerated Closed-Form SLP}
A complete FPGA accelerated design has been developed in \cite{lux-4} to efficiently solve the SLP problem $\mathcal{P}_{\text {CI-VP}}$ in \eqref{sec3:slpminproblem}. The design is built using Vivado High-Level Synthesis (HLS) to program a closed-form algorithm into HDL core and integrate the design into an FPGA. The estimations for the resource utilization and symbol throughput of the FPGA core are demonstrated in Table \ref{table:fpga_resource}. 

The FPGA core hits the target of the symbol rate of 83 MSymbols per second per each beam. The motivation behind the target is the new symbols rates, which are considered in the DVB-S2X standard \cite{lux-25}. The estimated resource consumption by the core design are calculated for various numbers of transmitting antennas and UEs, where $K = N_t = $ 2, 6, 12, 16 and 20. For all the scenarios, the core is optimized to operate at a 166 MHz clock ($\approx 6$ ns per cycle) with a cycle interval 2. The clock allows to operate at the $166 \text{ MHz} / 2 = 83 \text{ MSymbols per second}$ symbol rate per beam. For the case of a $20 \times 20$ MU-MISO case, the design utilizes around 50 percent of the DSP blocks available at the given FPGA model (xc7k410TFFG-2) and the effective baud rate of the core reaches 1.66~GSymbols per second.

\begin{figure}[!b]
\centering{\includegraphics[trim={0cm 0cm 0cm 0cm},width= 0.9\columnwidth]{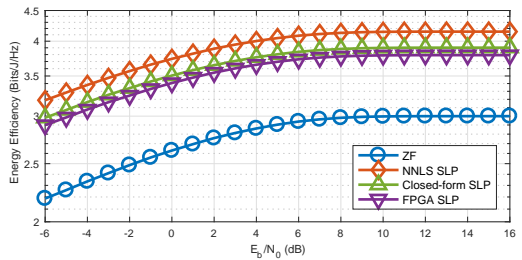}}
\caption{Energy efficiency of ZF, FPGA SLP, CVX SLP, and closed-form SLP, 8-PSK\label{fig:ee_8psk}}
\end{figure}

In Fig. \ref{fig:ee_8psk}, the energy efficiency is benchmarked as a function of the energy per bit to noise power spectral density ratio ($ {E_{\text b}/N_0} = 10\log_{10} (\frac{1}{3\sigma^2})$) for the ZF, Fast NNLS SLP and the closed-form SLP algorithm in \cite{lux-4} running in MATLAB and on FPGA core. The benchmark is performed on 8-PSK modulation symbols and averaged over 50 iterations of $6 \times 6$ channel matrix with a condition number, defined as $\kappa_2( \mathbf{H} ) = \left\|{\mathbf{H}} \right\| _2 \left\|{{\mathbf H}^{-1} } \right\|_2$, fixed to 18. The difference between the performance of the Fast NNLS and the closed-form algorithm running on MATLAB is around 2.5dB due to the approximation method used in the closed-form solution. The additional 1dB difference can also be observed between the MATLAB and FPGA implementations of the closed-form algorithm, which is due to the losses in fixed-point arithmetic, which are calculated on the hardware.

\subsubsection{Experimental Validation}
\begin{figure}[!t]
\centering
\subfloat[Transmit powers on the output antenna ports]{%
  \includegraphics[trim={0.7cm 0 1.3cm 0},width=0.8\columnwidth]{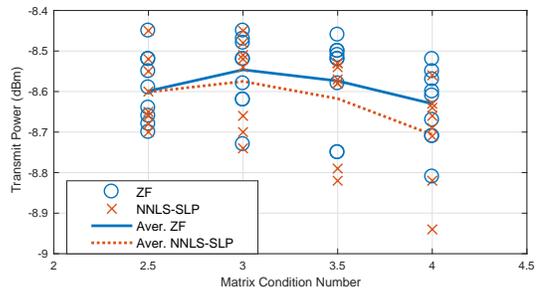}%
  \label{fig:slp_reductions}
}

\subfloat[Received powers on the input antenna ports]{%
  \includegraphics[trim={0.7cm 0 1.3cm 0},width=0.8\columnwidth]{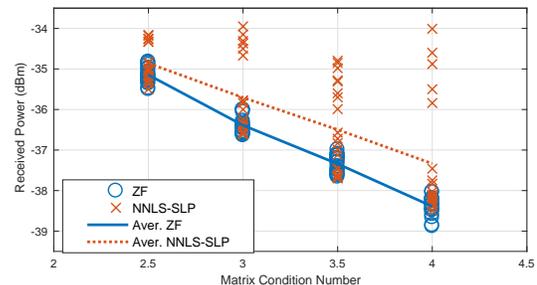}%
  \label{fig:slp_gains}
}

\caption{Transmit/Received powers in the testbed \cite{lux-26}}
\end{figure}

The experimental measurements of the output power are conducted using the testbed, as shown in Fig. \ref{fig:slp_reductions}. It is evident that ZF technique generates signals with a higher averaged total transmit power than the SLP technique. The reduction in the transmit power by SLP becomes more significant as the matrix condition number increases compared to ZF.

In Fig. \ref{fig:slp_gains}, we can observe that the received power for ZF precoding is not constant for a given channel condition number. These variations come from the imperfections in the actual hardware implementation. Some of these imperfections are the limited accuracy in the CSI estimation, and its quantization error. Nevertheless, since these imperfections have the same impact on the ZF and the SLP, we can observe that the SLP approach still has gains in the received power. These gains become more frequent as the matrix condition number is increased. There are particular channel realizations in which the SLP performs the same as ZF for both receivers, and others realizations in which the perturbation vector for the optimal symbols is not strong enough. 

\section{Open Problems and Challenges}
In this section, we discuss some remaining open problems and challenges that are to be addressed to motivate further research efforts in the field of CI exploitation and SLP techniques.

\subsection{Communication Theoretical Aspects}
While CI-based precoding techniques have been extensively studied based on optimization, the fundamental extent of performance improvements from exploiting the CI effects is unknown, since there still lack information-theoretic studies on this topic to provide performance benchmarks. As discussed in this paper, since the precoding matrix for CI-based techniques is modulation dependent, the Shannon channel capacity expression based on the assumption of Gaussian inputs is no longer valid. Instead, the capacity needs to be analyzed based on the complex finite-constellation approaches \cite{capacity-1}\nocite{capacity-2}-\cite{capacity-3}. Therefore, one open problem is the development of such an analytical framework that is able to offer a benchmark performance, towards which optimization-based CI techniques are designed in the future.

\subsection{Adaptive Modulation}
Another open problem is how CI approaches can be extended to adaptive modulation scenarios. In the current literature, it has been discussed how CI schemes currently available can be tailored for different modulations, and particularly how various techniques can enhance the CI effect based on the specific modulation format of the data. Nonetheless, typical communication systems in practice do not use a fixed modulation format, but rely on adaptive modulation schemes which adaptively adjust the modulation based on the estimated SINR (SINR estimation is further discussed below). This is eventually connected to the scheduling aspects of wireless communication systems as well. Therefore, an open challenge is to account for adaptive modulation schemes in the formulation of CI problems, which would enhance their flexibility and pave the way to practical applications. 

\subsection{SINR Estimation}
It has been mentioned how typical communication systems use adaptive modulation schemes based on the estimated SINR. Thus, it is particularly important to use a reliable SINR estimation scheme at the receivers \cite{lux-29}, in order to feed the information back to the transmitter and choose the appropriate modulation type. However, the SINR estimation is an additional challenge when CI schemes are employed. In fact, as mentioned earlier, CI-based designs lead to distorted received constellations, with the outer constellation points pushed further away from their respective detection regions. As a consequence, there is an imbalance between the instantaneous SINR of the outer constellation points (which can be very high due to CI) and that of the inner constellation points. Such imbalance is not taken into account by current SINR estimation schemes, which only evaluates an average SINR. Therefore, an open challenge in the context of CI precoding is to propose novel SINR estimation methods that are able to take into account the imbalance across different constellation points.

\subsection{Channel-Coded CI}
Another remaining open challenge is related to the optimal design of CI approaches accounting for channel coding, or forward error correction (FEC) schemes. While the use of FEC over symbol-level precoded waveforms has been assessed in some works (e.g. \cite{lux-11}), it is well known that the channel coding optimality (specifically when soft detection is employed) is directly related to the reference constellation considered to calculate the log-likelihood ratios. Since CI precoding is aimed at distorting the received constellation in order to gain benefits from CI effects, CI precoding will not only affect the performance of FEC, as extensively discussed, but also affect the distribution of the received signals, which then requires a redesign of the decoder. Therefore, an open problem is to optimize the transmission by jointly accounting for the design of CI precoding, modulation and FEC.

\subsection{Synchronization in DA Systems}
It is well known that DA systems require strict synchronization between the distributed nodes to achieve the promised performance \cite{syn-1}, \cite{syn-2}. When SLP techniques instead of traditional block-level precoders are applied at these DA systems, which require accurate synchronization on a symbol level, we conjecture that timing misalignment will have a much more pronounced impact on symbol-level compared to block-level approaches. Therefore, it is still unknown how current synchronization approaches should be adapted to emerging SLP techniques, which remains another open problem.

\subsection{Waveform Design}
Even though SLP has been applied in a wide range of wireless systems, use cases and architectures, its impact on waveform design is still not well understood. In principle, SLP enables the management of interference between symbols as long as we have an accurate information about the inter-symbol channel. This entails that SLP can be applied for a wide range and even a combination of interference types such as ISI, adjacent-channel interference (ACI) and multi-user interference. In some cases, the application becomes even more straightforward, since the inter-symbol channel depends on the digital transceiver (e.g. filtering, sampling) and is static in contrast to dynamic propagation channels. The first study in this direction looked at the combination of ISI and multi-user interference in FTN systems \cite{lux-11}, while SLP waveform design is still an open question for multi-carrier transmission schemes with ACI, time-frequency packing, channels with memory, etc.

\section{Conclusion}
In this paper, we have provided an extensive tutorial on interference exploitation techniques. The characterization of interference shows that interference classification into constructive or destructive is dependent on the information of both the channel and the data symbols, which implies that interference exploitation techniques need to operate on a symbol level. Moreover, the mathematical condition for achieving CI obtained via the geometry of the constellation points leads to convex formulations of PM and SB problems, which are typically more difficult to handle in the interference-reduction scenarios. The performance gains for interference exploitation in terms of transmit power savings and error rate are presented numerically. We have also discussed the extension of CI precoding to a variety of wireless communication scenarios and included the description of a proof-of-concept testbed, where CI precoding also exhibits notable performance improvements.

\ifCLASSOPTIONcaptionsoff
  \newpage
\fi

\bibliographystyle{IEEEtran}
\bibliography{refs_list}

\end{document}